\documentclass[iop,twocolumn,revtex4,numberedappendix,apj]{emulateapj}    

\usepackage{natbib}
\usepackage{amsmath}
\usepackage{color}
\usepackage{longtable}
\usepackage{graphicx}
\usepackage{lscape}
\usepackage{textcomp}

\usepackage{hyperref}

\shorttitle{YSOs in Orion.}

\newcommand{\kms}{\,km\,s$^{-1}$}

\newcommand{\ergps}{erg\,s$^{-1}$}
\newcommand{\mum}{\,$\mu$m}

\newcommand{\Msun}{$M_{\odot}$}

\newcommand{\rev}{ }
\newcommand{\newrev}{ }
\newcommand{\newnewrev}{ }

\newcommand{\fig}{Fig.\,}
\newcommand{\sect}{Sect.\,}

\newcommand{\LiI}{Li\,I\,$\lambda$6708\AA}

\begin{document}

\title{NGC~1980 is not a foreground population of Orion: Spectroscopic survey of young stars with low extinction in Orion~A}

\author{Min~Fang\altaffilmark{1}, Jinyoung~Serena~Kim\altaffilmark{1}, Ilaria~Pascucci\altaffilmark{2}, D\'aniel Apai\altaffilmark{1, 2}, Lan Zhang\altaffilmark{3, 4},  Aurora~Sicilia-Aguilar\altaffilmark{5}, Miguel~Alonso-Mart\'{i}nez\altaffilmark{6}, Carlos~Eiroa\altaffilmark{6}, Hongchi~Wang\altaffilmark{7}}
\altaffiltext{1}{Department of Astronomy, University of Arizona, 933 North Cherry Avenue, Tucson, AZ 85721, USA}
\altaffiltext{2}{Department of Planetary Sciences, University of Arizona, 1629 East University Boulevard, Tucson, AZ 85721, USA}
\altaffiltext{3}{Key Lab of Optical Astronomy, National Astronomical Observatories, CAS, 20A Datun Road, Chaoyang District, 100012 Beijing, China}
\altaffiltext{4}{CAS South America Center for Astronomy, Camino El observatorio \#1515, Las Condes, Santiago, Chile}
\altaffiltext{5}{SUPA, School of Physics and Astronomy, University of St Andrews, North Haugh, St Andrews KY16 9SS, UK}
\altaffiltext{6}{Departamento de Fisica Teorica, Facultad de Ciencias, Universidad Autonoma de Madrid, 28049 Cantoblanco, Madrid, Spain}
\altaffiltext{7}{Purple Mountain Observatory and Key Laboratory of Radio Astronomy, Chinese Academy of Sciences, 2 West Beijing Road, 210008 Nanjing, China}

\begin{abstract} 
We perform a spectroscopic survey of the foreground population in Orion~A  with MMT/Hectospec. We use these data, along with archival  spectroscopic data and photometric data, to derive spectral types, extinction values, and masses for 691 stars. Using the Spitzer Space Telescope data, we characterize the disk properties of these sources. We identify 37 new transition disk (TD) objects, one globally depleted disk candidate, and 7 probable young debris disks.{\rev We discover an object with a mass less than 0.018--0.030\,\Msun, which harbors a flaring disk.} Using the H$\alpha$ emission line, we characterize the accretion activity of the sources with disks, and confirm that fraction of accreting TDs is lower than that of optically thick disks (46$\pm$7\% versus 73$\pm$9\%, respectively). Using kinematic data from the Sloan Digital Sky Survey and APOGEE  INfrared Spectroscopy of Young Nebulous Clusters program (IN-SYNC), we confirm that the foreground population shows similar kinematics to their local molecular clouds and other young stars in the same regions. Using the isochronal ages, we find that the foreground population has a median age around 1--2\,Myr, which  is similar to the one of other young stars in Orion~A. Therefore, our results argue against the presence of a large and old foreground cluster in front of Orion~A. 
\end{abstract}

\section{Introduction}
The Orion complex is the most active star-forming region in the solar neighborhood. The age range  ($<$1--10\,Myr), the diversity of local environments, and  the distance \citep[$\sim$414\,pc,][]{2007A&A...474..515M} of this complex make it an ideal laboratory to study challenging questions in star formation, e.g., star formation history, initial mass function, disk evolution \citep[see e.g.][]{1997AJ....113.1733H,2007ApJ...671.1784H,2007ApJ...662.1067H,2009A&A...504..461F,2009ApJ...694.1423L,2010ApJ...722.1226H,2012ApJ...752...59H,2013ApJ...764..114H,2013ApJS..207....5F,2014ApJ...794...36H}. 

The entire Orion complex shows  evidence for multiple star formation episodes, and the current, most active star formation is located in Orion~A and B. Orion~A, located in the southern part, consists of the most massive cluster in this region, the Orion nebula cluster, and several medium-size clusters, e.g., OMC~2, OMC~3, and Lynds~1641 (L1641). In L1641, about half of the young stellar objects (YSOs) are formed in isolation and others are formed in small aggregates \citep{2013ApJS..207....5F}. Orion~B is in the northern part of Orion and comprised of four clusters NGC~2023, 2024, 2068, and 2071. There is also a large population of off-cloud stellar groups aged 3-30 Myr, including the Orion~OB1a and 1B associations \citep{2005AJ....129..907B,2007ApJ...661.1119B}.  The understanding of star formation in  the Orion complex was further complicated by the discovery of a foreground population aged 4--5\,Myr to Orion~A \citep{2012A&A...547A..97A,2014A&A...564A..29B}. This presumably older population is centered on NGC\,1980 ($\iota$~Ori), and extends northward to Orion nebula cluster and NGC~1981, and eastward to L1641, thus contaminating the young populations in these regions. 

{\rev In \citet{2012A&A...547A..97A}, the foreground population was proposed by searching for the stars with low or no extinction in the field of Orion~A. The surface densities of this population show a well-defined peak coinciding with NGC~1980, and several less distinct peaks around other star-forming regions in Orion~A \citep{2012A&A...547A..97A,2014A&A...564A..29B}. In $J-H$~vs.~$H-K_{\rm s}$ color-color diagram, the stars in the population show near-infrared colors consistent with B- to M-type stars \citep{2012A&A...547A..97A}. Therefore, \citet{2012A&A...547A..97A} propose that the NGC~1980 cluster is a foreground population seen in projection against the Orion A cloud.  \citet{2012A&A...547A..97A} estimate the age of the  NGC~1980 cluster on the basis of the median spectral energy distribution (SED) and the age of a massive star, $\iota$~Ori, which is 4--5\,Myr. However, the previous studies are all based only on photometric data, and lacks  spectroscopic follow-up. In this work, we provide a spectroscopic study of the putative foreground population, and address the issue of its age and other properties. Our results suggest that NGC~1980 is actually associated with the Orion~A cloud and has an age similar to other star-forming regions in Orion~A. We organized the paper as follows: in $\S$2 we will describe our observations and data reduction, in $\S$3 we will delineate our data analysis, we will present our results in $\S$4, followed by a discussion in $\S$5, and we will summarize our results in $\S$6.   

\begin{figure*}
\begin{center}
\includegraphics[angle=0,width=1.5\columnwidth]{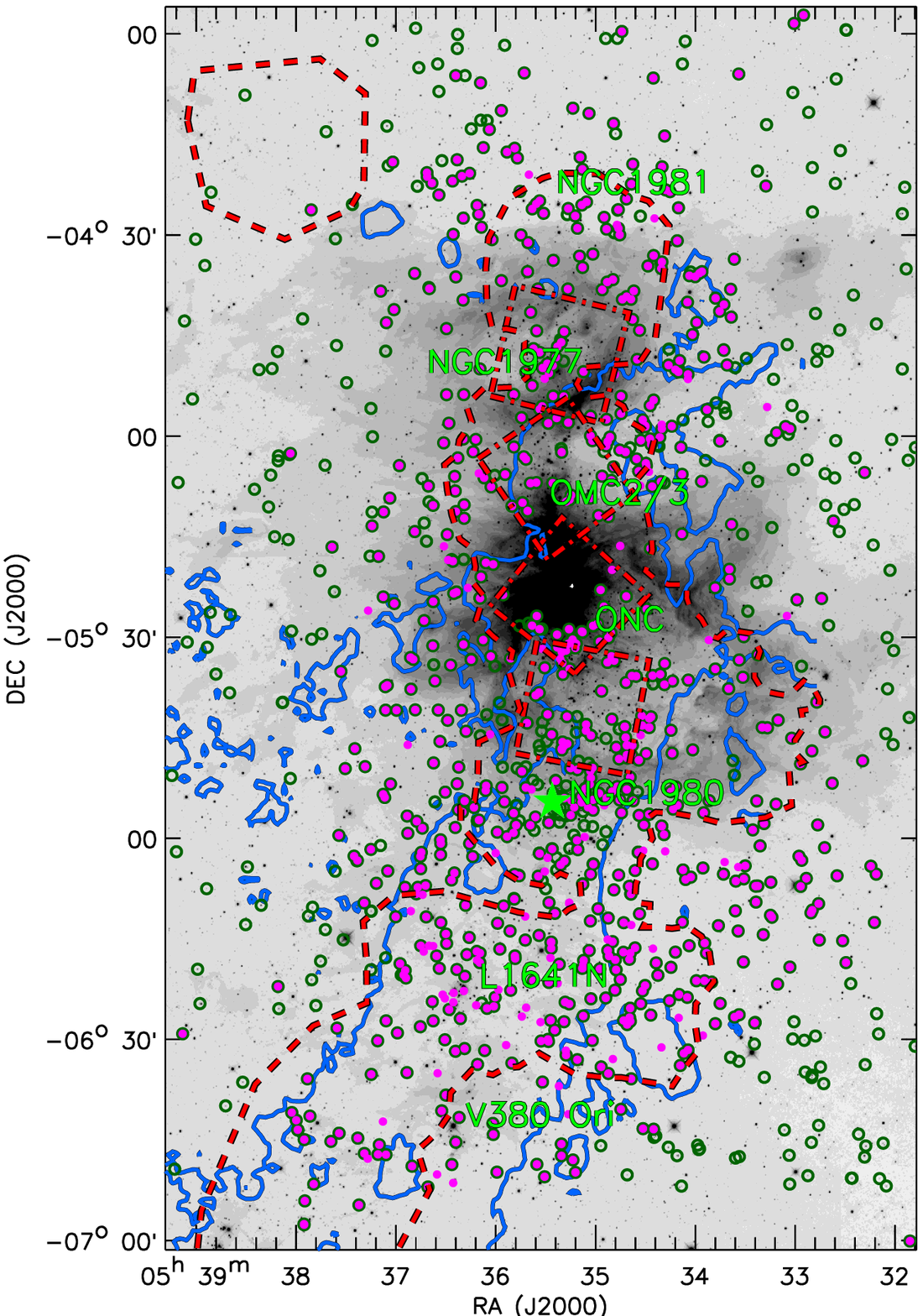}
\caption{The  WISE 3.4 micron image of Orion\,A. The circles ({\newnewrev open dark green and magenta filled}) show all the stars with probability more than 99\% of being members of NGC 1980 and brighter than 19\,mag in $r$ band. The {\newnewrev magenta} filled circles mark the stars with spectroscopic data. The {\newnewrev green} filled star symbol mark the position of the massive star $\iota$~Ori. The {\newnewrev blue} solid-line contours are the $^{13}$CO integrated intensity from \citet{1987ApJ...312L..45B} with a contour level of 5~K\,\kms ($5\sigma$). The {\newnewrev red} dash lines enclose the field of view (FOV) of the XMM observations, and the {\newnewrev red} dash-dotted line boxes are for the FOVs of the Chandra observations.}\label{Fig:Orion}
\end{center}
\end{figure*}

\section{Observations and data reduction}

\subsection{Spectroscopic observations}

\begin{table*}
\caption{Hectospec observation logs.}\label{Tab:obs_log}
\centering
\begin{tabular}{clccccccccccccccccccccccccccccccccc}
\hline\hline
        &          & RA      &DEC     & \multicolumn{2}{c}{Exposure}   \\
\cline{5-6}
        &          & (J2000) &(J2000)      &Obj      &sky           \\
Config. & Obs-date (UT) & (h:m:s) &(d:m:s) &(minutes)&(minutes)\\
\hline 
  1  & 2016 Feb 4&$+$05:33:54.6 & $-$06:01:19   &3$\times$15   &1$\times$15   \\
  2  & 2016 Feb 4&$+$05:35:51.5& $-$05:48:52   &3$\times$15   &1$\times$15  \\
  3  & 2016 Feb 6&$+$05:35:15.9 & $-$04:32:58   &3$\times$15   &1$\times$15  \\
  4  & 2016 Feb 8&$+$05:35:32.2 & $-$05:22:39   &3$\times$15   &1$\times$15 \\
\hline
\end{tabular}
\end{table*}

\subsubsection{Target selection}

\citet{2014A&A...564A..29B} lists 1895 sources with probability more than 99\% of being members of NGC~1980. These stars are distributed over several tens of square degrees with the stellar surface densities peaking around the Orion molecular cloud. Our targets for the spectroscopic study of NGC~1980 are selected from \citet{2014A&A...564A..29B}, and are limited to the vicinity of the Orion cloud. In Fig.~\ref{Fig:Orion}, we show the regions we studied. In this area, there are 1,275 sources with probabilities more than 99\% of being memebers of NGC~1980. Hereafter, these sources are considered to be high-confidence cluster members. We search for the available spectral types for these high-confidence cluster members in the literature \citep{1997AJ....113.1733H,2000AJ....119.3026R,2001AJ....121.1676R,2013AJ....146...85H,2009A&A...504..461F,2013ApJS..207....5F,2012ApJ...752...59H}, and in our unpublished data from our previously spectroscopic survey with MMT/Hectospec and WHT/AF2 (Fang, M., et al. 2017, in preparation). We obtain spectral types for 330 sources from published literature, and for 172 sources from our unpublished datasets. Due to limited observing time, we only selected a subsample of high-confidence cluster members for spectroscopic observations with MMT/Hectospec. In Fig.~\ref{Fig:CMD}, we show the $r~vs.~r-i$ color-magnitude diagram for high-confidence cluster members, as well for sources with known spectral types and stars we selected for the  MMT/Hectospec observations. {\rev We calculate the 3\,Myr isochrone in the $r-i~vs.~r$ color-magnitude diagram for the pre-main-sequence (PMS) stellar evolutionary models of \citet{2015A&A...577A..42B}, using the BT-Settl model atmospheres \citep{2011ASPC..448...91A}, with solar abundances from \citet{2009ARA&A..47..481A} \footnote{A discussion on the BT-Settl models with two types of solar abundances \citep{2009ARA&A..47..481A,2011SoPh..268..255C} can be found in \sect~\ref{Sect:colors}.}. As a comparison, we show the isochrone in Fig.~\ref{Fig:CMD}, assuming a distance 414\,pc \citep{2007A&A...474..515M}. Most of our targets are located above this isochrone, indicating that they are younger than 3\,Myr.} {\newrev However, the ages of our targets could be older, if they lie much closer than Orion~A. If we assume an age of $\sim$4--5\,Myr for our targets as suggested in \citet{2012A&A...547A..97A}, they should be located at $\sim$210\,pc to match their position in the color-magnitude diagram (see Fig.~\ref{Fig:CMD}). However, \citet{2012A&A...547A..97A} argue that the distances of these sources would not be substantially different from the ones in Orion, and might be around 400\,pc, since they show similar kinematics to the Orion~A molecular cloud. Therefore, the ages and the distance for these sources assumed in \citet{2012A&A...547A..97A}  are inconsistent.}

\subsubsection{Spectroscopic observations and data reduction}

We performed a low-resolution spectroscopic survey of the stellar population in NGC~1980  with the Hectospec multi-object spectrograph, capable of taking a maximum of 300 spectra simultaneously. We used  the 270 groove mm$^{-1}$ grating and obtained spectra in the 3700--9000\,$\AA$ range with a spectral resolution of $\sim$5\,$\AA$.  The targets are selected from \citet{2014A&A...564A..29B}, and have probabilities $\geq$99\% as members of NGC~1980. Our $\sim$300 sources are distributed at 4 pointings. The data were taken in February, 2016. Table~\ref{Tab:obs_log} lists the observational logs.

\begin{figure*}
\begin{center}
\includegraphics[angle=0,width=1.9\columnwidth]{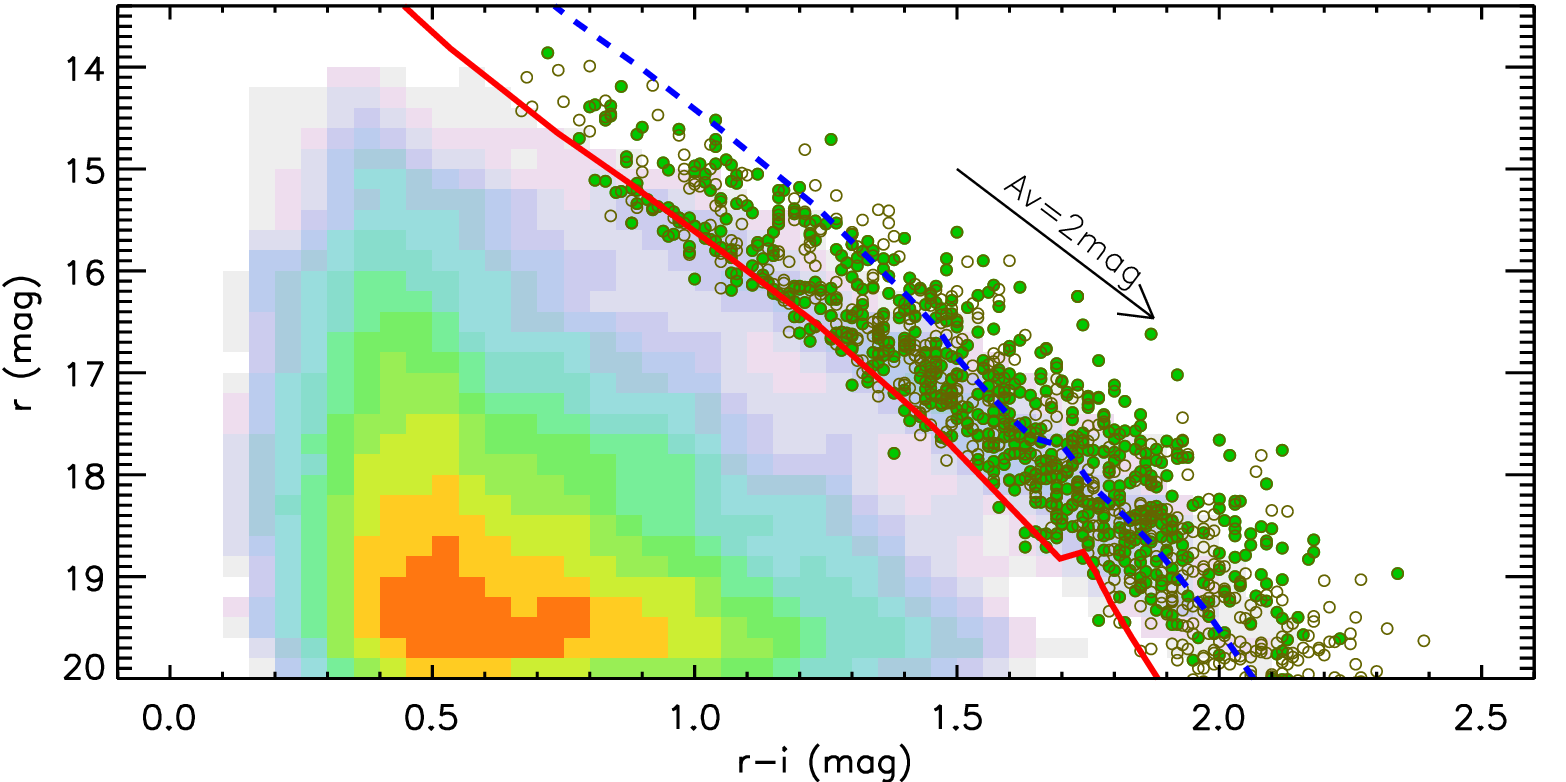}
\caption{Hess $r-i$ vs. $r$ color-magnitude  diagram for all the stars in the field of Orion from \citet{2014A&A...564A..29B}. The circles (open and filled) show all the high-confidence members of NGC~1980 in the field shown in Fig.~\ref{Fig:Orion}, and the filled circles are for those with spectroscopic data. The solid line shows the 3\,Myr isochrone at an distance of 414\,pc calculated for the PMS evolutionary models of \citet{2015A&A...577A..42B}, using the BT-Settl model atmospheres \citep{2011ASPC..448...91A}, with the solar abundances from \citet{2009ARA&A..47..481A}. {\newrev The dashed line is similar to the solid line, but for a 4.5\,Myr isochrone at a distance of 210\,pc.} The arrow show the extinction vector with visual extinction of 2~mag.}\label{Fig:CMD}
\end{center}
\end{figure*}

We use standard IRAF routines to reduce the Hectospec data according to standard procedures. We perform flat-field correction and extract the spectra using dome flats with the IRAF task \textit{\textbf{dofibers}} in the package \textit{\textbf{specred}}. The wavelength solution for Hectospec is obtained with HeNeAr comparison spectra, using the IRAF task \textit{\textbf{identify}} and \textit{\textbf{reidentify}} in the package \textit{\textbf{specred}}.  We calibrate the spectra with a wavelength solution constructed using the IRAF task \textit{\textbf{dispcor}} under the package \textit{\textbf{specred}}. For each pointing, we took 3 exposures for the science targets, and one exposure for the sky by shifting the telescope by $\sim$5 arcseconds. We extracted the spectra for each exposure. Finally, we obtained the spectra for each target and the corresponding sky spectra close to this target. We subtracted the sky from the spectra of each target, and combined the sky-subtracted spectra into one final spectrum.

\subsection{Photometric data}

The photometric data used in this work are mainly from  \citet{2014A&A...564A..29B}, in which they combine the photometric data from the Sloan Digital Sky Survey \citep[SDSS,][]{2000AJ....120.1579Y}, CTIO/DECam,  the Two-Micron All Sky Survey \citep[2MASS, ][]{2006AJ....131.1163S}, the UKIRT Infrared Deep Sky Survey \citep[UKIDSS,][]{2007MNRAS.379.1599L}, and the AAVSO Photometric All-Sky Survey \citep[APASS,][]{2016yCat.2336....0H}, and present a catalog with photometry in the $grizYJHKs$ bands. The photometric data from  CTIO/DECam are in $grizY$ bands, and have been calibrated in flux by cross-matching the common stars in the SDSS~DR9 catalog in $griz$ bands, and in the UKIDSS catalog in $Y$ band \citep[see][]{2014A&A...564A..29B}. We complement them with the Spitzer data from \citet{2012AJ....144..192M}, and Wide-field Infrared Survey Explorer \citep[WISE,][]{2010AJ....140.1868W}. These infrared data is used to characterize the circumstellar disks.

\subsection{X-ray data}
Low-mass young stars usually present strong magnetic activity, and show two to three orders of magnitude brighter X-ray emission than the field population \citep{1999ARA&A..37..363F,2007prpl.conf..313F}. Thus X-ray data can be used to distinguish young stars from field stars in star-forming regions, particularly diskless (Class III) stellar population \citep{2007prpl.conf..313F}. Part of the region studied here has been observed by the X-ray space telescopes XMM-Newton and Chandra. The sky coverage of the XMM (Proposal IDs: 004956, 008994, 009300, 011259, 011266, 013453, 040657, 050356, 060590, and 069020) and Chandra (Proposal IDs: 01200704, 03200289, and 04200331) observations are shown in Fig.~\ref{Fig:Orion}. The X-ray data will be used to characterize the youth of our sources.

\section{Data analysis}

\subsection{Spectral classification}
 The spectral types of young stars are usually obtained by classifying their observed spectra. This is typically done by building a relation between the strength of spectral features and the spectral types of dwarf stars, and apply them to the young stars \citep{1997AJ....113.1733H,2004AJ....127.1682H}. Uncertainties on the spectral classification can be several subclasses depending on the spectral features and/or ranges used for the determination. One typical example is TW\,Hya, of which spectral type range from K6 to M2.5 \citep{1989ApJ...343L..61D,2006A&A...460..695T,2013ApJS..208....9P,2011ApJ...732....8V,2014ApJ...786...97H}. \citet{2013ApJS..208....9P} recently studied a sample of young stars in nearby groups: the $\eta$\,Cha cluster, the TW~Hydra Association, the $\beta$\,Pic Moving Group, and the Tucana-Horologium Association. We found that the spectral types in \citet{2013ApJS..208....9P} for the same stars are typically one subclass different from those in the literature. Though it is not clear  which classification results are more accurate, we decide to classify our stars based on the same criteria as those in \citet{2013ApJS..208....9P}. 

We select X-Shooter spectra for 85 young stars from ESO Phase 3 spectral data archive. Among them, 27 young stars in the $\eta$\,Cha cluster and the TW~Hydra Association have been spectrally classified by \citet{2013ApJS..208....9P}. We use these spectra as the templates to classify the other sources with  X-Shooter spectra. Using these X-Shooter spectra, we construct new relations between the strengths of spectral features and spectral types for young stars in the range of K2-M9.5 (See the detailed description in Appendix~\ref{Appen:spectral_classification}). The new relations should be consistent with the classification criteria from  \citet{2013ApJS..208....9P}.  Thus, the conversion from spectral types to effective temperatures, intrinsic colors, and bolometric corrections for young populations in  \citet{2013ApJS..208....9P}  can be used for other young stars with spectral types classified based on our relations. In this work, we use these relations to classify the spectra from Hectospec. In Fig.~\ref{Fig:SPT}, we compare the spectral types from our relations and from the SPTCLASS code \citep{2004AJ....127.1682H}. It can be noted that both spectral types are  consistent with each other given the uncertanty within the range of M0--M6. When the spectral types are later than M6 or earlier than M0, our relations give about one subclass ealier spectral type than SPTCLASS.

\begin{figure}
\begin{center}
\includegraphics[angle=0,width=1\columnwidth]{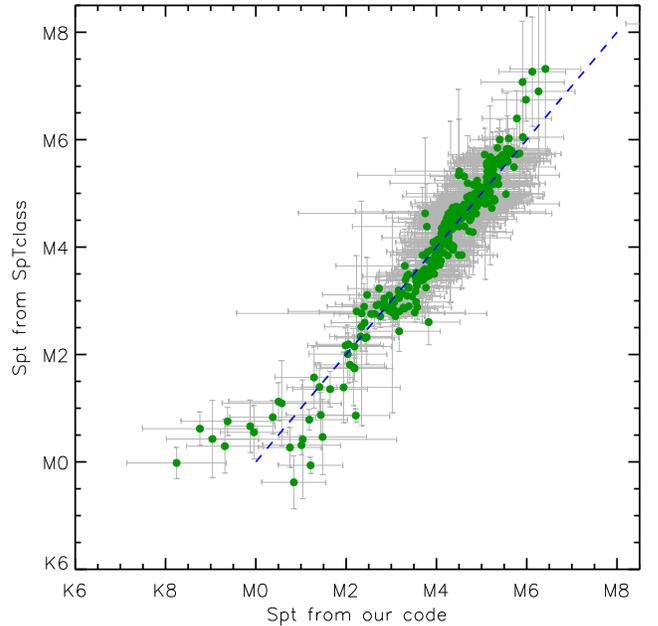}
\caption{Comparison of spectral types from our code and from the SPTCLASS code \citep{2004AJ....127.1682H}. The uncertainty for each spectral type are indicated. The dash line marks where the two spectral types are equal in M-type range. }\label{Fig:SPT}
\end{center}
\end{figure}

\subsection{YSO selection criteria}
The targets in this work are obtained from a sample of candidate pre-main sequence (PMS) stars selected from different sets of color-magnitude diagrams by \citet{2014A&A...564A..29B}. Some of them could be interlopers from main-sequence (MS) stars or giants. In this section, we will clarify our criteria to remove the contaminators in our sample.

\subsubsection{Spectral features}\label{Sect_YSO:spectral_features}
The youth of the stars can be characterized using several indicators. The typical one is the Li\,I absorption line at 6708\,\AA. In Fig.~\ref{Fig:Li}, we show the examples of \LiI\ detections for 3 sources with MMT/Hectospec. Among the 752 sources in our sample, there are 444 sources for which we have obtained spectra. Among them, 295 stars show  \LiI\ absorption, 11 do not  show clear  \LiI\ absorption, and others have spectra too noisy for \LiI\ to be identified. An additional 152 stars are found to show  Li\,I absorption in \citet{2012ApJ...752...59H}. In total, a group of 447 stars in our sample are  identified as young stars based on the  \LiI\ absorption. In Table~\ref{Tab:spt}, we list the equivalent width (EW) of the  \LiI\ absorption for those stars. The typical uncertainty of the \LiI\ EWs is around 0.1--0.2\AA.

The strength of the Na\,I doublets at 8183 and 8195\,\AA\ has proved to be a good indicator for stellar surface gravitiy  for M-type stars \citep{2009MNRAS.400L..29L}. MS stars usually show strong  Na\,I doublets, and giants present very weak ones. The PMS stars show the strengths of the Na\,I doublets between the main sequence stars and giants. In this work, we will use the $Index ({\rm Na\,I})$ as a second indicator to select the PMS stars. We calculate the strength of the Na\,I doublets at 8183 and 8195\,\AA\  as $Index ({\rm Na\,I})=F_{8135-8155}/F_{8180-8200}$, where $F_{8135-8155}$ and $F_{8180-8200}$ are fluxes between 8135 and 8155\,\AA, and between  8180 and 8200\,\AA, respectively. We also calculate the $Index ({\rm Na\,I})$ for the MS stars and giants using the spectra for MS stars and giants from \citet{1994PASP..106..382D} and \citet{2007AJ....134.2398C}, respectively. To clarify the criteria for the selection of PMS stars, we use a  well-studied sample of the PMS stars in L1641 \citep{2013ApJS..207....5F}. In Fig.~\ref{Fig:Na}(a), we show the $Index ({\rm Na\,I})$ for PMS stars and MS stars identified in L1641. In the figure, we also show the expected $Index ({\rm Na\,I})$ values for MS stars and giants, which are derived from the spectra for MS stars and giants in the IRTF spectral library \citep{2009ApJS..185..289R}, and the SDSS spectral templates for MS stars \citep{2007AJ....134.2398C}.   We note that using $Index ({\rm Na\,I})$ to separate the MS and PMS stars  only works for the M-type stars, which is consistent with the findings of \citet{2009MNRAS.400L..29L}. According to the distribution of the MS stars, PMS stars, and giants in  Fig.~\ref{Fig:Na}(a), we draw the boundaries for selecting PMS stars described as follows: For spectral types between M1 and M3.5, $1.03<Index ({\rm Na\,I})<1.22$, and for spectral types between M3.5 and M7, $1.03 <Index ({\rm Na\,I})<-3.61+0.0657\times Sptnum$, where \textit{Sptnum} is a number corresponding to a spectral type with 70 for M0, 75 for M5, and etc. As a comparison, in Fig.~\ref{Fig:Na}(b) we show the PMS stars, identified with the \LiI\ absorption, in this work. As expected most sources are at the PMS boundary but six of them are not. We have checked their spectra. For three of those stars, which show  the  $Index ({\rm Na\,I})$ values simiar to giants, the fringe patterns  on their spectra can explain lower $Index ({\rm Na\,I})$ values. The other three stars with large Index(NI) values show clear Li absorption as well as strong  Na\,I doublets at 8183 and 8195\,\AA. A detailed discussion of these sources is presented in \sect~\ref{Sect:Na_Li}. In  Fig.~\ref{Fig:Na}(c), we show the stars without estimated \LiI\ EW due to their noisy spectra. For these stars, we characterize their youth using $Index ({\rm Na\,I})$, in combination of X-ray emission and infrared excess emission.


\begin{figure}
\begin{center}
\includegraphics[angle=0,width=1\columnwidth]{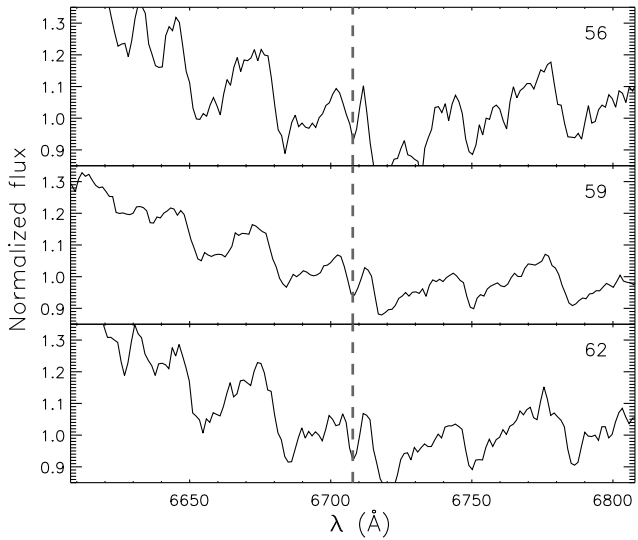}
\caption{Examples of \LiI\ detection with the MMT/Hectospec. The flux is in normalized in each panel.}\label{Fig:Li}
\end{center}
\end{figure}

\begin{figure*}
\begin{center}
\includegraphics[angle=0,width=2\columnwidth]{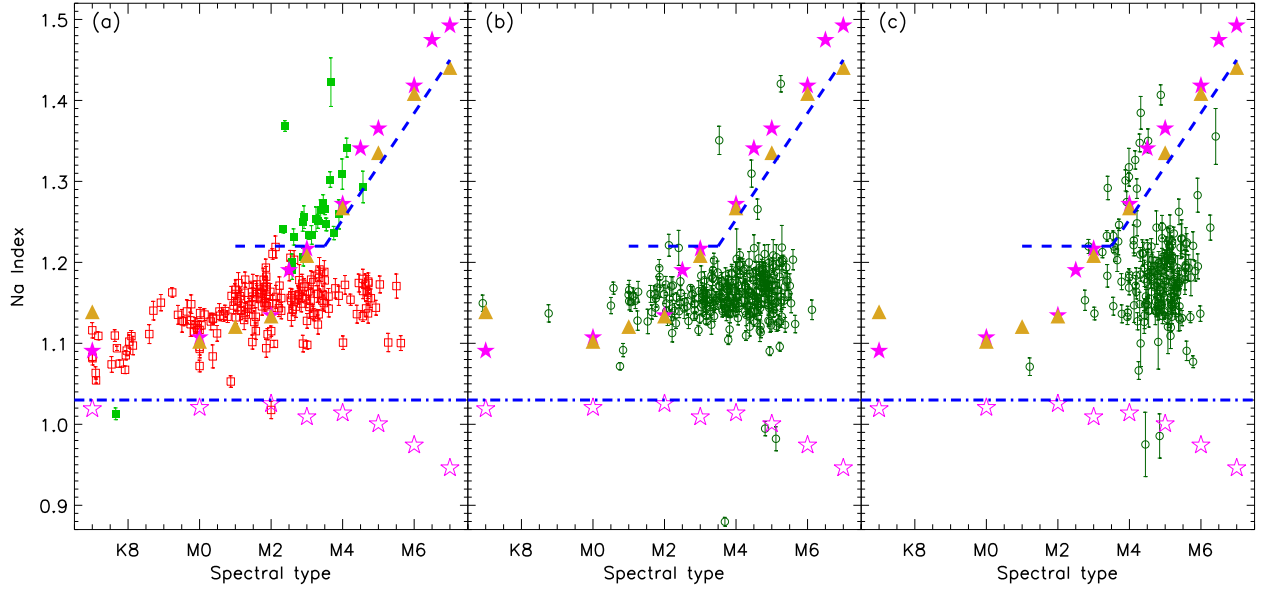}
\caption{(a) $Index ({\rm Na\,I})$ vs. spectral types for the stars in the field of L1641 \citep{2013ApJS..207....5F}. The filled squares are for the identified field MS or giant stars in the field of L1641, and the open squares show the PMS stars.  The dash and dash-dotted lines show the boundaries for MS stars and giants, respectively. The boundaries are defined  according to the expected $Index ({\rm Na\,I})$ of of MS stars and giants, which are derived using the spectra for MS (filled star symbols) and giants (open star symbols) in the IRTF spectral library \citep{2009ApJS..185..289R}, and the SDSS spectral templates for MS stars (filled triangles) \citep{2007AJ....134.2398C}. (b) Similar to Panel~(a), but for the PMS stars, which are identified using \LiI\ absorption, in this work. (c) Similar to Panel~(b), but for the stars in this work without knowledge of \LiI\ absorption due to their noisy spectra.}\label{Fig:Na}
\end{center}
\end{figure*}

\subsubsection{X-ray emission and Infrared excess emission}

 In the field shown in Fig.~\ref{Fig:Orion}, we extracted 3,688 XMM X-ray sources from the third XMM Serendipitous Source Catalogue \citep{2015arXiv150407051R}, and 1616 Chandra X-ray sources from \citet{2005ApJS..160..319G}. We matched the X-ray sources to our sources using 1$''$ tolerance, and found 171 counterparts for our targets. We note that there are more than 100  targets with X-ray sources within radii of 1--3$''$. We visually check the 2MASS images, and consider those as conterparts to the X-ray sources if there is only one source within 1--3$''$ from X-ray source. In this way, we found 107 counterparts in our sample for the X-ray sources. In total, we have 267 sources with X-ray emission from the XMM observations, and an additional 11 sources with  X-ray emission from Chandra observations.  There are two main contaminators to the YSO catalog selected based on the X-ray data: extragalactic sources and nearby foreground stars. Since we have spectra for each source,  extragalactic contaminators can be excluded.  

The foreground nearby main-sequence stars can show detectable X-ray emission. We used the tool, Flux Limits from Images from XMM-Newton (FLIX)\footnote{FLIX is a on-line tool provided by the XMM-Newton Survey Science Center (see http://www.ledas.ac.uk/flix/flix3). It provides robust estimates of the X-ray upper limit to a given point in the sky where there are no sources detected in the 3XMMi catalog.}, to estimate the 5-$\sigma$ upper limit of the 0.2--2\,keV X-ray luminosity in the studied field. We note that the 5-$\sigma$ upper limits vary from region to region with a typical value several~10$^{29}$\,\ergps\ at the vicinity of Orion. The typical X-ray luminosity of field main-sequence stars is several$\times10^{27}$\,\ergps\ for solar-type to M-type main sequence stars \citep{2004A&ARv..12...71G}. According to the 5-$\sigma$ upper limit of the 0.2--2\,keV X-ray luminosity in the XMM survey of Orion estimated with FLIX, the XMM observations can detect the field main-sequence stars within the distance$\lesssim$50\,pc. Based on the Besan\c{c}on\ model of stellar population synthesis of the Galaxy \citep{2003A&A...409..523R}, we expect only $4\pm2$ field main-sequence stars within a distance of 50\,pc in the direction of Orion, suggesting that the fraction of contaminators from foreground stars in our YSO catalog, selected from X-ray emission,  is negligible.

We also use infrared excess to identify the young stars. A detailed description of characterizing the disk properties of our sample is presented in \sect~\ref{Sec:ana_SED_classification}.

\begin{figure}
  \begin{center}
\includegraphics[angle=0,width=1\columnwidth]{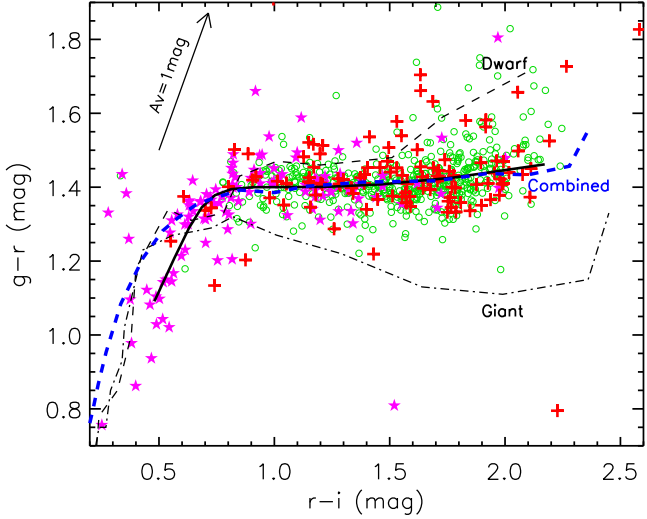}    
\caption{$g-r~vs.~r-i$ color-color diagram. The open circles show the young stars in this work. {\newrev The filled star symbols and pluses} show the dereddened colors of young stars with low extinction ($A_{\rm V}<0.3$) in Orion~OB1/$\sigma$~Ori, and L1641, respectively. The thick dark solid line show the empirical colors for young stars used in this work (see \sect~\ref{Sect:colors}). The thin dark dashed line and the dash-dotted line show the empirical colors for dwarfs and giants, respectively \citep{2007AJ....134.2398C}. The thick blue dashed line shows the synthetic colors using a set of the combined  BT-Settl models with  solar abundances from \citet{2009ARA&A..47..481A} (see \sect~\ref{Sect:colors}).}\label{Fig:CCD}
\end{center}
\end{figure}

\begin{figure*}
\centering
\includegraphics[width=2.0\columnwidth]{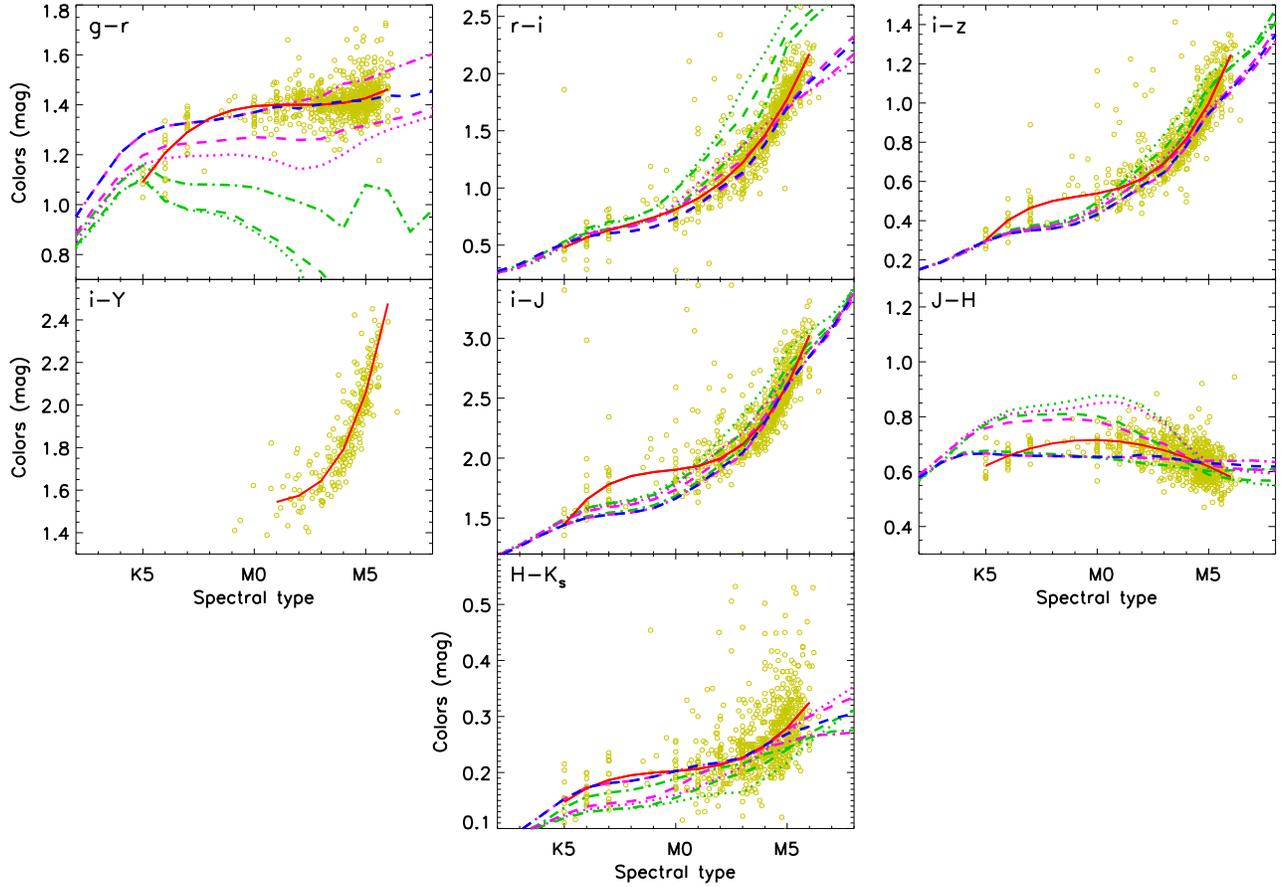}
\caption{Intrinsic colors for young stars in this work and those in Orion OB1, $\sigma$~Ori, and L1641. In each panel, The red solid line is the fit to the relation between the colors and the spectral types, the green lines show the synthetic colors from the BT-Settl atmospheric models with the AGSS2009 abundances with log~g=3.5 (dotted line), 4.0 (dashed line), and 5.0 (dash-dotted line), and the magenta lines are similar to the green lines but for the models with the CIFIST2011 abundances. The blue dashed lines show the synthetic colors from the combined BT-Settl models.} \label{Fig:synthetical_colors}
\end{figure*}

\begin{figure}
\begin{center}
\includegraphics[angle=0,width=1\columnwidth]{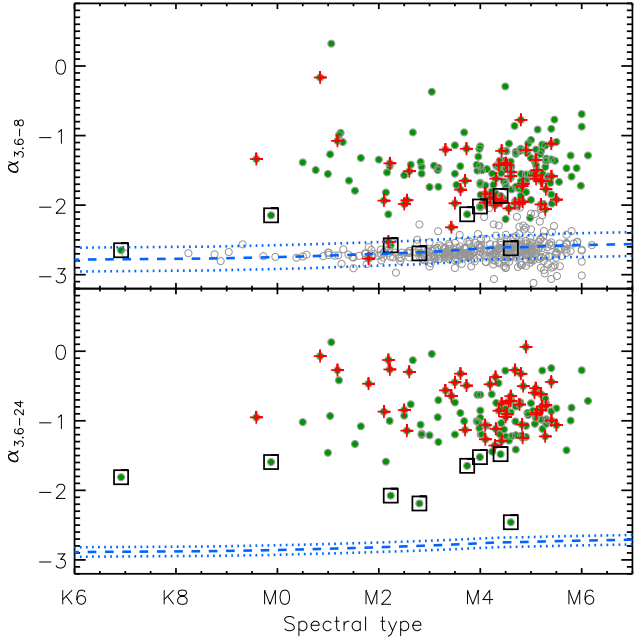}
\caption{Dereddened Spitzer infrared spectral slopes, $\alpha_{3.6-8}$ and $\alpha_{3.6-24}$ vs. spectral type. In each panel, the open gray circles show the diskless stars, and the filled circles are for the sources with disks. The plus symbols mark the transition disks, and the open squares are for the other evolved disks in our sample (see Sect.~\ref{Sec:TD} and Fig.~\ref{Fig:DB_SED}). The dashed line show the infrared spectral slope of the photospheric emission calculated with the BT-Settl atmospheric models, and the dotted lines are the 1$\sigma$ standard deviation, assuming a 10\% uncertainty in Spitzer photometry.}\label{Fig:alpha}
\end{center}
\end{figure}

\begin{figure}
  \begin{center}
 \includegraphics[angle=0,width=1\columnwidth]{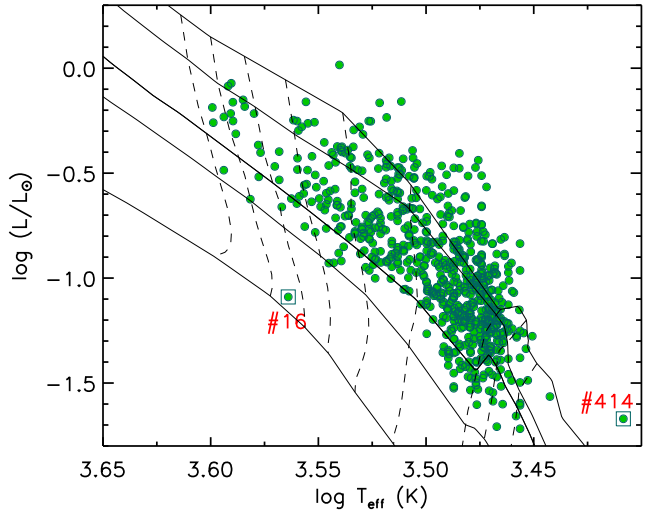}   
\caption{H-R diagram for the PMS stars (filled circles) in this work. The open squares mark the two interesting objects in our sample (see Sect.~\ref{Sect:sublum} and ~\ref{Sect:Planet}). The solid lines show the isochrones at ages of 0.5, 1, 3, 5, ,10, and 30\,Myr, respectively, from \citet{2015A&A...577A..42B}. The dash lines present the evolutionary tracks of young stellar/substellar objects with masses of 0.06, 0.08, 0.1, 0.2, 0.3, 0.4, 0.5, 0.6, and 0.7\,\Msun, respectively. }\label{Fig:HRD}
\end{center}
\end{figure}

\begin{figure}
  \begin{center}
 \includegraphics[angle=0,width=1\columnwidth]{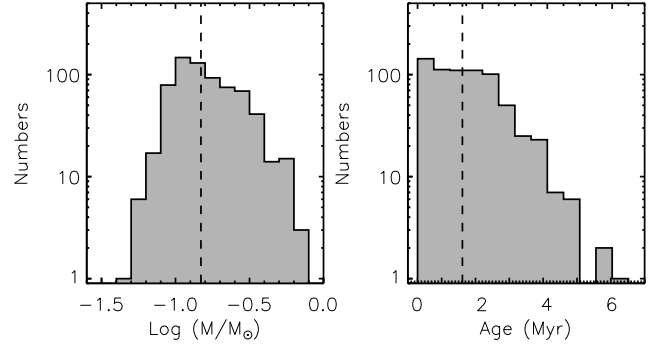}   
\caption{The mass (left) and age (right) distributions of the young stars. The dashed lines shows the median mass ($\sim$0.15\,\Msun) and  age ($\sim$1.4\,Myr) of our sample, respectively.}\label{Fig:Mass_Age_dis}
\end{center}
\end{figure}

\begin{figure}
  \begin{center}
  \includegraphics[angle=0,width=1\columnwidth]{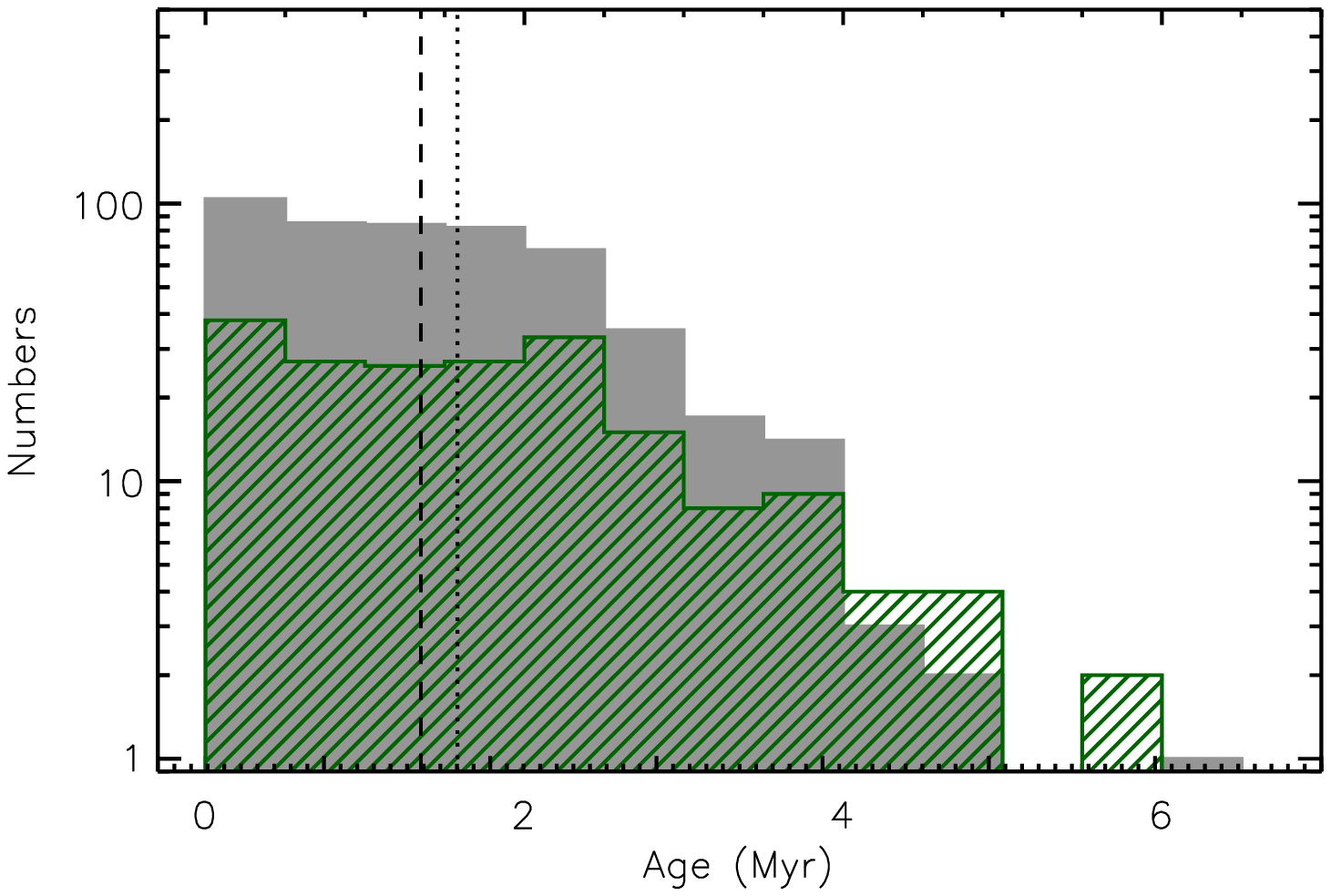}  
\caption{The age distributions of the young stars with disks (line-filled histogram) and without disks (gray-color filled histogram). The  median ages of the two populations, 1.6 and 1.3\,Myr, respectively, are shown with the dotted line and the dashed line.}\label{Fig:disk_age_dis}
\end{center}
\end{figure}

\subsection{Intrinsic colors of young stars}\label{Sect:colors}
{\newnewrev The extinction of young stars can be determined by comparing the observed colors with the intrinsic colors expected from their spectral types. In Fig.~\ref{Fig:CCD}, we show the $g-r~vs.~r-i$ color-color diagram for our sources. In the figure, we also show the young stars with low extinction ($A_{\rm V}<0.3$)   in  Orion OB1, $\sigma$~Ori, and L1641 \citep{2005AJ....129..907B,2007ApJ...661.1119B,2014MNRAS.444.1793D,2015MNRAS.450.3490D,2014ApJ...794...36H,2013ApJS..207....5F}. Their colors are dereddend using the extinction law from \citet{2011ApJ...737..103S}, a total to selective extinction value typical of interstellar medium dust ($R_{V}=3.1$), and the extinction values in the literature. These young stars in other regions show colors similar to our sources, indicating that the majority of our sources in this work have low or no extinction. As a comparison, in Fig.~\ref{Fig:CCD} we also show the empirical colors for dwarfs and giants \citep{2007AJ....134.2398C}. We note that there is a shift between the observed colors and the empirical colors for the  dwarfs {\newrev (thin dark dashed line)} and giants {\newrev (thin dark dash-dotted line)}. In order to derive the extinction properly, we need to construct intrinsic colors for our sources. We use the sources with negligible extinction shown in Fig.~\ref{Fig:CCD}, and extract their $griz$-band photometric data from SDSS,  $JHK_{\rm s}$-band photometric data from 2MASS. The Y-band photometry is only available for the sources in Orion~A, and come from \cite{2014A&A...564A..29B}. In Fig.~\ref{Fig:synthetical_colors}, we show their colors vs. their spectral types. We fit the relations between them using 3-order polynomial functions. In Table~\ref{Tab:color}, we list the empirical colors for young stars with spectral types ranged from K5 to M6 from the fitting.

We calculate the synthetic colors using two sets of  BT-Settl atmospheric models \citep{2011ASPC..448...91A,2012RSPTA.370.2765A}: one with the solar abundances (AGSS2009) from \citet{2009ARA&A..47..481A}, and the other with the solar abundances (CIFIST2011) from \cite{2011SoPh..268..255C}. Fig.~\ref{Fig:synthetical_colors} shows the synthetic $g-r$, $r-i$, $i-z$, $j-J$, $J-H$, and $H-K_{\rm s}$ colors from these models with  surface gravity log~g=3.5, 4.0, and 5.0. While both sets of models are consistent with each other in  $i-z$, $j-J$, $J-H$, and $H-K_{\rm s}$ colors, the models with the AGSS2009 abundances show the $g-r$ and $r-i$ colors, which are more consistent with the observations than the ones with the CIFIST2011 abundances. To reproduce the observed colors, we construct a set of BT-Settl atmospheric models with the AGSS2009 abundances and using different log~g values for different spectral types: log~g=5.0 for the spectral types earlier than M1, log~g=4.8 for spectral types between M1 and M3, log~g=4.6 for  spectral types between M3 and M6, and log~g=4.5 for spectral types later than M6. The synthetic colors from this set of models are shown in Fig.~\ref{Fig:CCD} and ~\ref{Fig:synthetical_colors}, and can better fit the observations than using the atmospheric models with one log~g value. Table~\ref{Tab:synthetical_color} lists the synthetic colors and bolometric correction in $J$ band from this set of models. In the table, the conversions from the spectral types to effective temperatures are from \citet{2013ApJS..208....9P} for stars earlier than M4 and  from  \citet{2014ApJ...786...97H} for stars later than M4. In Table~\ref{Tab:synthetical_color}, we also list the $J$-band bolometric correction \citet[P13]{2013ApJS..208....9P}, and the ones from \citet[H15]{2015ApJ...808...23H}. Our bolometric corrections are consistent with those in the literature. 
}

\begin{table*}
\caption{Emprical intrinsic colors for young stars}\label{Tab:color}
\centering
\begin{tabular}{clccccccccccccccccccccccccccccccccc}
\hline\hline
   &Teff  & $g-r$ & $r-i$ &$i-z$ &$i-Y$ &$i-J$ &$J-H$ &$H-Ks$\\
Spt&(K)   &(mag)  & (mag) &(mag) &(mag) &(mag) &(mag) &(mag)\\
\hline
K5&4140 & 1.09&0.48&0.30&\nodata&1.45&0.62&0.15\\
K6&4020 &1.21&0.57&0.40&\nodata&1.65&0.66&0.17\\
K7&3970 &1.29&0.63&0.46&\nodata&1.78&0.68&0.19\\
K8&3940 &1.35&0.69&0.50&\nodata&1.85&0.70&0.20\\
K9&3880 &1.38&0.74&0.52&\nodata&1.88&0.71&0.20\\
M0&3770 &1.39&0.81&0.54&\nodata&1.90&0.72&0.20\\
M1&3630 &1.40&0.91&0.57&1.54&1.93&0.71&0.21\\
M2&3490 &1.40&1.04&0.61&1.57&2.00&0.70&0.21\\
M3&3360 &1.40&1.22&0.69&1.64&2.11&0.68&0.23\\
M4&3160 &1.41&1.46&0.82&1.79&2.31&0.65&0.25\\
M5&2980 &1.43&1.78&1.00&2.06&2.60&0.62&0.28\\
M6&2860 &1.46&2.17&1.25&2.48&3.02&0.58&0.32\\
\hline
\end{tabular}
\end{table*}

\begin{table*}
\caption{Synthetic colors for young stars}\label{Tab:synthetical_color}
\centering
\begin{tabular}{clccccccccccccccccccccccccccccccccccccc}
\hline\hline
 &      &\multicolumn{7}{c}{This work} & &\multicolumn{2}{c}{P03} & &\multicolumn{2}{c}{H04, H05}\\
       \cline{3-9}                                                 \cline{11-12} \cline{14-15}
 &Teff  & $g-r$ & $r-i$ &$i-z$ &$i-J$ &$J-H$ &$H-Ks$ & $BC_{\rm J}$& &Teff & BCJ& &Teff & BCJ\\
 &(K)   & (mag) & (mag) &(mag) &(mag) & (mag) & (mag) & (mag)      & &(K) &(mag)& &(K) &(mag)\\
\hline
F0&7280& 0.15&-0.03&-0.08& 0.57& 0.15& 0.02& 0.60&&7280&0.57&&...&...\\
F1&6990& 0.21&-0.01&-0.06& 0.62& 0.18& 0.02& 0.70&&6990&0.68&&...&...\\
F2&6710& 0.26& 0.02&-0.04& 0.67& 0.21& 0.02& 0.79&&6710&0.75&&...&...\\
F3&6660& 0.27& 0.02&-0.04& 0.68& 0.22& 0.02& 0.80&&6660&0.76&&...&...\\
F4&6590& 0.29& 0.03&-0.03& 0.69& 0.23& 0.03& 0.83&&6590&0.79&&...&...\\
F5&6420& 0.33& 0.05&-0.02& 0.72& 0.25& 0.03& 0.89&&6420&0.85&&6600&0.79\\
F6&6250& 0.37& 0.06&-0.01& 0.76& 0.27& 0.03& 0.94&&6250&0.91&&...&...\\
F7&6140& 0.41& 0.08& 0.00& 0.78& 0.29& 0.04& 0.98&&6140&0.95&&...&...\\
F8&6100& 0.42& 0.08& 0.01& 0.79& 0.29& 0.04& 1.00&&6100&0.96&&6130&0.95\\
F9&6090& 0.42& 0.08& 0.01& 0.79& 0.30& 0.04& 1.00&&6090&0.97&&...&...\\
G0&6050& 0.43& 0.08& 0.01& 0.80& 0.30& 0.04& 1.01&&6050&0.98&&5930&1.02\\
G1&5970& 0.45& 0.09& 0.02& 0.81& 0.31& 0.04& 1.04&&5970&1.00&&...&...\\
G2&5870& 0.49& 0.10& 0.03& 0.84& 0.33& 0.04& 1.08&&5870&1.03&&5690&1.10\\
G3&5740& 0.53& 0.12& 0.04& 0.87& 0.35& 0.05& 1.12&&5740&1.08&&...&...\\
G4&5620& 0.57& 0.13& 0.05& 0.90& 0.37& 0.05& 1.16&&5620&1.12&&...&...\\
G5&5500& 0.61& 0.15& 0.06& 0.94& 0.40& 0.05& 1.20&&5500&1.16&&5430&1.18\\
G6&5390& 0.65& 0.16& 0.07& 0.97& 0.42& 0.06& 1.23&&5390&1.19&&...&...\\
G7&5290& 0.69& 0.17& 0.08& 1.00& 0.44& 0.06& 1.27&&5290&1.23&&...&...\\
G8&5210& 0.72& 0.18& 0.09& 1.02& 0.46& 0.06& 1.29&&5210&1.25&&5180&1.26\\
G9&5120& 0.76& 0.20& 0.10& 1.05& 0.48& 0.07& 1.32&&5120&1.27&&...&...\\
K0&5030& 0.80& 0.22& 0.11& 1.08& 0.50& 0.07& 1.34&&5030&1.30&&4870&1.36\\
K1&4920& 0.87& 0.24& 0.13& 1.12& 0.53& 0.08& 1.38&&4920&1.34&&...&...\\
K2&4760& 0.95& 0.27& 0.15& 1.18& 0.58& 0.08& 1.42&&4760&1.40&&4710&1.41\\
K3&4550& 1.08& 0.33& 0.19& 1.26& 0.63& 0.10& 1.48&&4550&1.44&&...&...\\
K4&4330& 1.21& 0.43& 0.24& 1.36& 0.66& 0.12& 1.54&&4330&1.52&&...&...\\
K5&4140& 1.28& 0.51& 0.30& 1.44& 0.67& 0.15& 1.58&&4140&1.58&&4210&1.56\\
K6&4020& 1.31& 0.58& 0.33& 1.50& 0.66& 0.17& 1.61&&4020&1.61&&...&...\\
K7&3970& 1.33& 0.60& 0.35& 1.53& 0.66& 0.18& 1.63&&3970&1.63&&4020&1.62\\
K8&3940& 1.34& 0.62& 0.36& 1.54& 0.66& 0.18& 1.63&&3940&1.63&&...&...\\
K9&3880& 1.35& 0.66& 0.38& 1.58& 0.65& 0.19& 1.65&&3880&1.66&&...&...\\
M0&3770& 1.37& 0.73& 0.43& 1.66& 0.65& 0.20& 1.69&&3770&1.69&&3900&1.66\\
M1&3630& 1.39& 0.85& 0.50& 1.78& 0.65& 0.21& 1.73&&3630&1.74&&3720&1.73\\
M2&3490& 1.39& 1.00& 0.58& 1.92& 0.66& 0.22& 1.78&&3490&1.80&&3560&1.78\\
M3&3360& 1.40& 1.14& 0.65& 2.04& 0.65& 0.23& 1.82&&3360&1.84&&3410&1.84\\
M4&3160& 1.41& 1.38& 0.78& 2.30& 0.64& 0.25& 1.88&&3160&1.91&&3190&1.93\\
M5&2980& 1.42& 1.70& 0.95& 2.61& 0.63& 0.27& 1.95&&2880&2.01&&2980&1.99\\
M6&2860& 1.44& 1.93& 1.07& 2.85& 0.63& 0.28& 1.98&&...&...&&2860&2.03\\
M7&2770& 1.43& 2.10& 1.19& 3.08& 0.62& 0.30& 2.02&&...&...&&2770&2.06\\
M8&2670& 1.46& 2.28& 1.35& 3.37& 0.62& 0.30& 2.05&&...&...&&2670&...\\
M9&2570& 1.55& 2.36& 1.60& 3.77& 0.57& 0.33& 2.11&&...&...&&2570&...\\
\hline
\end{tabular}
\end{table*}

\subsection{Determining stellar properties}\label{Sect:extinction}

{\newnewrev We derive the extinction for our sources using the empirical colors listed in Table~\ref{Tab:color} for the ones with spectral types earlier than M6, and the synthetic colors in Table~\ref{Tab:synthetical_color} for those with spectral types later than M6.}  Using the intrinsic colors of $g-r$, $r-i$, $i-z$, $i-j$, we derive mean visual extinctions ($A_{V}$) for individual sources employing the extinction law from \citet{2011ApJ...737..103S} and a total to selective extinction ratio typical of interstellar medium dust ($R_{\rm V}=3.1$).  We converted the spectral types to effective temperatures using the relation in  Table~\ref{Tab:synthetical_color}, and derive the stellar luminosities for our sources using the $J$-band  bolometric correction for the corresponding spectral types calculated in this work. The stellar luminosities ($L_{\star}$) are then calculated as follows:

\begin{eqnarray*}
M_{bol}=BC_{J}+m_{J}-5\times log(\frac{d}{\rm 10~pc}) \\
\frac{L_{\star}}{L_{\odot}}=10^{\frac{M_{\rm bol, \odot}-M_{bol}}{2.5}}  
\end{eqnarray*}

\noindent where $BC_{J}$ is the the bolometric correction in $J$ band, $m_{J}$ is the dereddened apparent magnitude in $J$ band, $d$ is the distance in the unit of pc, $M_{\rm bol, \odot}$ is the bolometric magnitude of the Sun. Here we take $M_{\rm bol, \odot}=4.755$ \citep{2012ApJ...754L..20M}.

\begin{figure*}
  \begin{center}
   \includegraphics[angle=0,width=2\columnwidth]{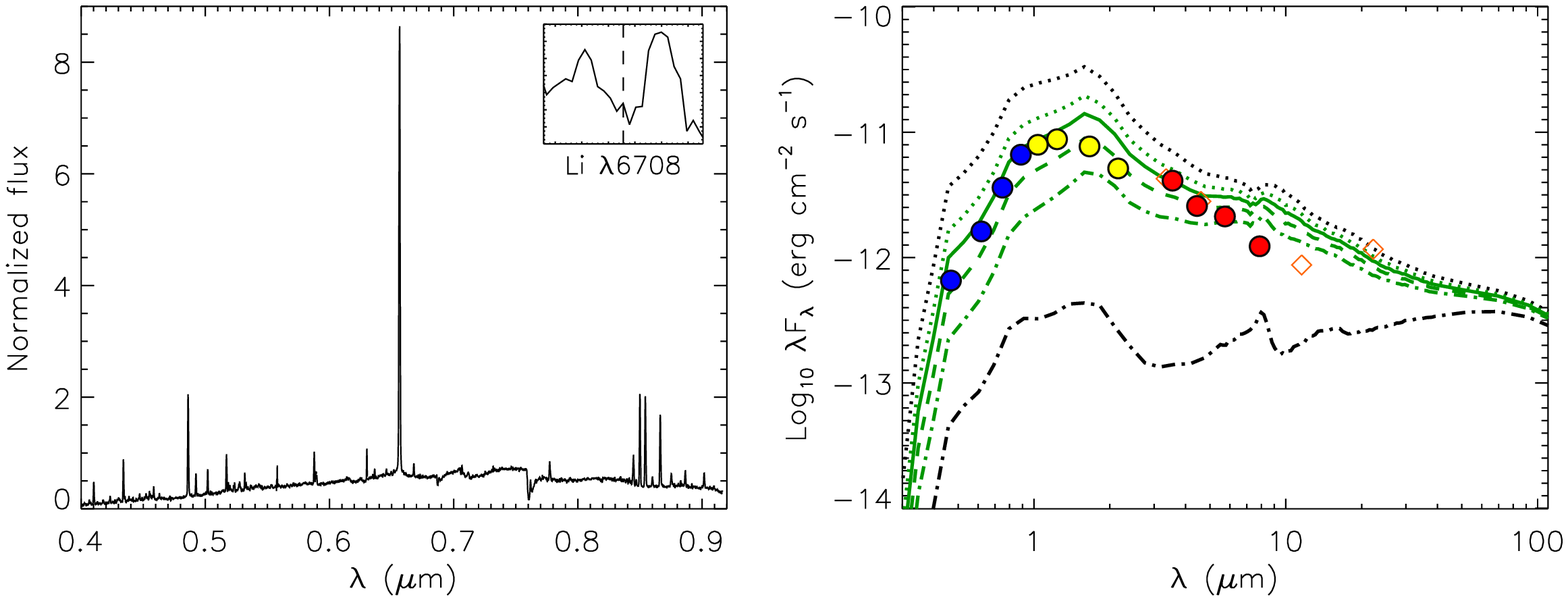} 
\caption{Left panel: the optical spectra of source 16. Right panel: the SED of source 16. The filled circles show the photometry from SDSS, 2MASS, and Spitzer. The open diamonds present the photometry from WISE. {\newrev The model SEDs are calculated for a disk model with $M_{\rm d}$=5$\times 10 ^{-4}M_{\star}$, $R_{\rm in}$=0.2\,AU, and $H_{out}/R_{\rm out}=$0.1, and for disk inclinations of 78$^{\circ}$ (black doted line), 79$^{\circ}$ (green doted line), 79.5$^{\circ}$ (green solid line), 80$^{\circ}$ (green dash line), 80.5$^{\circ}$ (green dash-dotted line), and 82.5$^{\circ}$ (black dash-dotted line), respectively.}}\label{Fig:sublum}
\end{center}
\end{figure*}

\begin{figure}
  \begin{center}
   \includegraphics[angle=0,width=1\columnwidth]{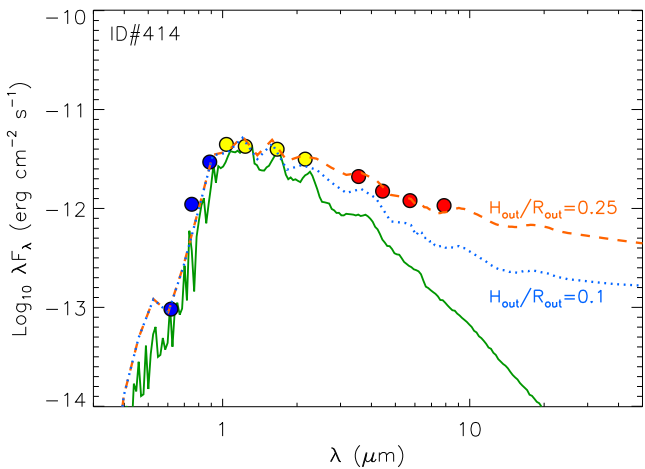} 
\caption{SED of an extremely low-mass object (Source 414). The filled circles show the photometry from SDSS, 2MASS, and Spitzer. {\newnewrev The model SEDs are calculated for a disk model with $M_{\rm d}$=1$\times 10 ^{-2}M_{\star}$, $H_{out}/R_{\rm out}=$0.25 (dashed line) and 0.1 (dotted line), and for a disk inclination of 40$^{\circ}$.}}\label{Fig:planet}
\end{center}
\end{figure}

\subsection{Characterizing disk properties}\label{Sec:ana_SED_classification}

We characterize the disk properties by comparing the model atmosperes from the BT-Settl models with the observed spectral energy distributions. The typical way is to compare the infrared spectral slopes with the expected slopes at the same wavelength ranges from  purely photospheric emissions \citep{2008ApJ...675.1375L}. The infrared spectral slopes, defined as $\alpha=d log(\lambda F_{\lambda})/d log(\lambda)$, are calculated with the dereddened photometry in each band.  We compute two sets of infrared spectral slopes, $\alpha_{3.6-8}$ and $\alpha_{3.6-24}$, corresponding to the spectral range of [3.6] to [8.0] and [3.6] to [24], respectively. Table~\ref{Tab:spt} lists the two infrared spectral slopes of the sources with detection in the corresponding infrared bands. {\rev In Fig.~\ref{Fig:alpha}, we show the infrared spectral slopes, $\alpha_{3.6-8}$ and $\alpha_{3.6-24}$ versus the spectral types, for our sources. As a comparison, we calculate the spectral slopes (see Fig.~\ref{Fig:alpha}) of the BT-Settl atmospheric models \citep{2012RSPTA.370.2765A}, assuming a 10\% uncertainty in Spitzer photometry. Stars with the infrared spectral slopes steeper than the slopes of photospheric emissions are considered to be diskless. For sources with shallower infrared spectral slopes, we visually examine their SEDs. The sources that show  infrared excess at  more than 3$\sigma$ confidence level are considered to have disks. A total of 185 sources belong to the disk population. In addition, there are 119 sources without detection in all the four IRAC bands. We visually compare their SEDs, constructed from the available Spitzer and WISE photometric data, to their photospheric emission. Among the 119 sources, 9 sources show infrared excess at  more than 3$\sigma$ confidence level, and are considered to have disks. In Table~\ref{Tab:spt}, we list the disk property of each source.}

The young stars can be grouped into weak-line T~Tauri stars (WTTS) or classical T Tauri stars (CTTS)  based on their H$\alpha$ $EW$. WTTSs have stopped accretion, and show weak and narrow H$\alpha$ emission line in the spectra, while CTTSs are still accreting, and present strong and broad  H$\alpha$ emission lines. We divide the YSOs into WTTS or CTTS using the criteria described in \citet{2009A&A...504..461F}, in which a star is classified as a CTTS if $EW$(H$\alpha$)$\geq3$\,\AA\ for K0--K3 stars, $EW$(H$\alpha$)$\geq$5\,\AA\ for K4 stars, $EW$(H$\alpha$)$\geq7$\,\AA\ for K5--K7 stars, $EW$(H$\alpha$)$\geq9$\,\AA\ for M0--M1 stars, $EW$(H$\alpha$)$\geq11$\,\AA\ for M2 stars, $EW$(H$\alpha$)$\geq15$\,\AA\ for M3--M4 stars, $EW$(H$\alpha$)$\geq18$\,\AA\ for M5--M6 stars, and $EW$(H$\alpha$)$\geq20$\,\AA\ for M7--M8 stars. Table~\ref{Tab:spt} list the H$\alpha$ $EWs$ of each source and its  accretion property. However, we must stress that using  $EW$(H$\alpha$) to distinguish WTTSs and CTTSs can fail to distinguish some stars that have low accretion rates, strong chromospheric activity, sky contamination, or self-absorptions of the H$\alpha$ line.

\section{results}

\subsection{A census of PMS stars}\label{Sect:Na_Li}
We identify PMS stars mainly based on the \LiI\ absorption. With this criterion, a sample of 447 stars are classified as PMS stars. For others without estimate of the EWs of \LiI\ absorption line, we use the $Index ({\rm Na\,I})$ to select the PMS stars as described in \sect~\ref{Sect_YSO:spectral_features}, which works for stars with spectral types later than M1. With this criterion, an additional 164 stars are classified as PMS stars. We also include 64 additional stars which show X-ray emission, and  27 additional stars with infrared excess emission. A total of 691 sources are classified as PMS stars according to the above criteria. Table~\ref{Tab:spt} list these stars, as well as the criteria to classify them as the PMS stars.  

{\newrev We use the $Index ({\rm Na\,I})$ to assess the contamination in our PMS sample from young field stars which may show weak \LiI\ absorption line.  In our sample, there are 447 sources with estimated EWs of \LiI\ absorption line, and 63\% of them (283/447) have the $Index ({\rm Na\,I})$ in Table~\ref{Tab:spt}. As noted in \sect~\ref{Sect_YSO:spectral_features}, there are three sources, ID~156, 438, 572 in Table~\ref{Tab:spt},  that are located outside the PMS boundary in Fig.~\ref{Fig:Na}(b) and show $Index ({\rm Na\,I})$ similar to the field dwarfs. However, for two sources, ID~156 and 438, the strengths of their \LiI\ absorption lines are consistent with other young stars in Orion \citep[see, e.g.,][]{2013ApJS..207....5F}. Hence, it is unclear why they show the strong Na\,I doublets. For the source 572, its \LiI\ absorption line is weak ($EW\sim$0.2), and could be a young field dwarf. Therefore, we expect that the contamination in our sample from the field dwarfs is not significant.}

\begin{figure}
  \begin{center}
   \includegraphics[angle=0,width=1\columnwidth]{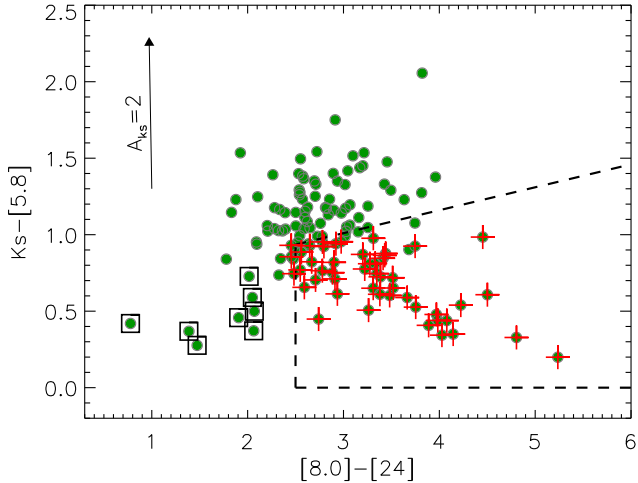} 
\caption{ [8.0]$-$[24] vs. $K_{\rm s}$$-$[5.8] color-color diagram. The dashed lines enclose the region for selecting transition disks. The pluses mark the transition disks, and the open squares show the other evolved disks in our sample (see Sect.~\ref{Sec:TD} and Fig.~\ref{Fig:DB_SED})  The large arrow show the reddening vector with $K_{\rm s}$-band extinction of 2\,mag.}\label{Fig:TDCCD}
\end{center}
\end{figure}

\begin{figure*}
\begin{center}
\includegraphics[angle=0,width=2.\columnwidth]{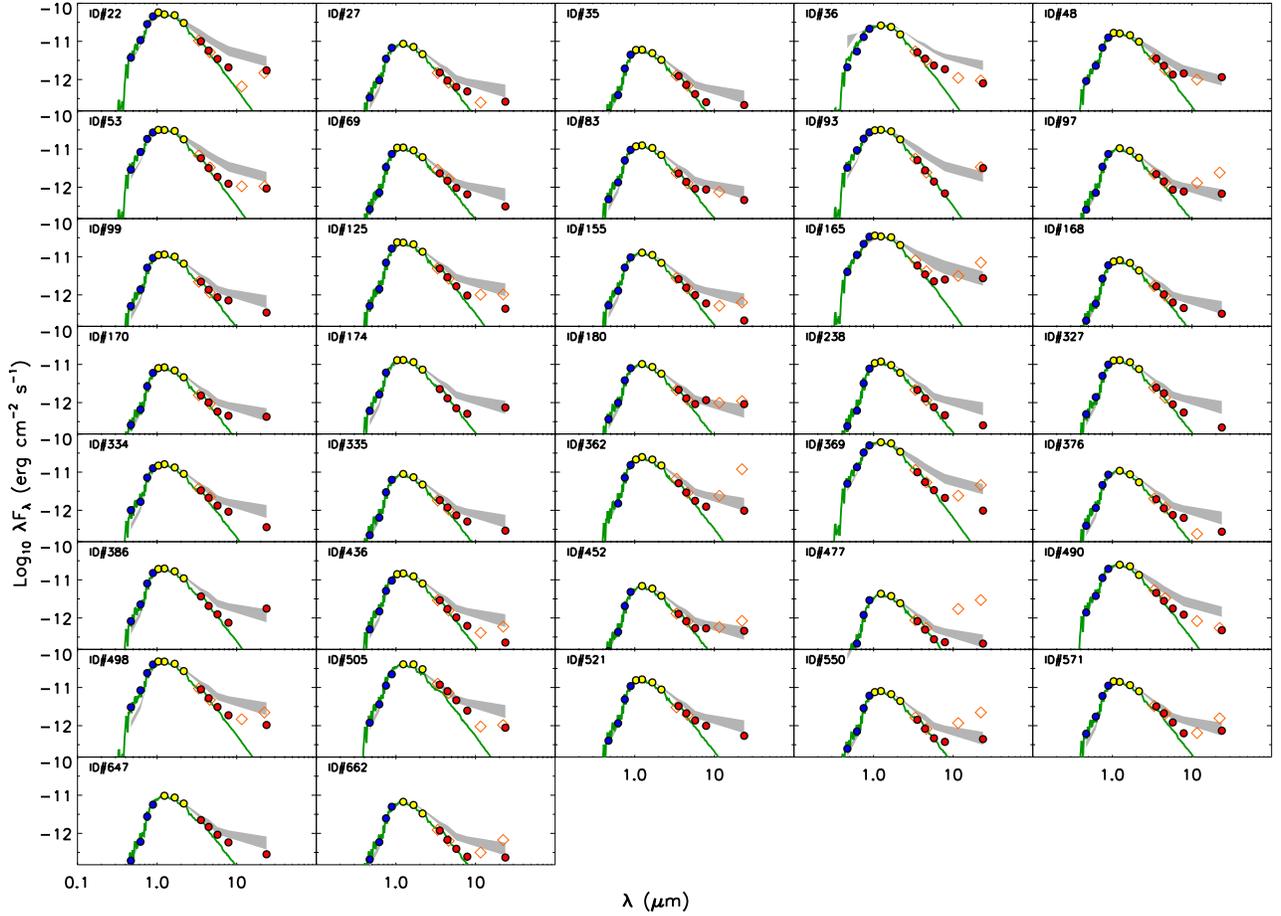}
\caption{The SEDs of newly confirmed YSOs with transition disks. The circles show the photometry in different bands. The open diamonds are the photometry from WISE. In each panel, the  green solid curve indicates the photospheric emission level, and the gray color filled region show the upper and lower quartiles of the median SEDs of L1641 CTTSs with the spectral types similar to our sources.}\label{Fig:TD_SED}
\end{center}
\end{figure*}

\subsection{Stellar properties}
We derive the extinction of individual sources in the way described in Section~\ref{Sect:extinction}. The typical uncertainty for our $A_{V}$ measurements is around 0.3\,mag, estimated from the sources without extinction in our sample. The resulting visual extinctions are listed in Table~\ref{Tab:spt}. With the derived effective temperatures and bolometric luminosities using the method described in \sect~\ref{Sect:extinction}, assuming a distance of 414\,pc, we place the stars in the H-R diagram in Fig.~\ref{Fig:HRD}. Most of our sources lie between the 0.1 and 3 Myr isochrones. We use a  distinct symbol (open square symbols) for an ``exotic'' object that is apparently subluminous and a extremely low-mass object (see Section~\ref{Sect:sublum} and Section~\ref{Sect:Planet}). We derive the masses and ages of the stars using the PMS evolutionary tracks from \citet{2015A&A...577A..42B}. For the stars  above the youngest isochrone ($\sim$0.5\,Myr) from the evolutionary  models, their masses are estimated using the 0.5\,Myr isochrone since the low-mass stars are evolving vertically in H-R diagram during the first several Mys. In Table~\ref{Tab:spt}, we list the stellar masses and ages of the stars. Seventeen objects in our sample have masses less than 0.075\,\Msun, and are brown dwarfs. Among them, the source 414 has the minimum mass ($\sim$0.018\,\Msun). In Fig.~\ref{Fig:Mass_Age_dis}, we show the mass and age distributions of our sample. The median mass and age of our sample is $\sim$0.15$M_{\odot}$ and $\sim$1.4\,Myr, respectively. In  Fig.~\ref{Fig:disk_age_dis}, we show the age distributions of the young stars with and without disks. Both populations show a similar age distribution with most of stars younger than 2.5\,Myr. The median ages of the two populations are 1.3, and 1.6\,Myr for diskless and disk populations, respectively.
 
\begin{figure*}
\begin{center}
\includegraphics[angle=0,width=2.0\columnwidth]{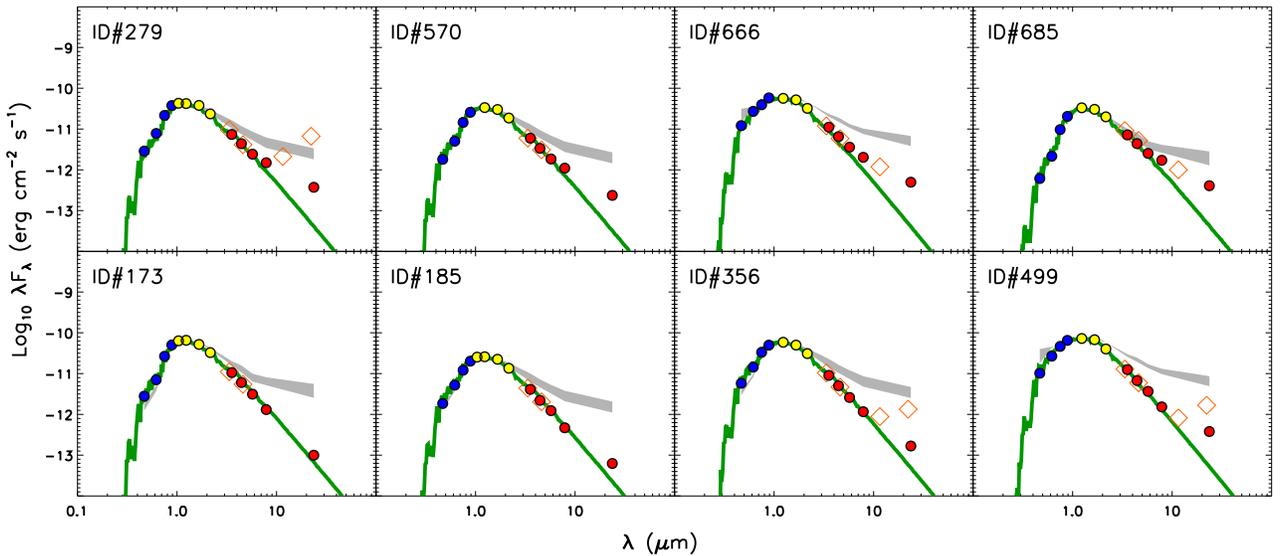}
\caption{Same as in Fig.~\ref{Fig:TD_SED}, but for the SEDs of other evolved disks.}\label{Fig:DB_SED}
\end{center}
\end{figure*}

{\rev We must stress that the stellar ages derived from the H-R diagram are dependent on the distance that we use. Here, we assume that the foreground population is associated with the Orion molecular cloud, and use the same distance (414\,pc) as Orion, which is supported by the fact that the foreground population and the Orion molecular cloud show similar kinematics (see Sect~\ref{Sec:Kinematics})}.

\subsection{Disk properties}
\subsubsection{A subluminous object}\label{Sect:sublum}

In the H-R diagram (see Fig.~\ref{Fig:HRD}), one source (ID~16 in Table~\ref{Tab:spt}) appears to be subluminous compared with others with the similar spectral type.  In its spectrum, we detect \LiI\ absorption line, indicating that it is a PMS star. The isochrone age of this object is $\sim$23\,Myr. However, its optical spectrum show numerous strong emission lines. Among them (see the left panel in Fig.~\ref{Fig:sublum}), the H$\alpha$  are the strongest one with the EW of $-134\,\AA$. Such strong  emission lines in the spectrum of Source~16 indicate that it should be much younger than its isochronal age ($\sim$23\,Myr). Its SED  shows the strong infrared excess emission  (see the right panel in Fig.~\ref{Fig:sublum}), suggesting that it is surrounded by a disk. Similar young stars have been discovered in our previous spectroscopic surveys in Orion \citep{2009A&A...504..461F,2013ApJS..207....5F}, and also found in other regions, e.g.,  the Lupus 3 dark cloud, Taurus, and $\epsilon$~Cha \citep{2003A&A...406.1001C,2004ApJ...616..998W,2013A&A...549A..15F}.  One promising hypothesis for these exotic objects is that they are harboring disks with high inclinations, and the light seen mainly comes from photons scattered off the disk surface, and is therefore much reduced. The optical emission lines may arise in outflows or disk winds with emitting areas much larger than the central star, which allows at least part of the line fluxes to reach us relatively unattenuated.  We employ the radiative transfer code RADMC-3D \citep{2012ascl.soft02015D} to model the SED of Source~16. We set the stellar effective temperature to 3,664\,K (M0.8) and stellar radius 1.3\,$R_{\odot}$, corresponding to a PMS star with mass $\sim$0.4\,\Msun\ and an age $\sim$2\,Myr \citep{2015A&A...577A..42B}. We set the outer disk radius ($R_{\rm out}$) to  100\,AU, and assume a pressure scale height ($H_{\rm P}$) that varies as a power law with the radius ($R$), $H_{\rm P}/R\propto R^{1/7}$. We vary the disk mass ($M_{\rm d}$) from $10^{-4}M_{\star}$,  5$\times 10^{-4}M_{\star}$,  1$\times 10 ^{-3}M_{\star}$,  5$\times 10^{-3}M_{\star}$, to  $10 ^{-2}M_{\star}$, the inner disk radius ($R_{\rm in}$) from 0.1, 0.2, 0.3, to 0.4\,AU, $H_{out}/R_{\rm out}$($H_{out}$: the pressure scale height at $R_{\rm out}$) from 0.1, 0.2, to 0.3, and {\newnewrev disk inclinations from  60$^\circ$ to 83$^\circ$.} We find that the SED of Source~16 can be reproduced by a model with parameters $H_{out}/R_{\rm out}$=0.1, $M_{\rm d}$=$10^{-4}M_{\star}$--1$\times 10^{-3}M_{\star}$, and $R_{\rm in}$=0.1--0.4\,AU,  and an inclination {\newnewrev $\sim$79--82$^\circ$} (see Fig.~\ref{Fig:sublum} for a representative model). {\newnewrev  A lower or higher disk inclination can lead to significantly over-predicted or under-predicted fluxes at short wavelengths, respectively (see Fig.~\ref{Fig:sublum}).}

 {\newnewrev The prominent emission lines in the spectrum of ID~16 include Balmer lines, He~I lines at 5876\,\AA\ and 6678\,\AA, [O~I] lines at 5577\,\AA\ and 6300\,\AA, Ca\,II infrared triplet (8498, 8542, 8662\,\AA), as well as many other lines. Such a spectrum rich in emission lines is similar to the one of EX~Lup,  an M0-type young star and the prototype of EXor variable star \citep{2015A&A...580A..82S}. A detailed identification of all the lines and a comparison of ID~16 to EX~Lup would require spectra with high spectral resolution. If the underluminosity of ID~16 is due to the occultation and scattering of photospheric emission by a highly inclined disk, we would expect that the EWs of the  accretion-related lines are simialr to the ones typical to T~Tauri stars since they are formed in the magnetospheric infall flows, which are close to the stellar surface and should be similarly occulted. We notice that the source shows the normal He\,I\,$\lambda$6678\,\AA\ line, but $\sim$4 times higher EWs of He\,I\,$\lambda$5876\,\AA\ line, and $\sim$10 times higher EWs of  Ca\,II infrared triplet, compared with other CTTSs in Orion\citep[see e.g.][]{2009A&A...504..461F}. Both He\,I emission lines and  Ca\,II infrared triplet are related to accretion activities. The large EWs of these lines could be explained if ID~16 is actively accreting. The spectrum of ID~16 also shows [O~I] lines at 5577\,\AA\ and 6300\,\AA\ with $EWs$  which are also $\sim$10 times higher than the ones typical to T~Tauri stars \citep{2016ApJ...831..169S}. However, these [O~I] lines could be contaminated by telluric [O~I] emission lines since our spectral resolution is low. In the spectrum of ID~16 we did not clearly detect the the [S\,II] emission lines at 6716\,\AA\ and 6731\,\AA, but may see two peaks near them. The [S\,II] lines are thought to arise in jets emerging from accreting YSOs. New spectral data with high spectral resolution and signal-to-noise ratio is needed to identify them explicitly. Here, we must stress that the source ID~16 could be also a young star with a distance much larger than Orion. In this case, we do not need a highly inclined disk to reproduce its SED, and the distance of ID~16 could be $\sim$1\,kpc, if its age is around 1\,Myr.}



\subsubsection{A disk around an extremely low-mass object}\label{Sect:Planet}

{\rev In our sample, Source 414 has the latest spectral type. Its spectral type in this work is $\sim$M9$\pm$1, which is consistent with the one (M8) in \citet{2013AJ....146...85H} given the uncertainties. In the H-R diagram, this object is above the youngest PMS isochrone ($\sim$0.5\,Myr) from \citet{2015A&A...577A..42B}. Assuming a 0.5\,Myr isochrone,  the mass of a source with a spectral type M8--M9 is around 0.018-0.030\,\Msun. However, the mass of  0.018-0.030\,\Msun\ should be considered as the upper limit for  Source 414, since the effective temperatures of young extremely low-mass objects decrease during their evolution \citep{2015A&A...577A..42B}. In Figure~\ref{Fig:planet} we show the SED of this object. Its SED shows infrared excess emission in all four IRAC bands, suggesting that it possesses a disk.  We employ the radiative transfer code RADMC-3D to model the SED of the source 414. We set the stellar effective temperature to 2,570\,K (M9) and the stellar radius to 0.84\,$R_{\odot}$, and a disk mass to 1$\times 10 ^{-2}M_{\star}$, and the inner edge to the dust sublimation radius, set $R_{\rm out}$ to 50\,AU, and  assume $H_{\rm P}/R\propto R^{1/7}$. {\newrev We vary ($M_{\rm d}$) from $10^{-4}M_{\star}$,  5$\times 10^{-4}M_{\star}$,  1$\times 10 ^{-3}M_{\star}$,  5$\times 10^{-3}M_{\star}$, to  $10 ^{-2}M_{\star}$, $H_{out}/R_{\rm out}$ from 0.1, 0.15, 0.2, 0.25,to 0.3, and and {\newnewrev disk inclinations from  0$^\circ$ to 80$^\circ$.} We find that a flaring disk model with $H_{out}/R_{\rm out}$=0.25--0.3, $M_{\rm d}$=5$\times 10 ^{-4}M_{\star}$--1$\times 10 ^{-2}M_{\star}$ and an disk inclination $\lesssim$50$^{\circ}$ can  reproduce the SED of Source~414.  In Fig.~\ref{Fig:planet}, we compare the model SED with  $H_{out}/R_{\rm out}=$0.25 (a flaring disk model) with the one with $H_{out}/R_{\rm out}=$0.1 (a flat disk model).} Both flat and flared disks have been found around brown dwarfs \citep[][]{2005Sci...310..834A,2009ApJ...696..143P}.} 

\subsubsection{Evolved disks}\label{Sec:TD}

In our sample, 689 sources have infrared photometry in at least three IRAC or WISE bands and can be classified to be diskless or harboring disks. Among them, 72\% (495/689) are diskless, and the rest are  disk sources. Among the disk population, we select transition disk (TD) candidates based on colors as shown in Fig.~\ref{Fig:TDCCD}: [8.0]$-$[24]$\geq$2.5 and $K_{\rm s}$$-$[5.8]$\leq$0.56$+$([8.0]$-$[24])$\times$0.15. For individual candidates, we compare their SEDs with their model photospheric emissions, as well as the median SEDs of L1641 CTTSs \citep{2013ApJS..207....5F}, corresponding to their spectral types. A total of 49 sources are confirmed as TD objects which show  very weak or no infrared excess at near-infrared wavelengths and shorter IRAC bands, but strong excess emission at mid-infrared and longer wavelength. Among them, twelve TDs have already been described in the literarure \citep{2013ApJS..207....5F,2013ApJ...769..149K}. In Fig.~\ref{Fig:TD_SED}, we show the SEDs of the 37 new TDs in this work.

\begin{figure}
\begin{center}
\includegraphics[angle=0,width=1.0\columnwidth]{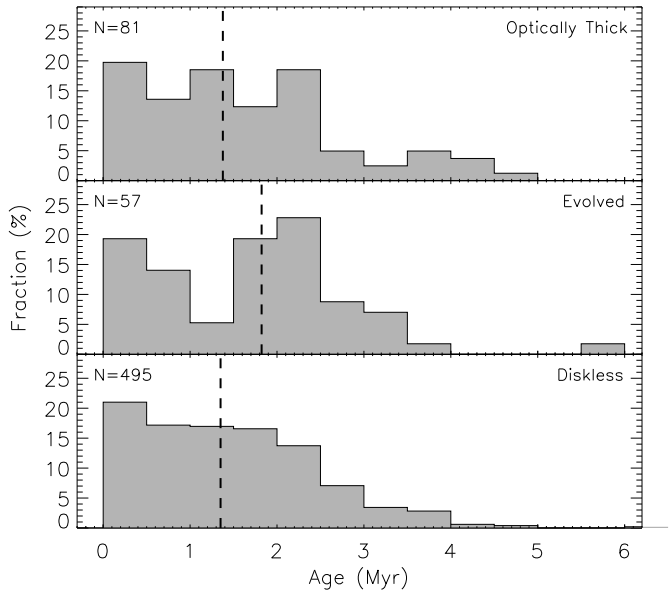}
\caption{{\newrev Histograms showing the age distribution for the young stars with optically thick disks, evolved disks, and without disks, respectively. The dashed line in each panel show the median age of each population.}}\label{Fig:Age_disk}
\end{center}
\end{figure}

{\rev In our sample, we also found 8 sources which are detected in 24\,\mum, and show much weaker infrared excess emission than a typical TD. Their infrared spectral slopes and colors are smaller than other sources (see Fig.~\ref{Fig:alpha} and ~\ref{Fig:TDCCD}). In Fig.~\ref{Fig:DB_SED}, we show the SEDs of those sources. Among these sources, one object (ID~666) shows the depleted infrared excess emission through the four IRAC band to 24\,\mum, and thus could be a globally depleted disk candidate, which can be produced when disks are deficient in small dust grains \citep{2009ApJ...698....1C,2011ApJ...742...39S}. In contrast to  Source 666, the SEDs of the objects 279, 570, and 685 start to show the infrared excess emission from 8\,\mum, while the others show infrared excess only at 24\,\mum. The weak infrared excess emission of the seven sources indicate that their inner disks have been more strongly dissipated  than a typical TD in the same region. } Without the data at far infrared bands, it is unknown if these sources are TDs with a big inner hole, or young debris disks. Actually, we also found two similar sources in L1641 \citep{2009A&A...504..461F,2013ApJS..207....5F}, one of which  shows strong excess emission at 70 and 160\,\mum\ \citep[Source\,069001 in][]{2013ApJ...767...36S}, suggesting that this object is a TD with a big inner hole. Similar objects have been found in the Tr~37 cluster \citep{2013A&A...559A...3S,2015A&A...573A..19S}.

{\newrev In  \fig\ref{Fig:Age_disk}, we show the age distributions of the three populations. The median ages of the sources with optically thick disks($\alpha_{3.6-8}\ge-1.8$), evolved disks, and without disks are $\sim$1.4\,Myr, 1.8\,Myr, and 1.4\,Myr, respectively. We note that the median age of the evolved disk systems is slightly larger than the other two populations. However, since each population shows a very broad distribution, the difference among their median ages is not significant.}


}

\begin{figure}
\begin{center}
\includegraphics[angle=0,width=1.0\columnwidth]{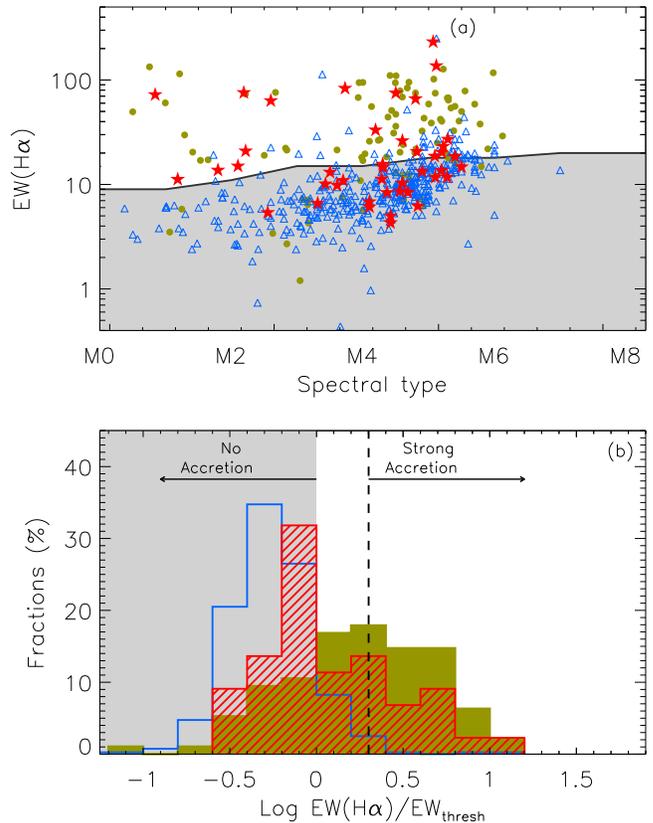}
\caption{ H$\alpha$ $EW$s vs. spectral type for PMS stars in this work. The filled circles are for YSOs with optical thick disks($\alpha_{3.6-8}\ge-1.8$), and triangles for YSOs without disks. The  star symbols show the transition disks in our sample. (b): the distribution of the ratio between the observed H$\alpha$ $EW$ and the $EW$ threshold, separating CTTSs or WTTSs, for the three populations in the left panel: YSOs with optically thick disks (filled histogram), YSOs without disks (open histogram), and transition disks (line-filled histogram). The dash line marks  $EW$(H$\alpha$)/$EW$$_{thresh}$=2. }\label{fig:EW_acc}
\end{center}
\end{figure}

\subsubsection{Accretion in disks at different evolutionary stages}
In Fig.~\ref{fig:EW_acc}(a), we compare the H$\alpha$~$EW$s for sources with TDs, optically thick disks($\alpha_{3.6-8}\ge-1.8$), or without disks. It can be noted that the YSOs with optically thick disks  usually present strong  H$\alpha$ emission, while diskless YSOs show weak H$\alpha$ emission. However, for TDs, about half of TDs show no accretion, while a small fraction of them show strong accretion.  In  Fig.~\ref{fig:EW_acc}(b), we display the distribution of logarithmic ratio between the observed H$\alpha$ $EW$ and the $EW$ threshold for the three populations shown in Fig.~\ref{fig:EW_acc}(a). Here, $EW$ threshold is the one  used to classify the YSOs into CTTSs or WTTSs, and is the spectral type dependent (see \sect~\ref{Sec:ana_SED_classification}).  According to these $EW$ thresholds,  73$\pm$9\% of YSOs with optically thick disks are accreting,  while only  46$\pm$7\% of TDs are accretors. If we define ``strong accretion'' sources to have H$\alpha$~$EW$s greater than twice the $EW$ thresholds, $25\pm7$\% of TDs belong to this group. For the YSOs with optically thick disks, this fraction is  45$\pm$10\%. {\newrev The accretion properties among the different types of disks are consistent with our previous studies \citep{2009A&A...504..461F,2013ApJS..207....5F,2010ApJ...710..597S,2013A&A...559A...3S}.} {\rev For the eight evolved disks shown in Fig.~\ref{Fig:DB_SED}, seven of them have estimates of H$\alpha$ $EWs$. Six of them show no accretion, and one (499) shows H$\alpha$ $EW$ which is just above the threshold to be classified as a CTTS. Therefore, the fraction of accretors among these sources should be less than 14\%, suggesting these objects are much more evolved than TDs.}

{\newrev }

\begin{figure*}
\begin{center}
  \includegraphics[angle=0,width=1\columnwidth]{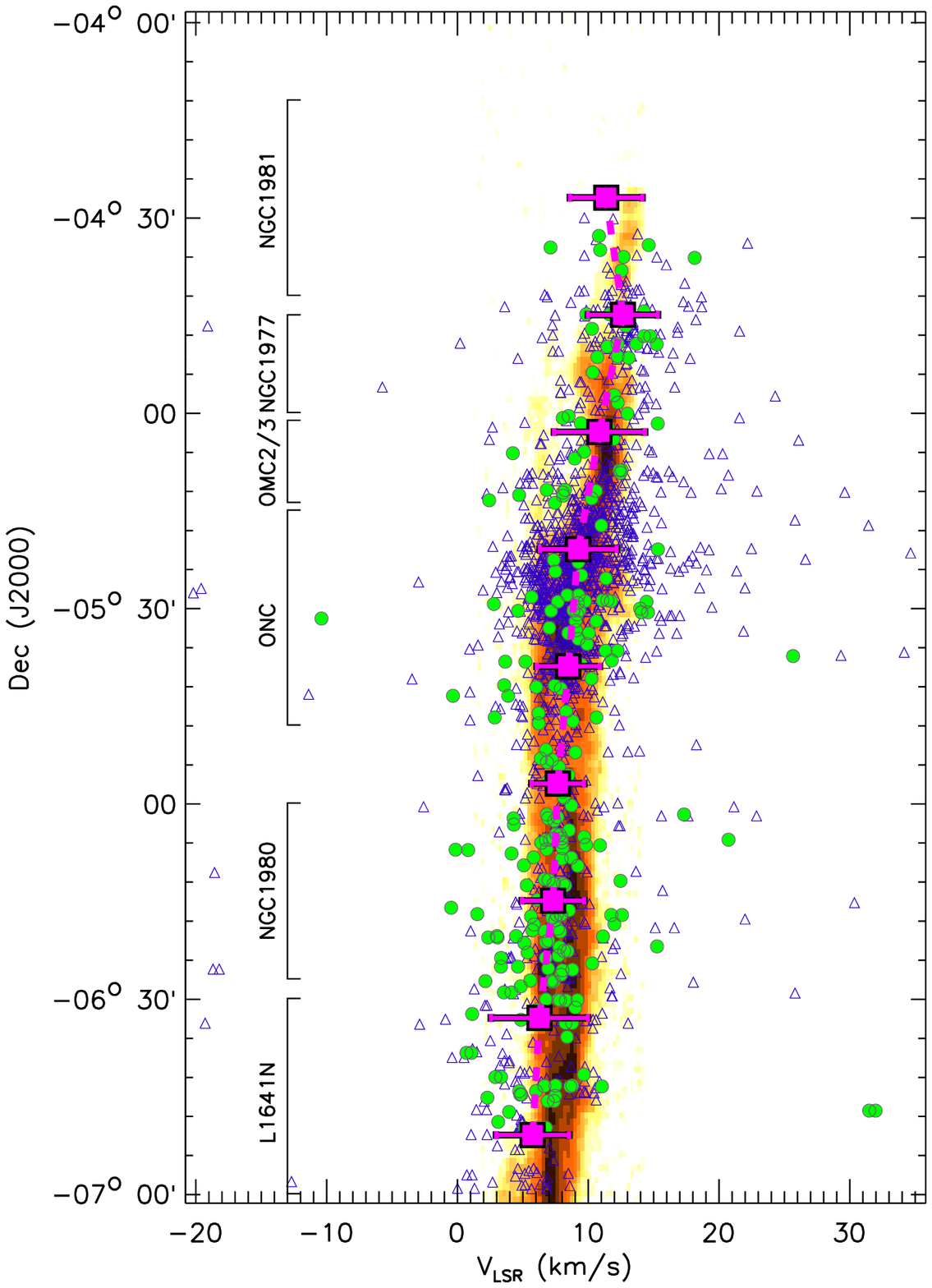}
  \includegraphics[angle=0,width=1\columnwidth]{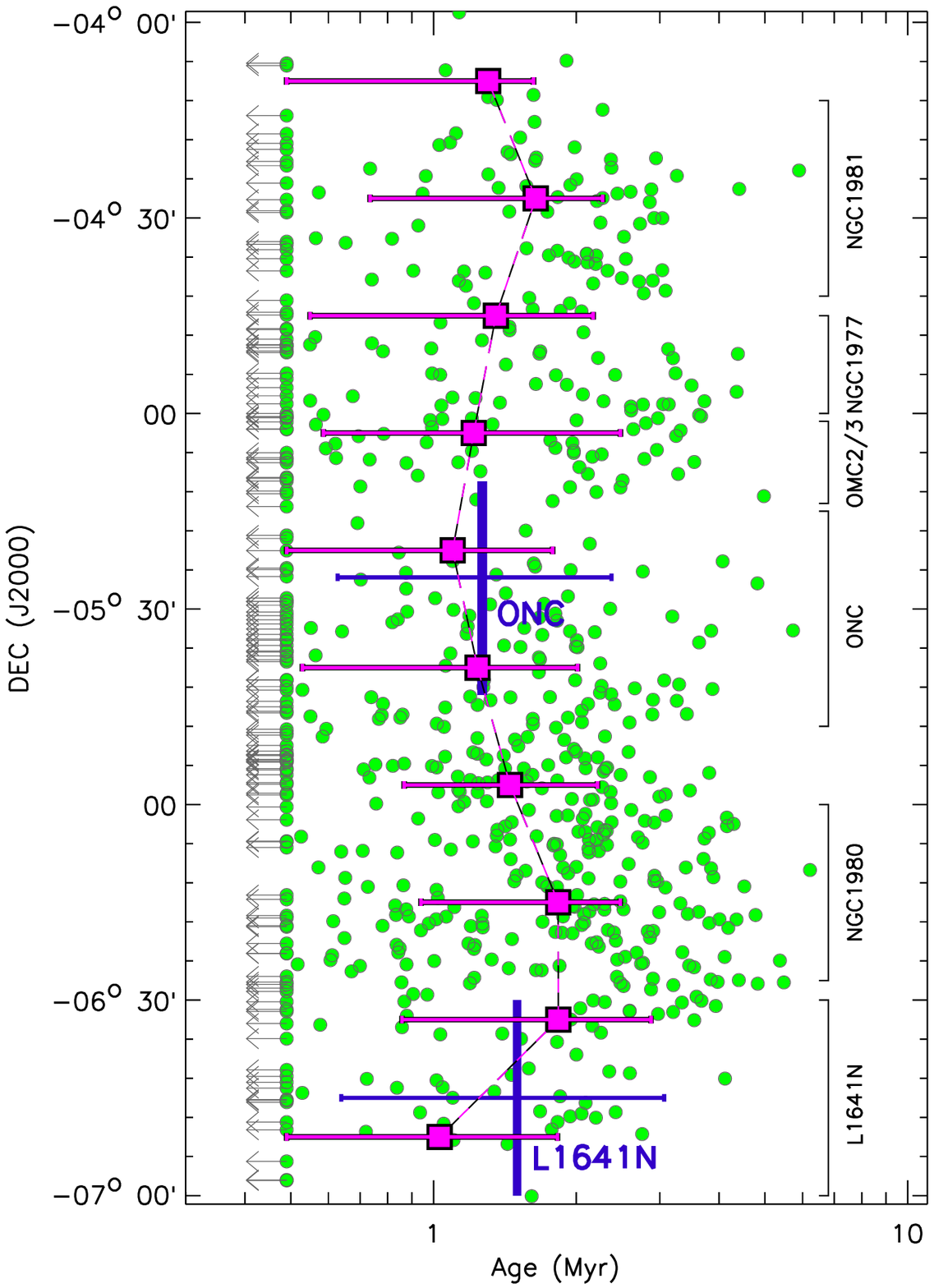}
\caption{Left:  Position-velocity (PV) of both stellar radial velocities and $^{13}$CO (J=1--0) emission (background color map) along the Orion A cloud. The filled circles are for the foreground population in \citet{2014A&A...564A..29B},  and triangles for young stars in Orion. The $^{13}$CO  emission is from \citet{1987ApJ...312L..45B}. The PV plot for the  $^{13}$CO  emission is summed over the range of R.A. shown in Fig.~\ref{Fig:Orion}. The filled squares connected with the dashed line show the mean radial velocities for individual declination bins with the err-bars showing the standard deviations. Right: the ages of the young stars in this work vs. their Declination. The left pointing arrow indicates the up-limits of ages for the stars with ages above the youngest isochrone ($\sim$0.5\,Myr) in the PMS evolutionary models of \citet{2015A&A...577A..42B}. The filled squares connected with the dashed line show the median ages for individual declination bins, and the err-bars show age ranges including 50\% of the sources near the median ages. The median ages of ONC and L1641 are also shown in the figure as a comparison.}\label{Fig:Orion_Kin_age}
\end{center}
\end{figure*}

\begin{figure}
  \begin{center}
 \includegraphics[angle=0,width=1\columnwidth]{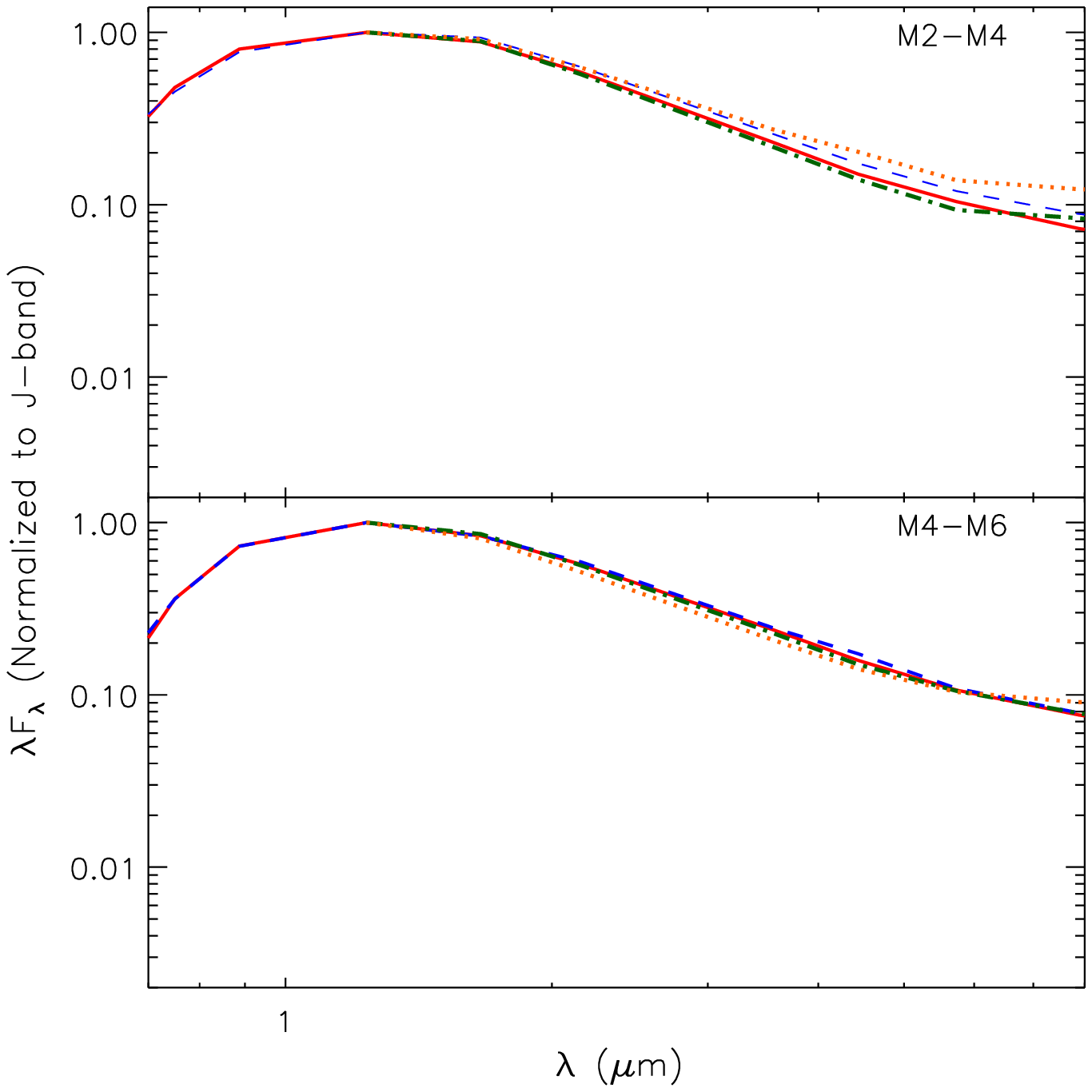}   
\caption{Median SEDs of class II sources with spectral type M2--M4 (upper) and M4--M6 (lower) in clusters of ONC  (dotted line, 1\,Myr), L1641 (dash line, 1.5\,Myr), NGC~1980 (solid line), and $\sigma$~Ori (dash-dotted line, 3\,Myr).
}\label{Fig:Median_SED}
\end{center}
\end{figure}

\section{Discussion}

\subsection{Is the foreground population associated with the Orion~A cloud?}\label{Sec:Kinematics}

\citet{2009ApJ...697.1103T} performed a kinematic survey of 1,613 stars which span from NGC\,1977 to L1641N. Their study suggests that the kinematics of young stars are consistent with the ones of their local clouds. Using the kinematic data from \citet{2009ApJ...697.1103T}, \citet{2012A&A...547A..97A} show that the velocity dispersion of young stars near NGC\,1980 is much smaller than those in other parts of Orion, though their kinematics are consistent with the gas. Therefore, they concluded that NGC~1980 is a distinct population from others. In this work, we use a better dataset to re-visit this issue. Our data are collected from the SDSS APOGEE INfrared Spectroscopy of Young Nebulous Clusters program (IN-SYNC) survey of the Orion A molecular cloud \citep[see detailed description in ][]{2016ApJ...818...59D}. In the region studied here, we find more than 2,200 sources, which have been observed with SDSS and have velocity uncertainties less than 0.5\,\kms. The size of this sample is much larger than the one (287 sources) with the same accuracy from \citet{2009ApJ...697.1103T}.  Among the APOGEE sample, more than 1,650 sources can be classified as young stars from the X-ray emission or infrared excess emission. In this dataset, we find 280 sources in the foreground population. In Fig.~\ref{Fig:Orion_Kin_age} (left), we show their radial velocity data as well as the ones of the other young stars in Orion~A. As a comparison, we also show the PV diagrams for the gas emission in Orion~A cloud. We do not find any significant difference between the kinematics of the foreground population, other young stars, and the gas material in Orion, which confirms the previous result in \citet{2009ApJ...697.1103T}. Futhermore, we derive the velocity dispersions of all the young stars in Fig.~\ref{Fig:Orion_Kin_age} (left) as did in \citet{2009ApJ...697.1103T}. However, we do not find significant difference between the the velocity dispersions of young stars at the different locations of Orion~A. Therefore, our study suggests that the foreground population, as other young stars in this region, are associated with the Orion~A cloud. {\newrev Here, we must stress that the kinematic study is only based on the radial velocities of the stars, and can be improved in the future with their proper motions from {\it Gaia}}

\subsection{Is the foreground population older than other regions in Orion~A?}
As shown in Fig.~\ref{Fig:Mass_Age_dis} (right), the median age of the foreground population with spectroscopic data is around 1.4\,Myr, which is younger than the age (4--5 Myr) proposed by \citet{2012A&A...547A..97A} for this population. In the previous work, the age is deduced based on two arguments: (1) the age of $\iota$~Ori, (2) the median SED of the disk population in NGC\,1980.  \citet{1986ApJS...61..419G} revealed that $\iota$~Ori is eccentric binary (O9~III+B1~III), and proposed that it could be formed as a result of binary-binary encounter that ejected the two runaways (AE~Aur, $\mu$~Col), which is confirmed by  \citet{2000ApJ...544L.133H} using the $Hipparcos$ data supplemented with the available radial velocities. The trapezium cluster is the most likely parent for the three sources. Therefore it is still unclear if the  $\iota$~Ori is related to the foreground population, and therefore uncertain to use the age of $\iota$~Ori as the one of the whole foreground population. When comparing the shapes of SEDs, \citet{2012A&A...547A..97A} normalized the SEDs of  different regions to the $J$-band flux. We note there are systematic differences between the SEDs of these regions at wavelength shorter than $J$-band, which may suggest that the extinctions of different regions have not been corrected when they constructed the SEDs. We estimate how the extinction affects the spectral slopes. For $A_{\rm V}$ = 5, the dereddened spectral slope between $J$-band and Spitzer 8\,\mum\ band can be 0.5 less than the observed spectral slopes. In Fig.~\ref{Fig:Median_SED}, we compare  the median SED of class II sources in this work with those in  Orion nebula cluster (ONC, 1\,Myr),   L1641 ( 1.5\,Myr), and $\sigma$~Ori (3\,Myr), using the data from the literature \citep{1997AJ....113.1733H,2010ApJ...722.1092D,2012AJ....144..192M,2009A&A...504..461F,2013ApJS..207....5F,2007ApJ...662.1067H,2014ApJ...794...36H}. These regions have been extensively surveyed with spectroscopy, which provides a reliable estimate of extinction for each source. In this work, most of sources are mid-M spectral types. Therefore, we only include the M-type stars for constructing the median SEDs. We divide the sources into two groups according to their spectral types: M2--M4 and M4--M6, since the luminosity of central stars can affect the infrared spectral slopes. For individual sources, their SEDs are first extinction-corrected before combination.  The extinction law is from \citet{2011ApJ...737..103S} for the SDSS bands adopting a total to selective extinction value typical of interstellar medium dust ($R_{V}=3.1$), \citet{1985ApJ...288..618R} for the 2MASS bands, and \citet{2007ApJ...663.1069F} for the Spitzer bands. Figure~\ref{Fig:Median_SED} show the median SEDs of different clusters. After correcting the extinction, we do not see any significant difference between the foreground population and other regions in Orion, especially for the M4-M6 group. {\rev The low disk fraction (28\%) of this population may indicate that this population may be older than its median isochronal age according to the relation between disk fractions and ages \citep{2001ApJ...553L.153H,2006ApJ...638..897S,2007ApJ...662.1067H,2012A&A...539A.119F,2013A&A...549A..15F}.} {\newrev However, this population is mostly likely biased against sources with hot inner disks, those showing $K_{\rm s}$-band excess emission, since their sample selection criteria select stars with  $i-K_{\rm s}$ or $H-K_{\rm s}$ colors similar to the intrinsic photospheric colors \citep{2012A&A...547A..97A,2014A&A...564A..29B}. The typical fraction of sources with hot inner disks is $\sim$40\%--50\% for a young population with an age of 1--2\,Myr \citep{2005astro.ph.11083H,2010ApJ...723L.113Y}. The low disk fraction (28\%) of our sample is most likely due to the exclusion of sources with hot inner disks.}

In Fig.~\ref{Fig:Orion_Kin_age} (right), we show the ages of sources in the foreground population vs. their declination. In different declinations, the ages of the stars all show large age spread with ages ranged from $\lesssim$0.5\,Myr to $\sim$6\,Myr. We divide the sources into different groups according to their declination, and derive the median ages. In the figure, we show these  median ages as well as the bars which show the ranges including 50\% of sources near the median ages. The median ages of the foreground population in the whole region are all around 1--2\,Myr, and clearly younger than the age (4--5\,Myr) proposed in \citet{2012A&A...547A..97A}. Along the declination, {\rev though it is not statistically significant}, the median ages of different regions seem to show a slight gradient, and the median ages of the stars around NGC\,1980 are slightly older (1.8\,Myr) than other regions, which is consistent with the result in \citet{2016ApJ...818...59D}. As a comparison, we show the median ages of ONC (1.3\,Myr) and L1641N (1.5\,Myr). For both regions, we re-estimate their ages using the same PMS evolutionary tracks from \citet{2015A&A...577A..42B} as we did for the foreground population. The effective temperature and luminosity of each star in ONC is collected from \citet{2010ApJ...722.1092D}, and those in L1641N are from \citet{2013ApJS..207....5F}. In addition, we only include the sources with spectral types between K7--M6, which is range of spectral types for the foreground population in this work. As shown in Fig.~\ref{Fig:Orion_Kin_age} (right), the median ages (1--2\,Myr) of the foreground population are consistent with the ones of ONC and L1641N. 

\subsection{Is the foreground population from one cluster?}
In \citet{2012A&A...547A..97A} and \citet{2014A&A...564A..29B}, they proposed that the foreground population in Orion is from one cluster and centered on NGC\,1980. However, as discussed in the above two sections, we can see that the kinematics of the foreground population are consistent with their local clouds and other young stars in the same regions. In Fig.~\ref{Fig:Orion_Kin_age} (left), the gradient in the kinematics of Orion~A cloud can be clearly shown in the kinematics of the foreground population, which is a strong evidence that the foreground population is associated with the local clouds instead of being one cluster. Furthermore, we may see an age gradient along the declination, and the median age of the sources near NGC~1980 is slightly higher than others. Such age gradient should not be seen if the foreground population are from the same cluster. Therefore, our results argue against the presence of an old, large foreground cluster in front of Orion~A. We propose that the foreground population seen in \citet{2012A&A...547A..97A} and \citet{2014A&A...564A..29B} is a combination of young stars with low or no extinction in NGC~1980 and other regions in Orion~A.

\section{Summary}

We performed a spectroscopic survey of the foreground population in Orion~A with MMT/Hectospec.  We combine the Hectospec data with optical and infrared photometric data to estimate the stellar effective temperatures, luminosities, and extinction values of individual sources, and derive  masses and ages of individual sources by their placement in the H-R diagram.  The disk properties of  individual sources are characterized using Spitzer and WISE data, and their accretion properties  are characterized using the H$\alpha$ line. We also use archival APOGEE radial velocity data to study the kinematics of the foreground population and other young stars  in the Orion A cloud. The main results are summarized as follows.

\begin{itemize}
\item[1.] We present a catalog of 691 young stars including their spectral types, line of sight extinction, stellar masses and ages, disk properties, and accretion properties.
\item[2.] We find one new subluminous object in our spectroscopic sample, and explain it as a star with a nearly edge-on disk.
\item[3.] We discovered an object with extremely low mass ($<$0.018--0.030\,\Msun). The SED modeling indicates this object possesses a flaring disk.
\item[4.] We identify 37 new transition disk objects, one globally depleted disk candidate, and 7 young debris disk candidates. We investigate the accretion properties of  YSOs with disks in our sample based on H$\alpha$ $EWs$. We find that the fraction of accretors among transition disks is much lower than among the YSOs with optically-thick disks (46$\pm$7\% versus 73$\pm$9\%, respectively), which confirm our previous results in L1641.
\item[5.] We confirm that the kinematics of the foreground population is consistent with their local clouds and other young stars in the same regions in Orion~A. The median age of the foreground population is also similar to those of other young stars in Orion~A. Therefore, our results suggest that the foreground population is associated with Orion~A, and prove that it is not a distinct large and old (4--5\,Myr) cluster in front of Orion~A. 
  
\end{itemize}

\acknowledgements
Many thanks to the anonymous referee for comments that help to improve this paper. This material is based upon work supported by the National Aeronautics and Space Administration under Agreement No. NNX15AD94G for the program ``Earths in Other Solar Systems''. The results reported herein benefitted from collaborations and/or information exchange within NASA’s Nexus for Exoplanet System Science (NExSS) research coordination network sponsored by NASA's Science Mission Directorate. L.Z. acknowledges supports from National Science Foundation of China (NSFC) grants 11303037 and 11390371/2 and support from the Chinese Academy of Sciences (CAS) through a CAS-CONICYT Postdoctoral Fellowship administered by the CAS South America Center for Astronomy (CASSACA) in Santiago, Chile. This research uses data obtained through the Telescope Access Program (TAP), which has been funded by the National Astronomical Observatories of China, the Chinese Academy of Sciences (the Strategic Priority Research Program "The Emergence of Cosmological Structures" Grant No. XDB09000000), and the Special Fund for Astronomy from the Ministry of Finance. Observations reported here were obtained at the MMT Observatory, a joint facility of the University of Arizona and the Smithsonian Institution.

\bibliographystyle{apj}
\bibliography{references}

\appendix

\section{Spectral classification of young stars}\label{Appen:spectral_classification}
\begin{figure*}
\centering
\includegraphics[width=0.8\columnwidth]{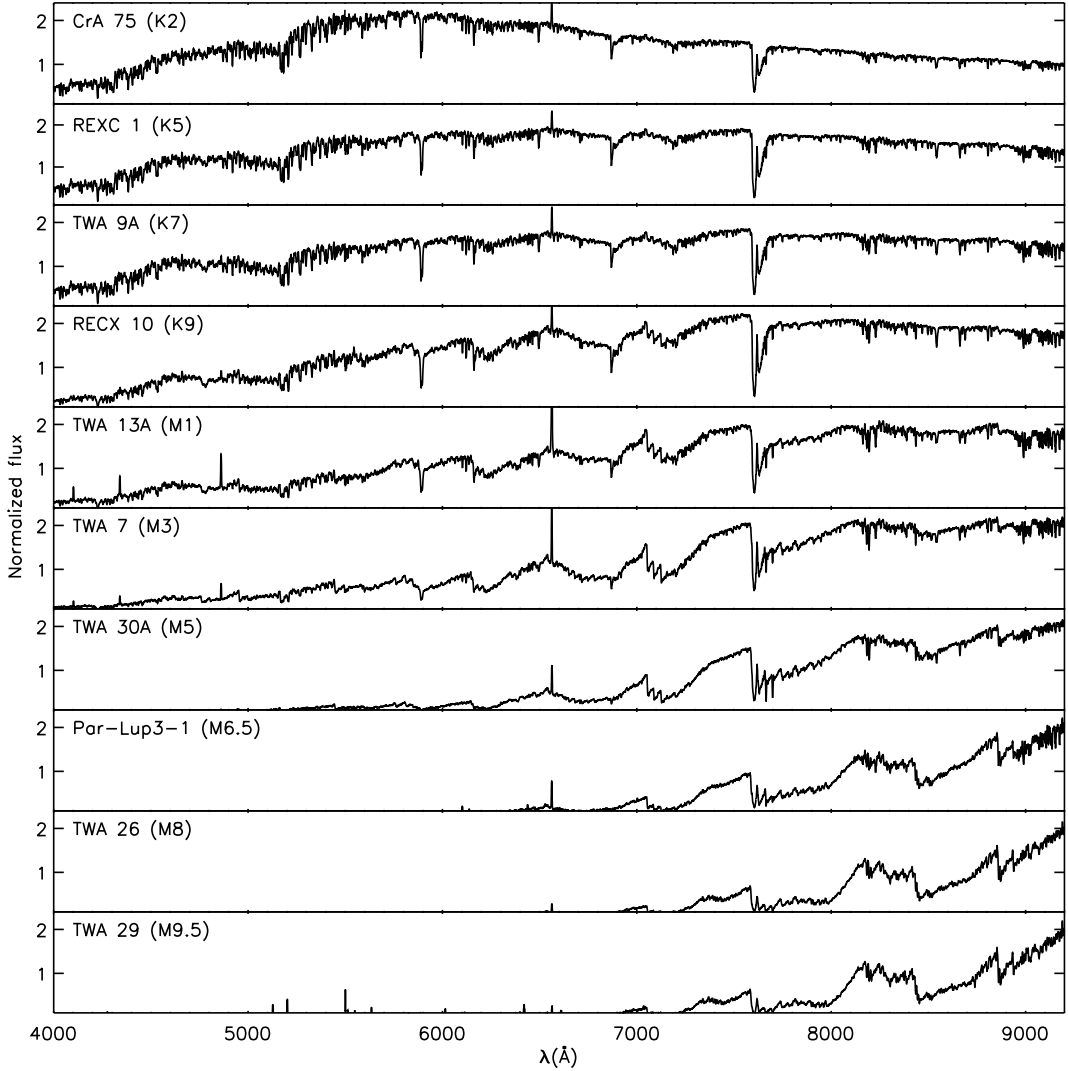}
\caption{Examples of X-Shooter spectra which are used to refine our spectral classification scheme} \label{Fig:exaple_spectra}
\end{figure*}

In Table~\ref{Tab:source}, we list the young stars with X-Shooter spectra. These sources are mainly from the $\eta$\,Cha cluster, the TW~Hydra Association, the Lupus star-forming region, the $\sigma$~Ori cluster, and the Cha\,I star-forming region.  We extract the spectra of these sources from the X-Shooter phase~III data archive. In Fig~\ref{Fig:exaple_spectra}, we show the example of spectra in our sample with spectral types from K2 to M9.5. In the figure, there is an obvious variation in the spectral type with the spectral type, which is mainly due to an change in the strength of molecular lines including the TiO, VO, and CaH bands. Thus these features can be used to do spectral classification. 

We use the  spectral features and define the indices of individual features in a similar way as do in the spectral classification code ``SPTCLASS'' \citep{2004AJ....127.1682H}. The index of each spectral feature is calculated by defining the central wavelengths of the feature band (FB), blue continuum band (BCB), and  red continuum band (RCB). The continuum flux ($F_{\rm FCB}$) at the central wavelengths ($\lambda_{\rm FB}$) of a FB is calculated as.

\begin{equation}
   F_{\rm FCB}=F_{\rm BCB}+\frac{\lambda_{\rm FB}-\lambda_{\rm BCB}}{\lambda_{\rm RCB}-\lambda_{\rm BCB}}\times(F_{\rm RCB}-F_{\rm BCB})
\end{equation}

Where $\lambda_{\rm BCB}$ and $\lambda_{\rm RCB}$ are the central wavelengths of BCB and RCB, respectively, and $F_{\rm BCB}$ and $F_{\rm RCB}$ are the flux of BCB and RCB, respectively.  $F_{\rm BCB}$ and $F_{\rm RCB}$ are the average fluxes over the widths of BCB and RCB. The index of a FB is calculated as:

\begin{equation}
   Index (FB)=\Delta\lambda_{\rm FB}\times(1-F_{\rm FB}/F_{\rm FCB})
\end{equation}

\begin{figure*}
\centering
\includegraphics[width=0.45\columnwidth]{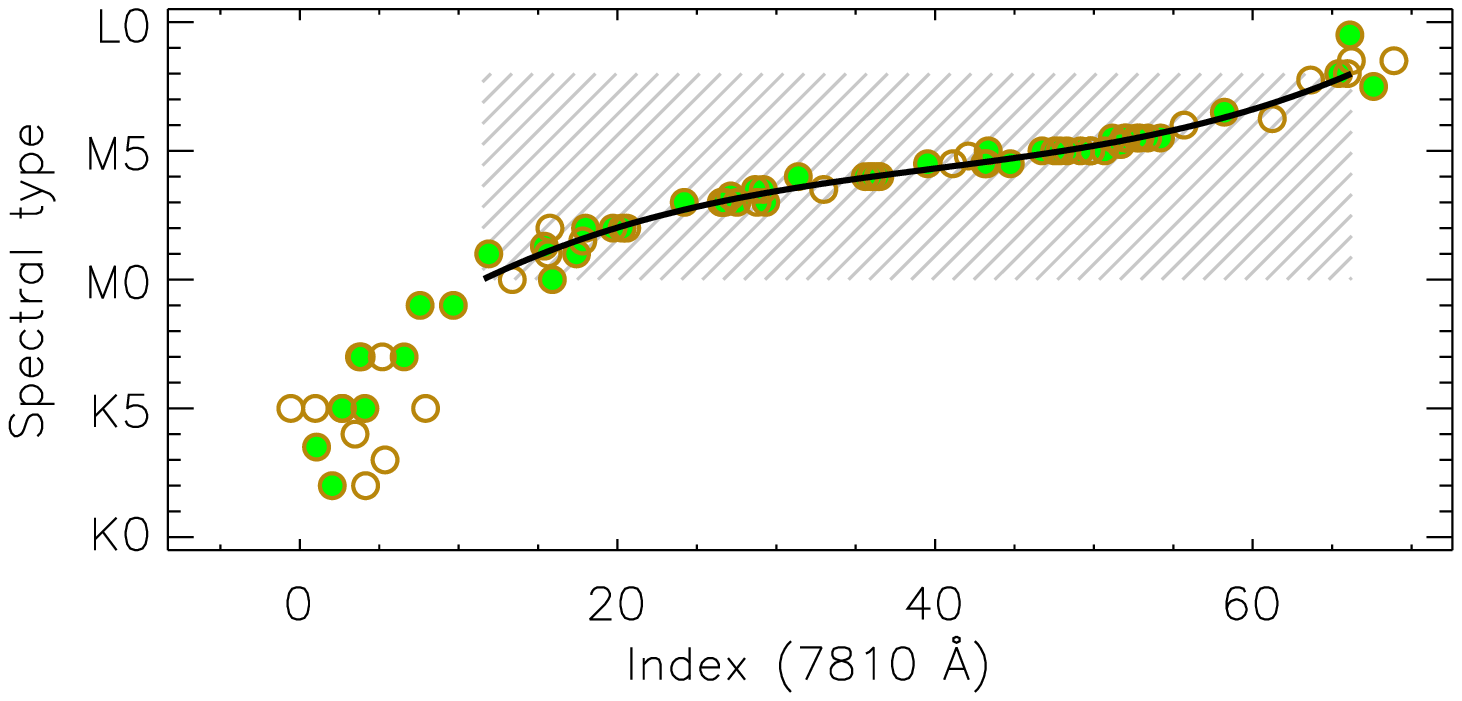}
\includegraphics[width=0.45\columnwidth]{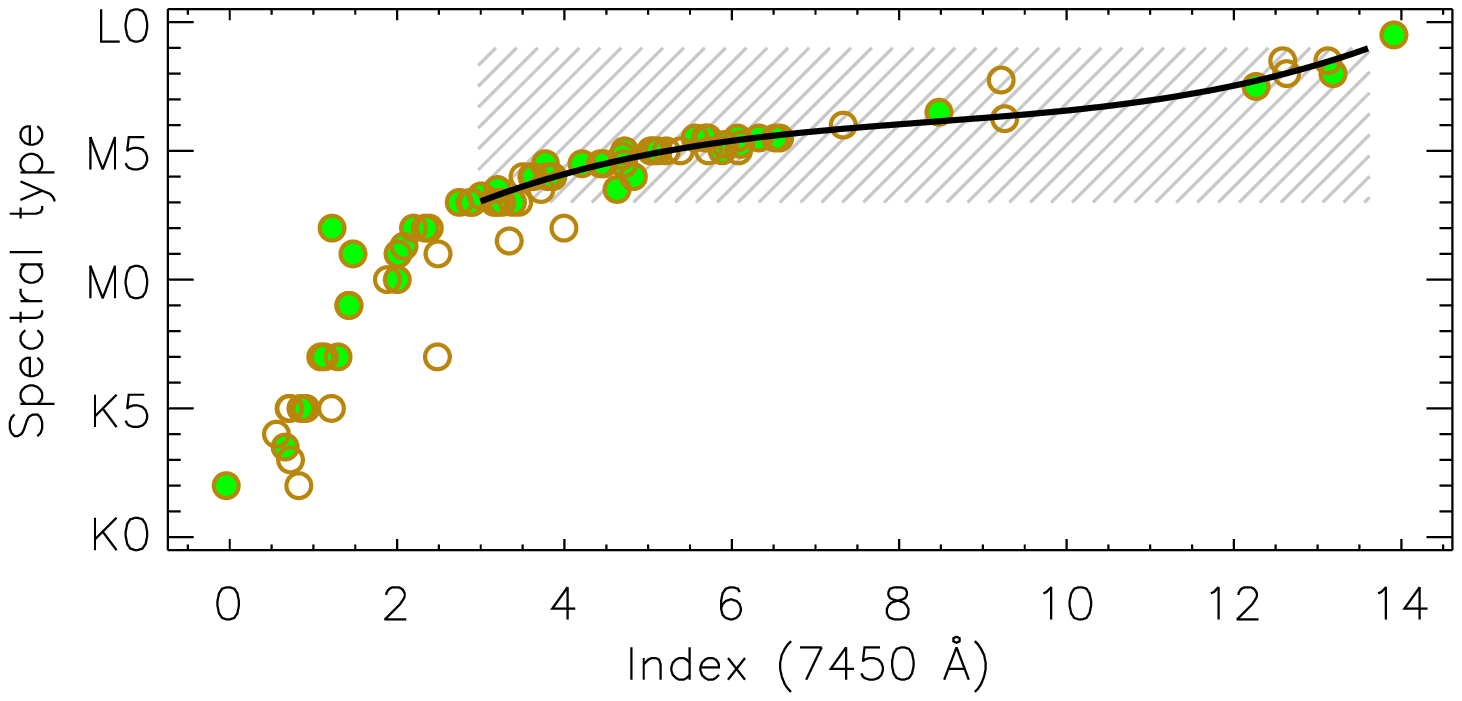}
\includegraphics[width=0.45\columnwidth]{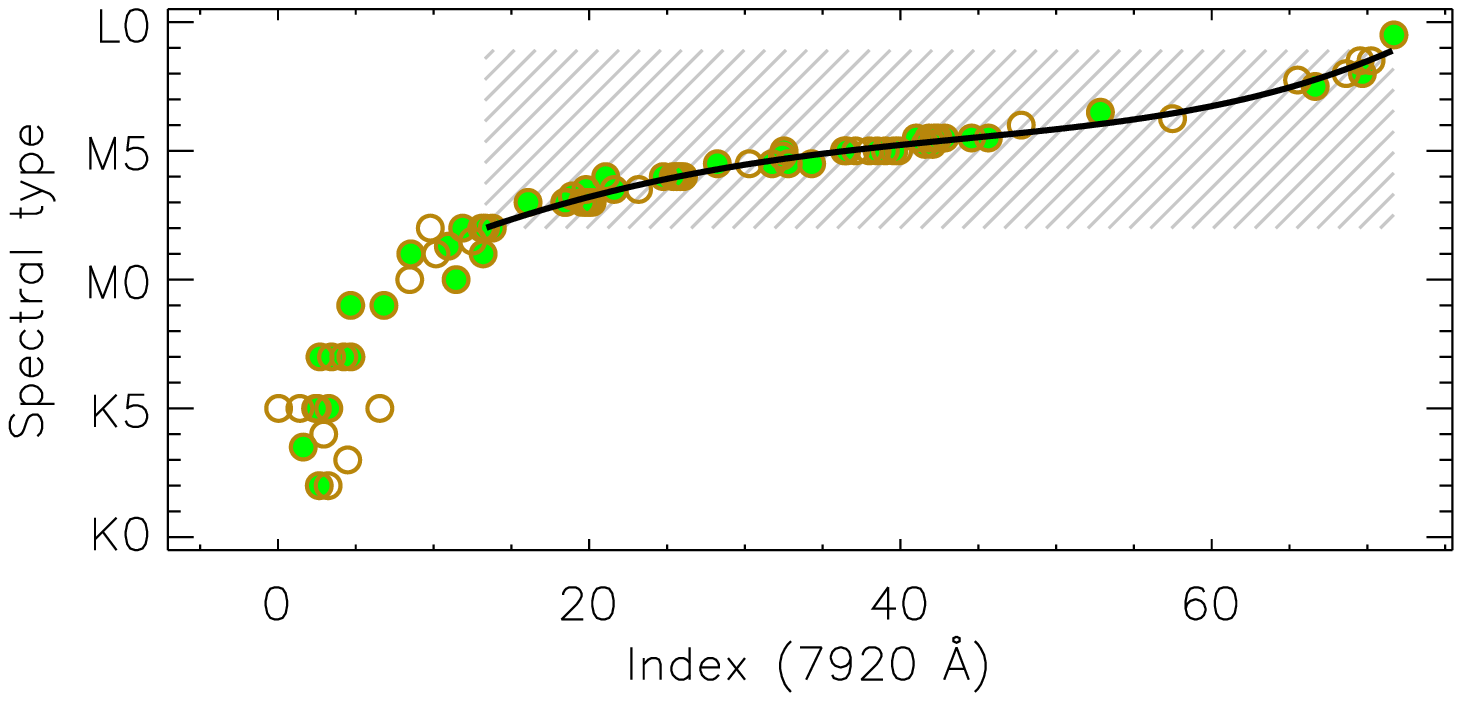}
\includegraphics[width=0.45\columnwidth]{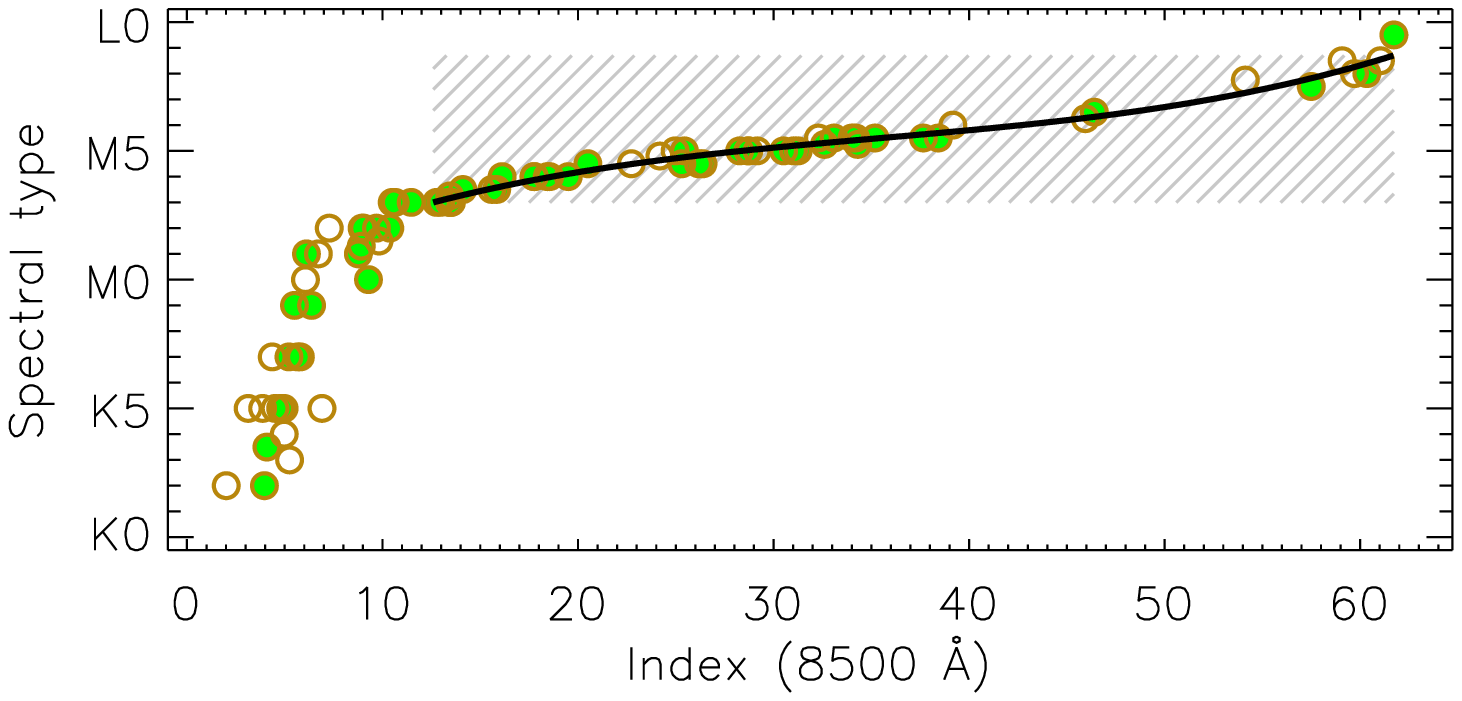}
\includegraphics[width=0.45\columnwidth]{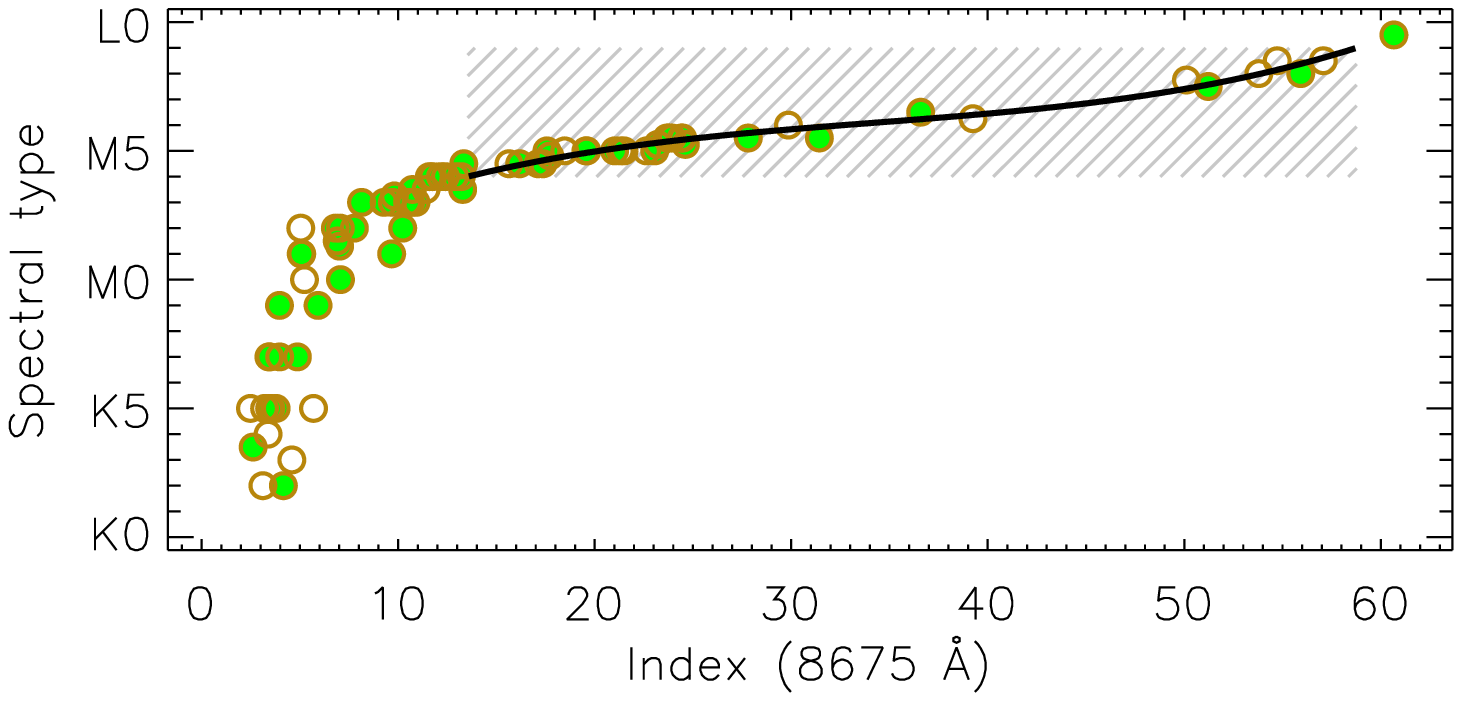}
\includegraphics[width=0.45\columnwidth]{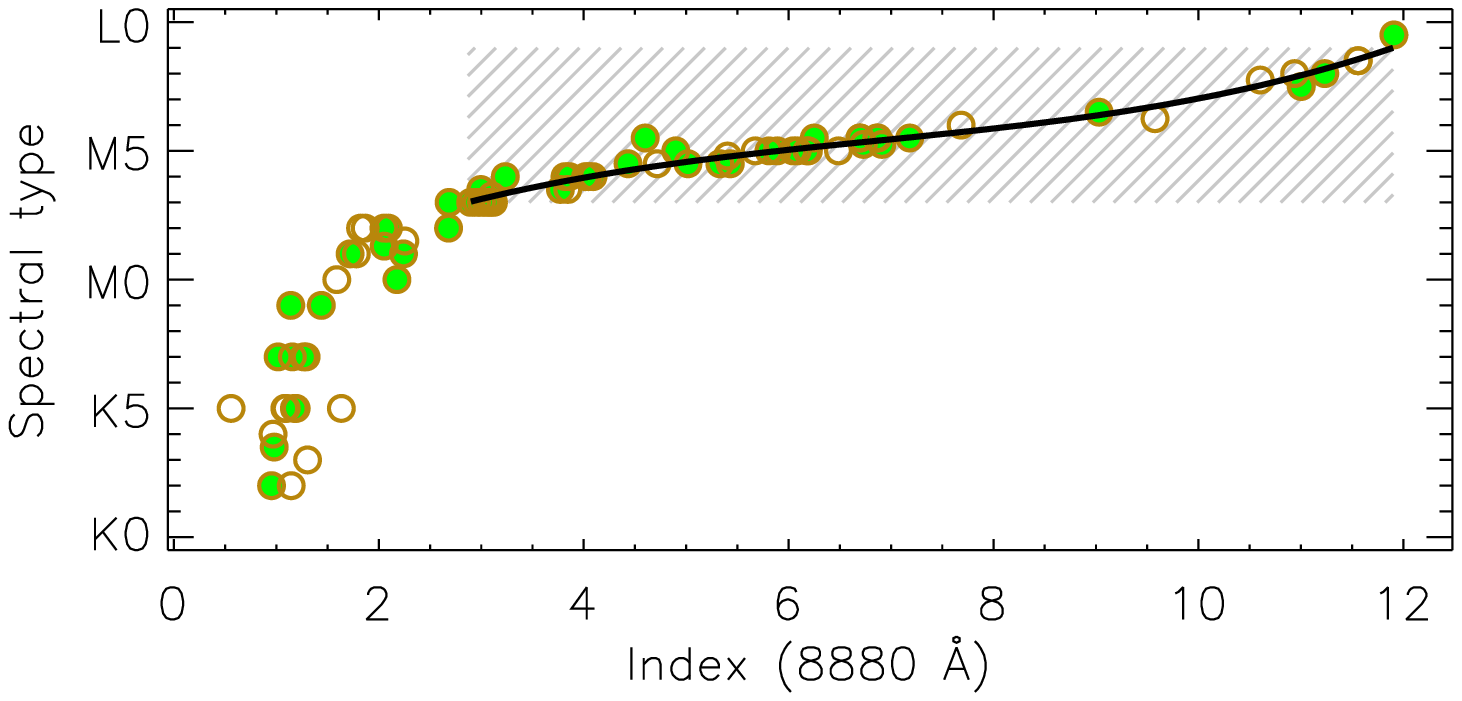}
\includegraphics[width=0.45\columnwidth]{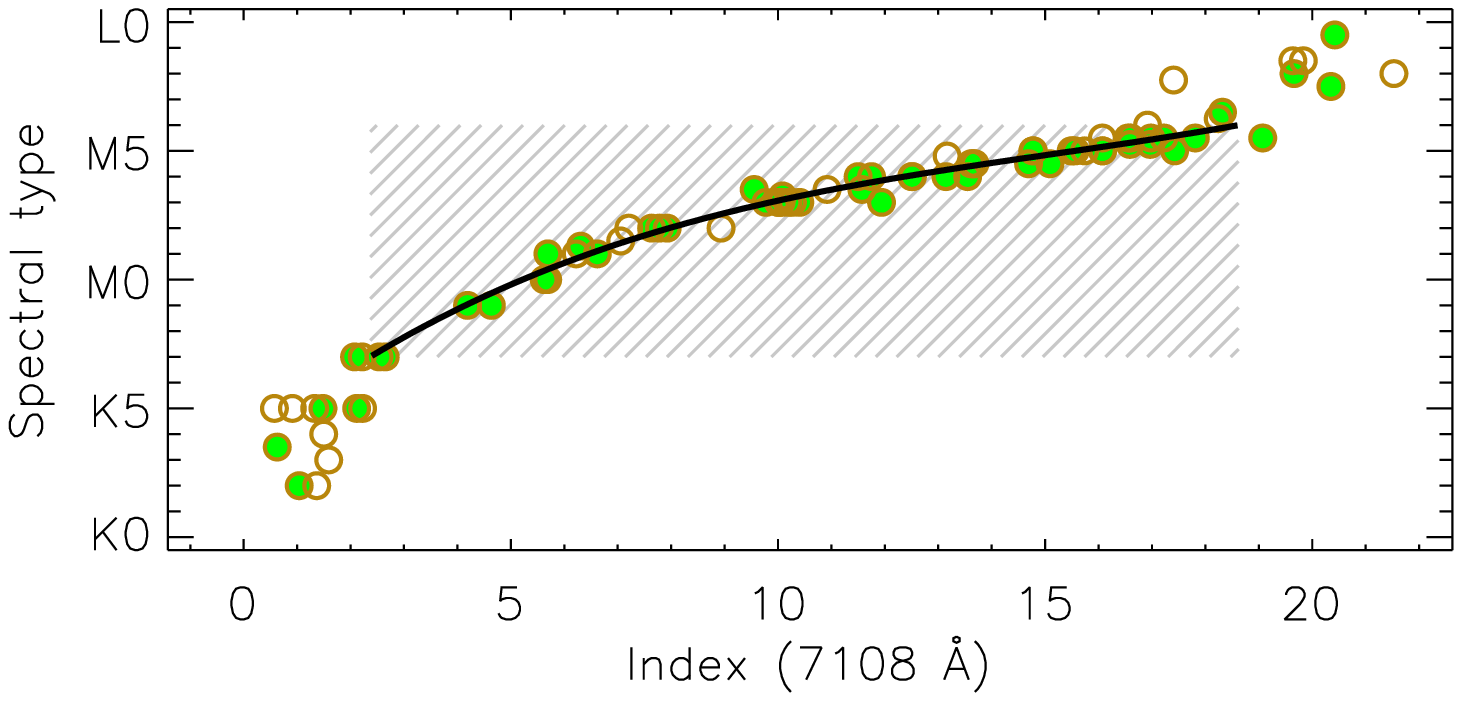}
\includegraphics[width=0.45\columnwidth]{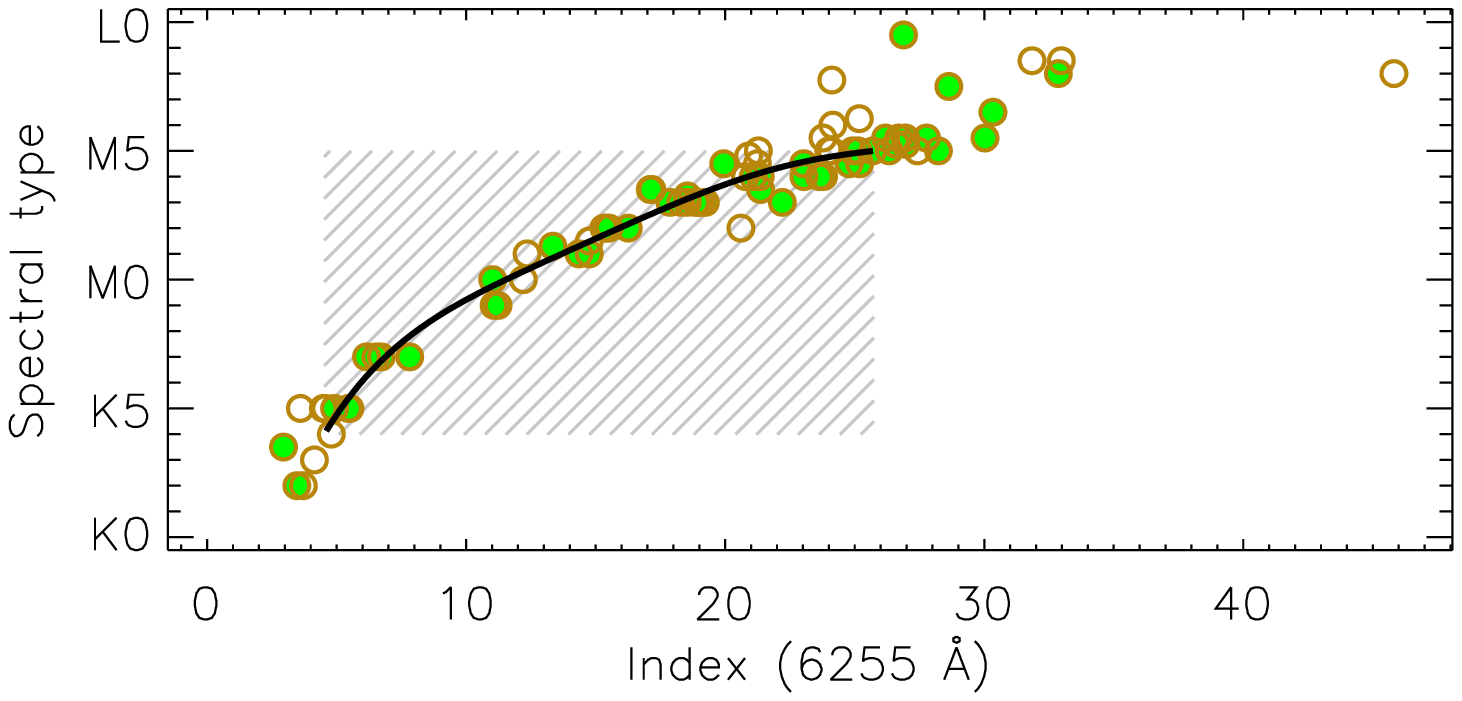}
\includegraphics[width=0.45\columnwidth]{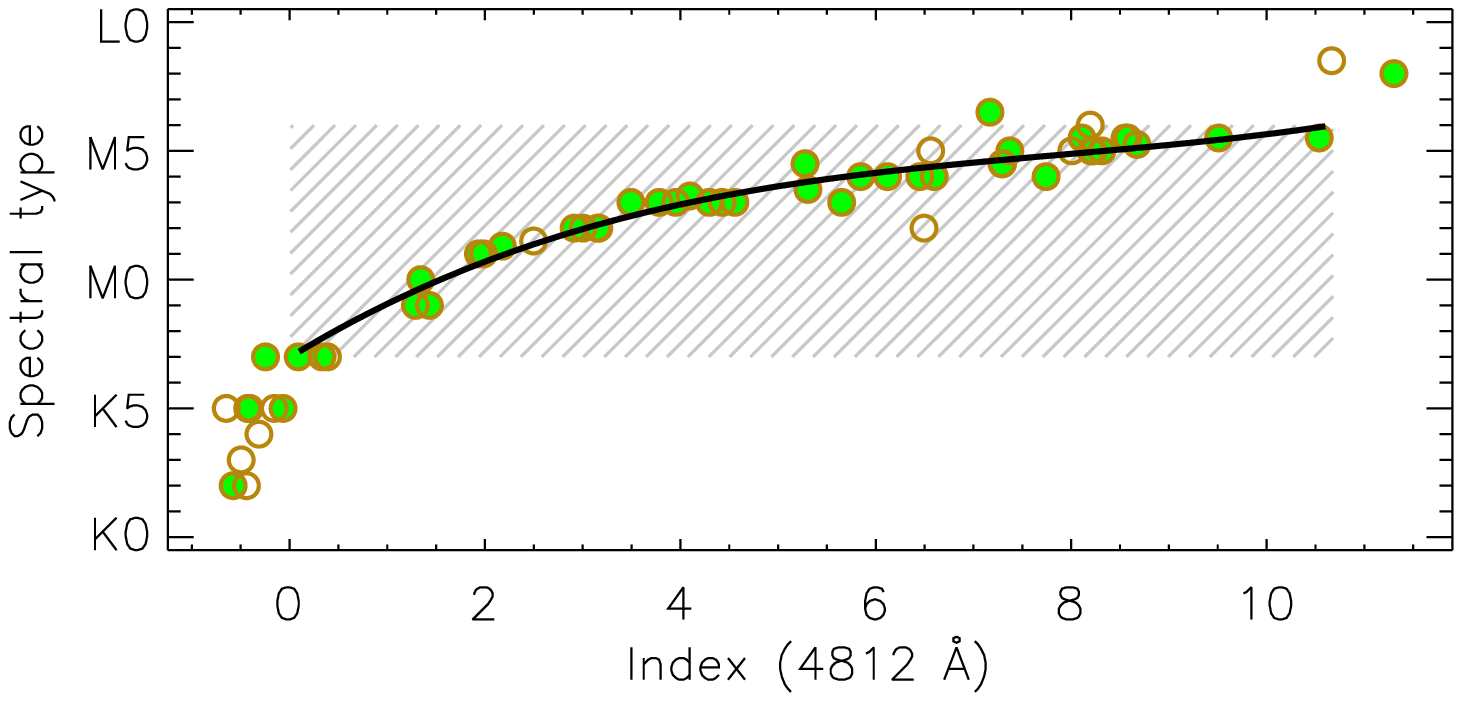}
\includegraphics[width=0.45\columnwidth]{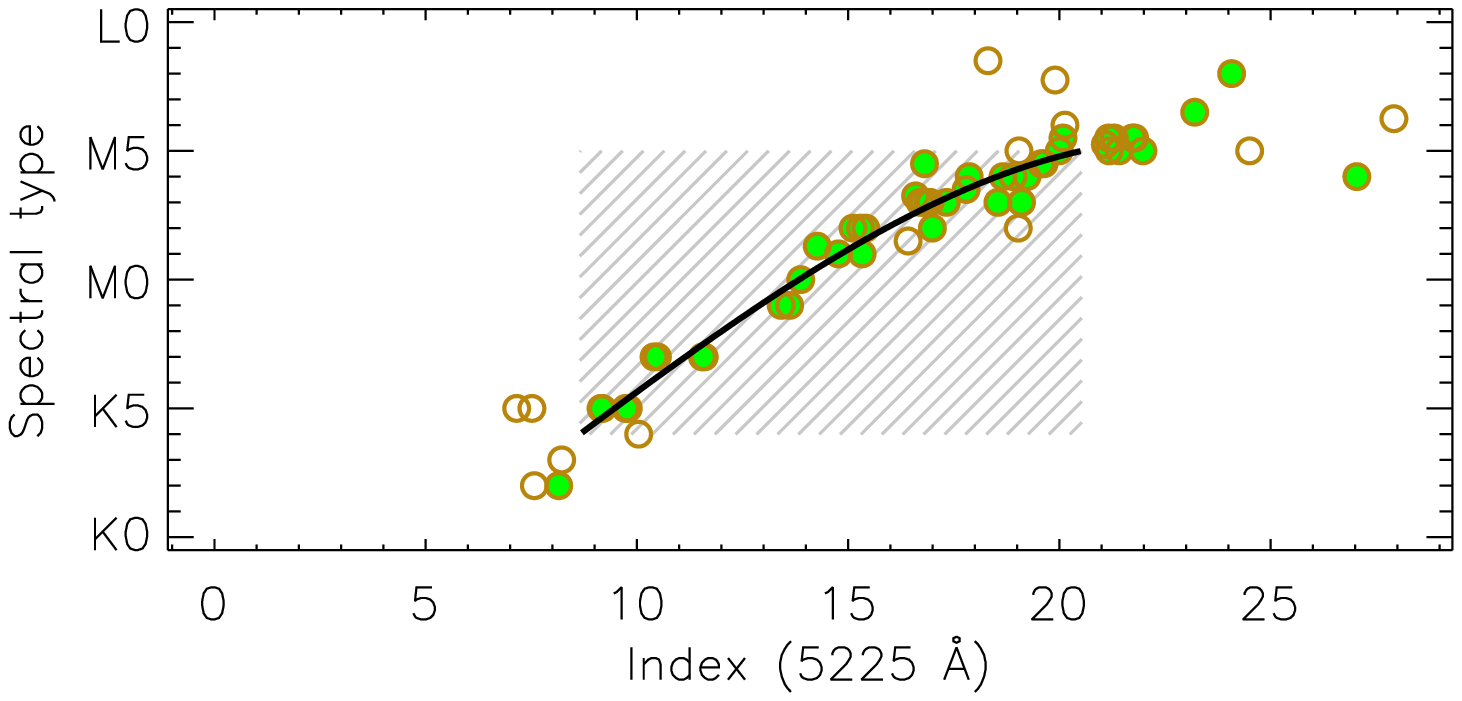}
\includegraphics[width=0.45\columnwidth]{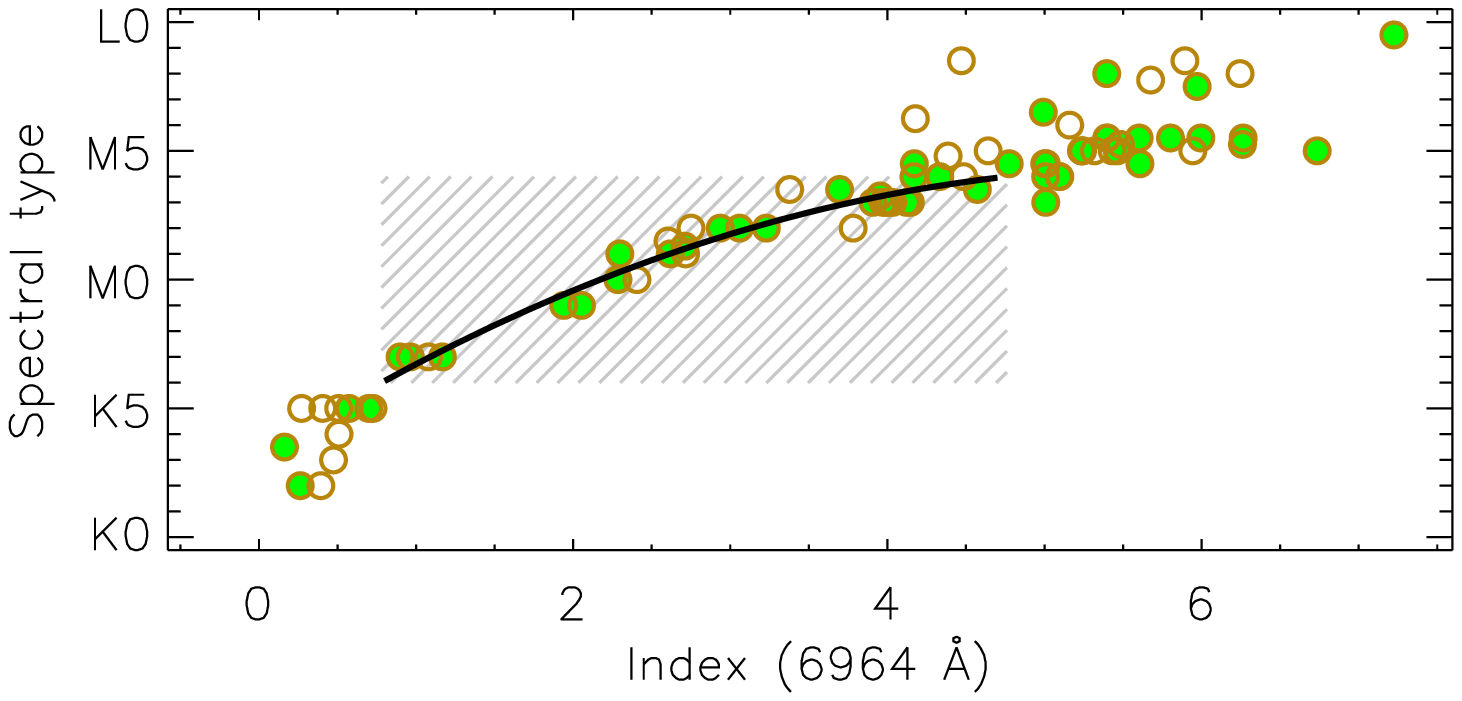}
\includegraphics[width=0.45\columnwidth]{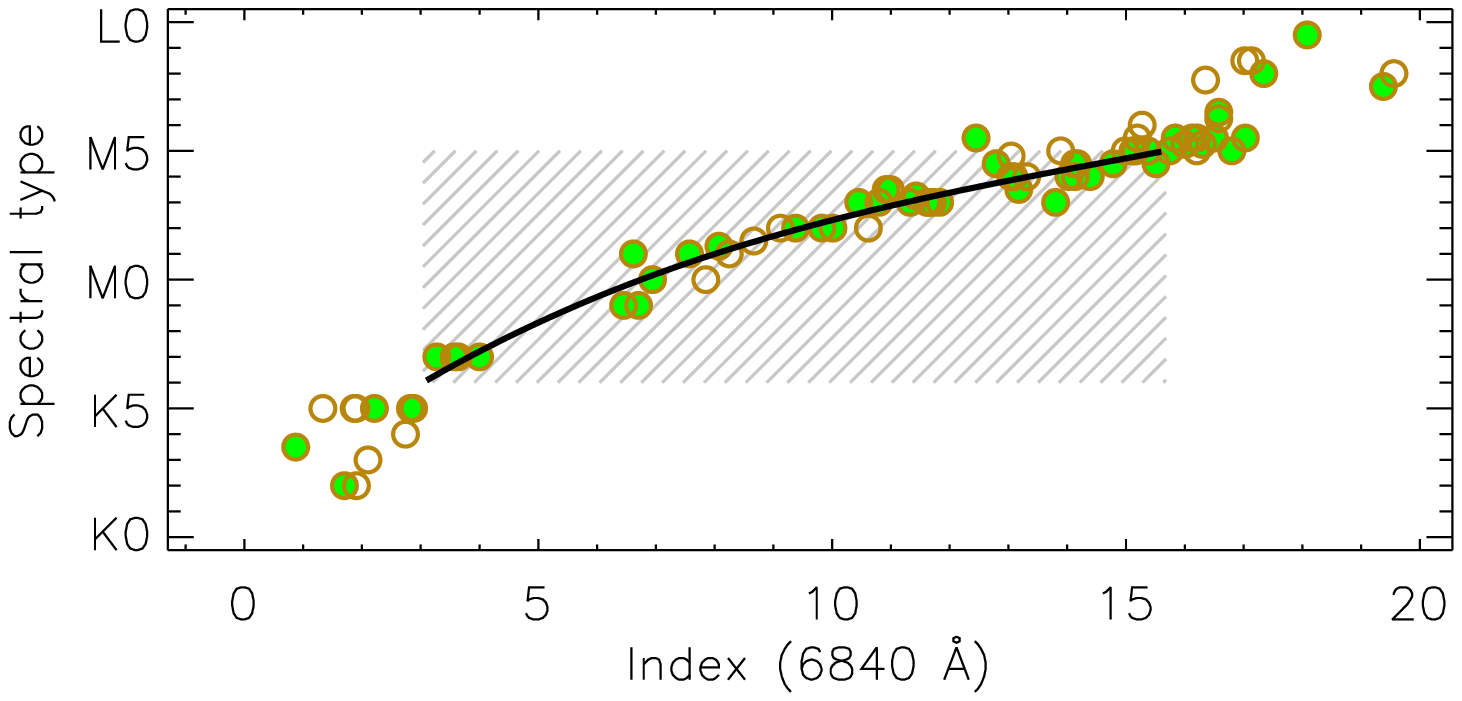}
\caption{The relations between indices of individual spectral features and their spectral types for the young stars listed in Table~\ref{Tab:source}.
The filled circles show the sources without accretion activity, and the open circles show the ones with weak accretion activity. The solid lines show the fits to the relations. The line-shaded regions mark the ranges for the fitting. \label{Fig:Index1}}
\end{figure*}

\begin{figure*}
\centering
\includegraphics[width=0.45\columnwidth]{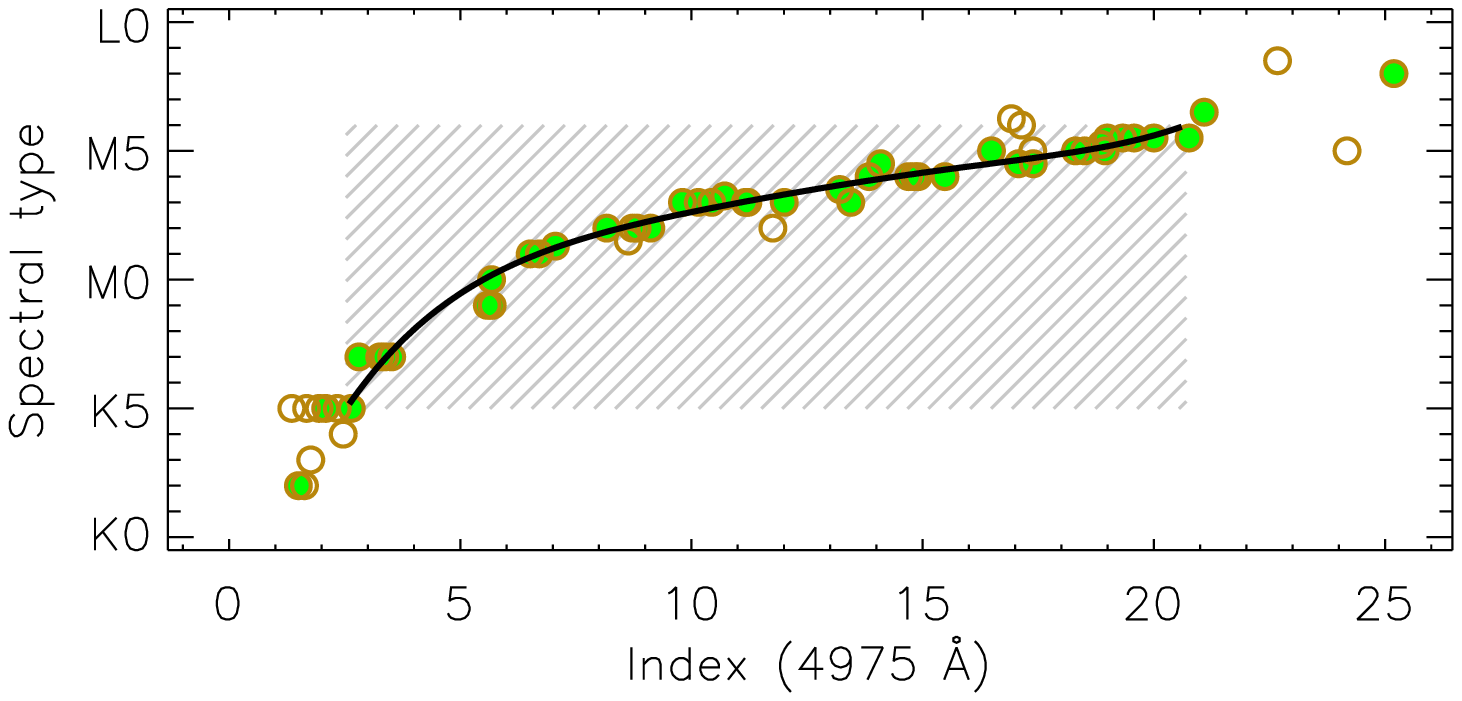}
\includegraphics[width=0.45\columnwidth]{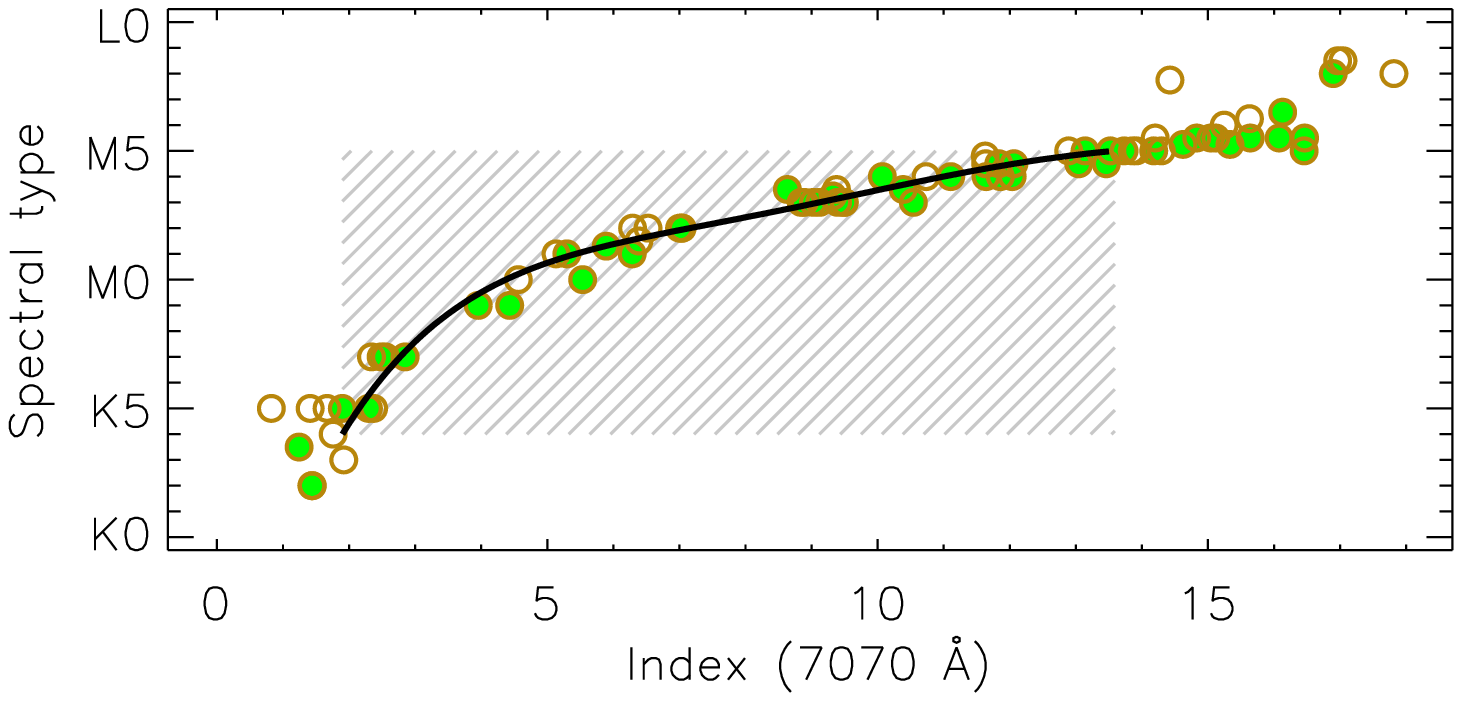}
\includegraphics[width=0.45\columnwidth]{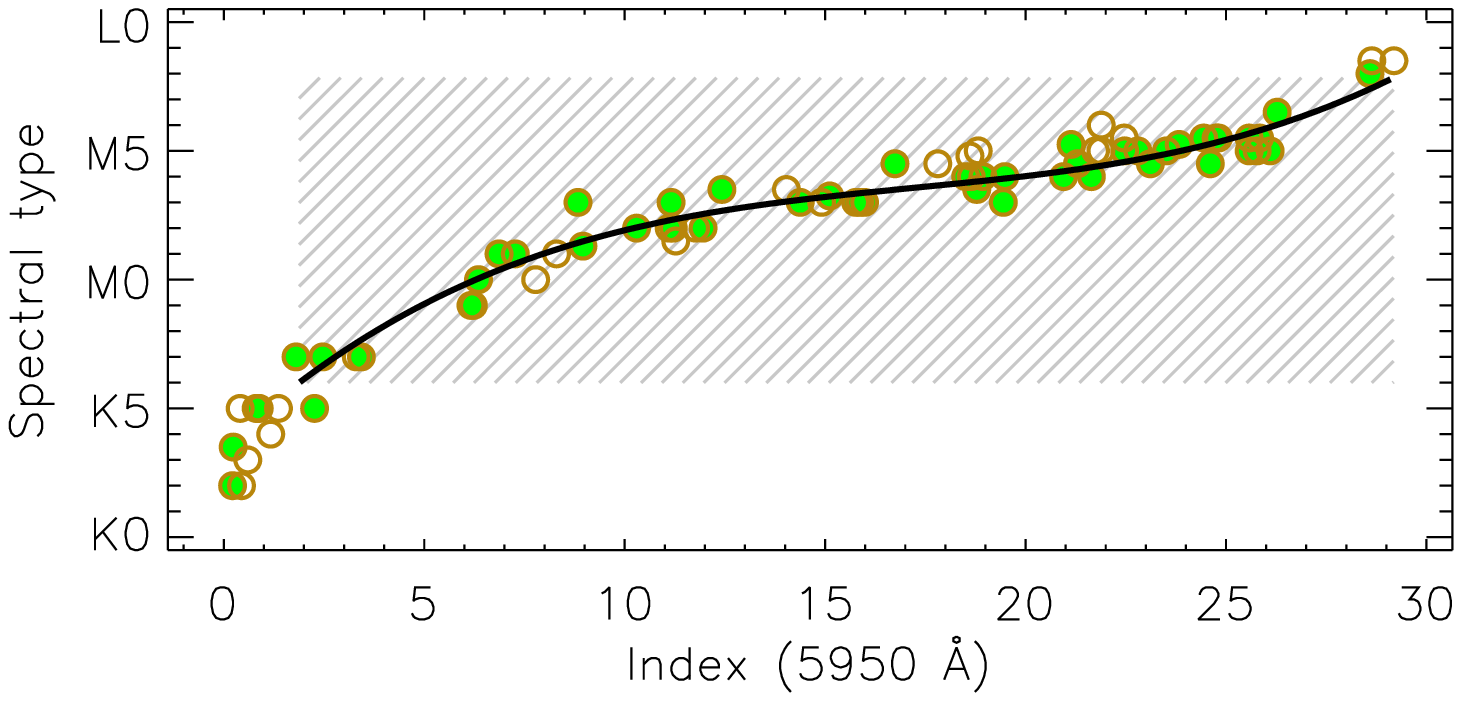}
\includegraphics[width=0.45\columnwidth]{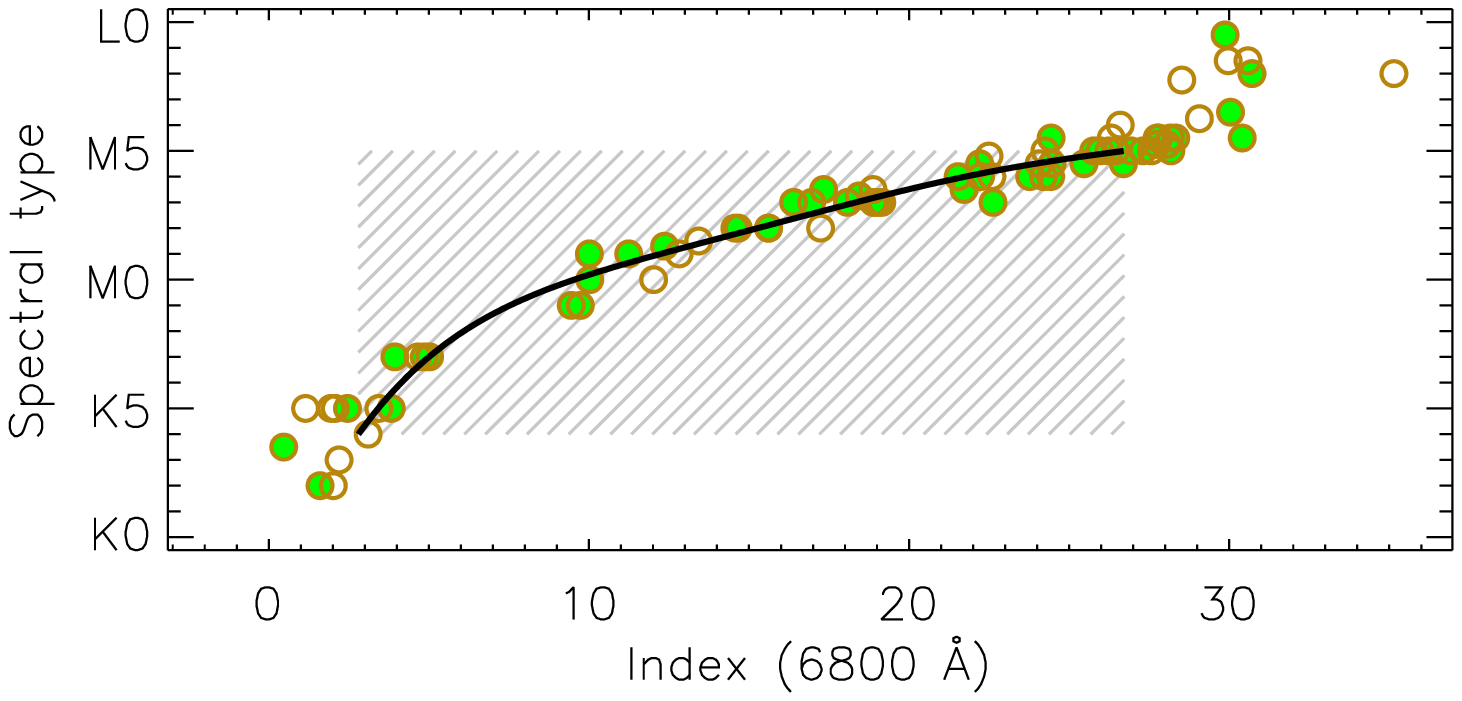}
\includegraphics[width=0.45\columnwidth]{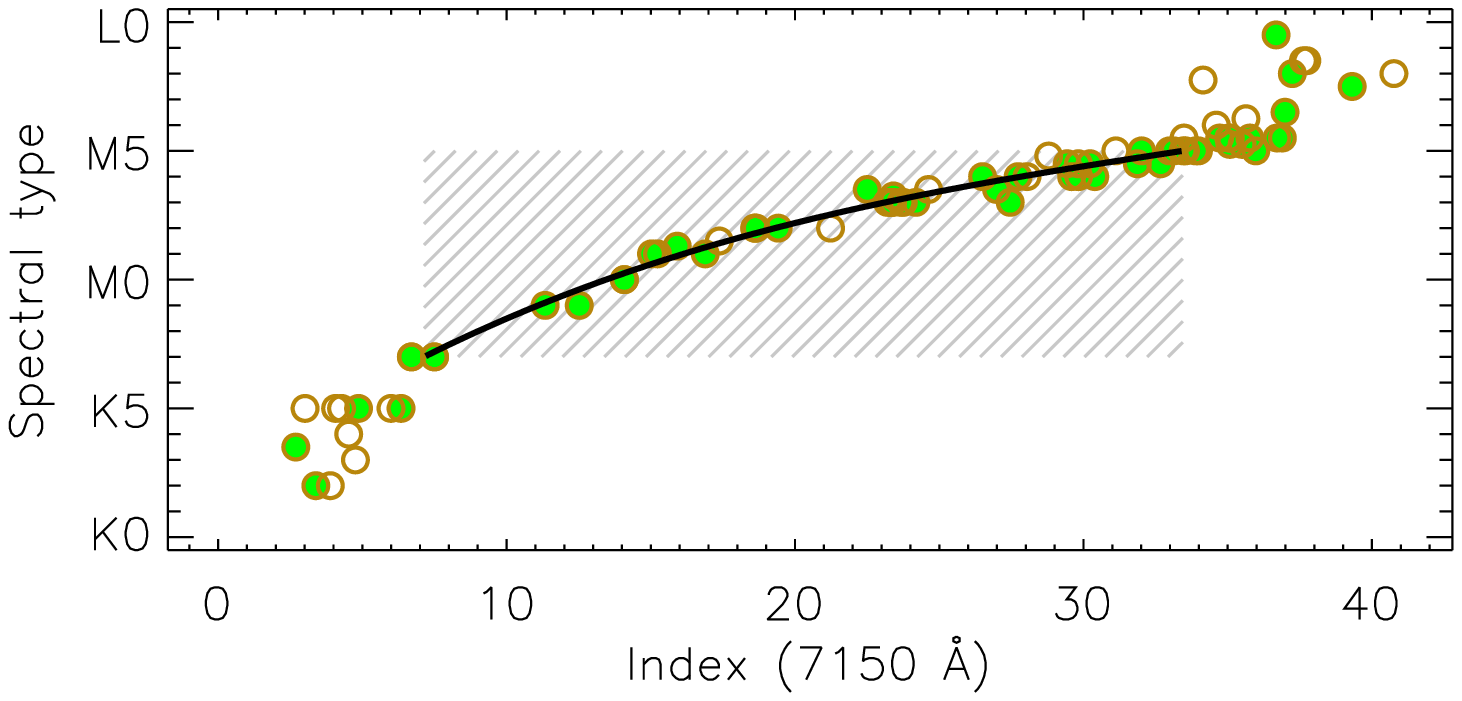}
\includegraphics[width=0.45\columnwidth]{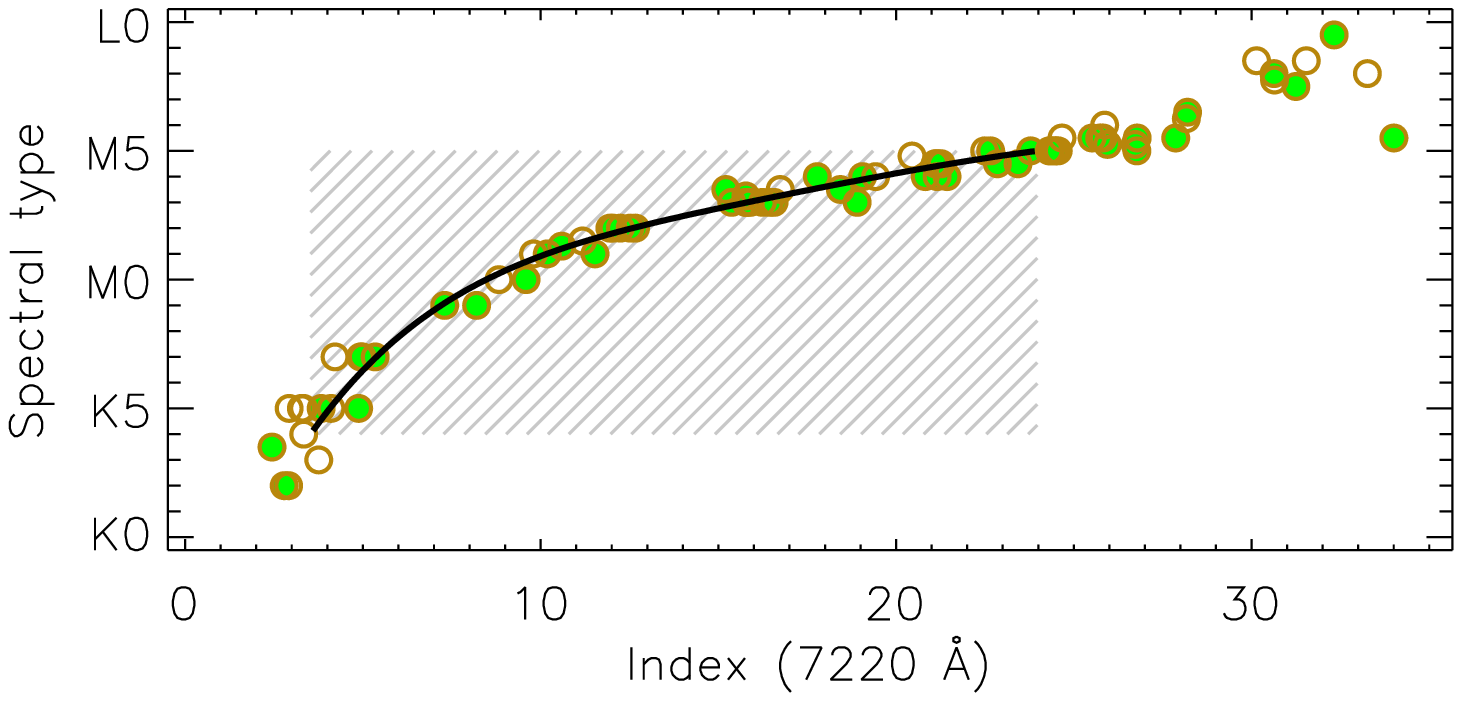}
\includegraphics[width=0.45\columnwidth]{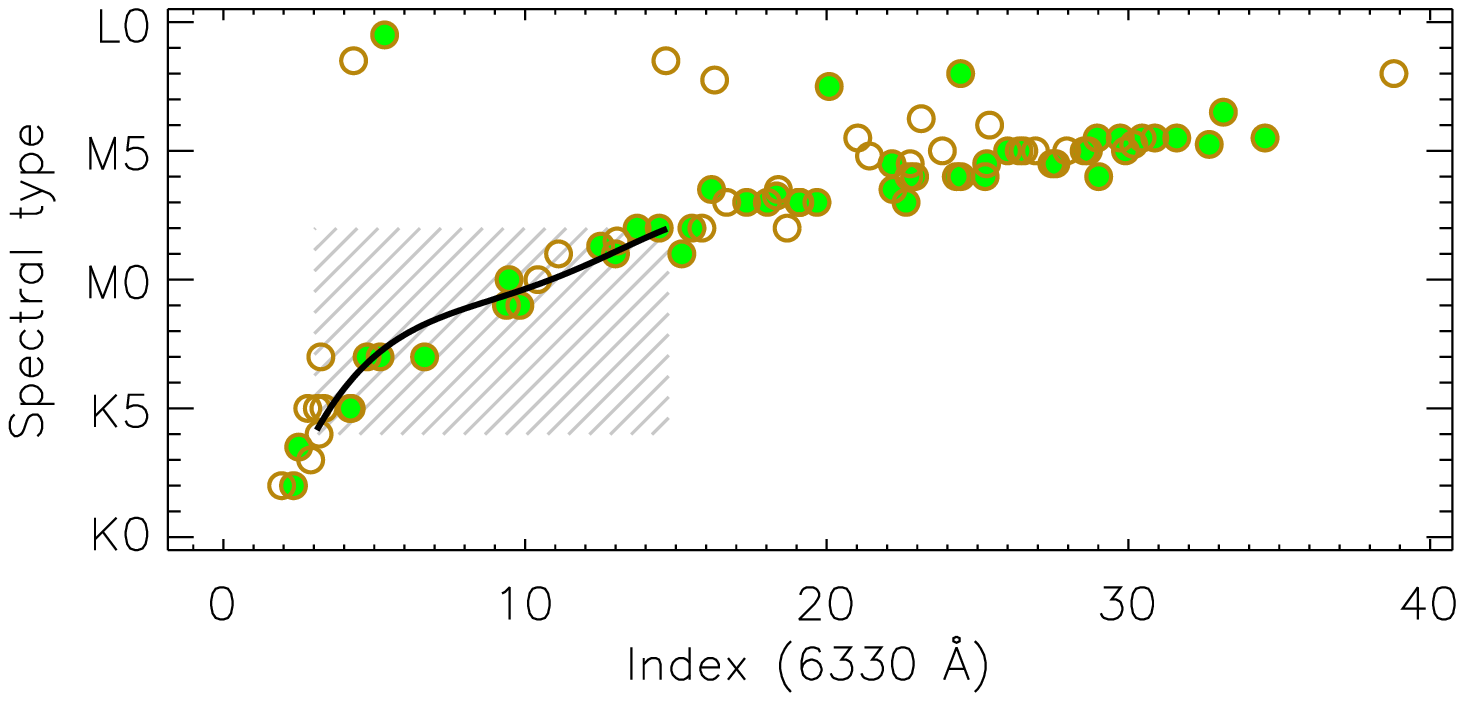}
\includegraphics[width=0.45\columnwidth]{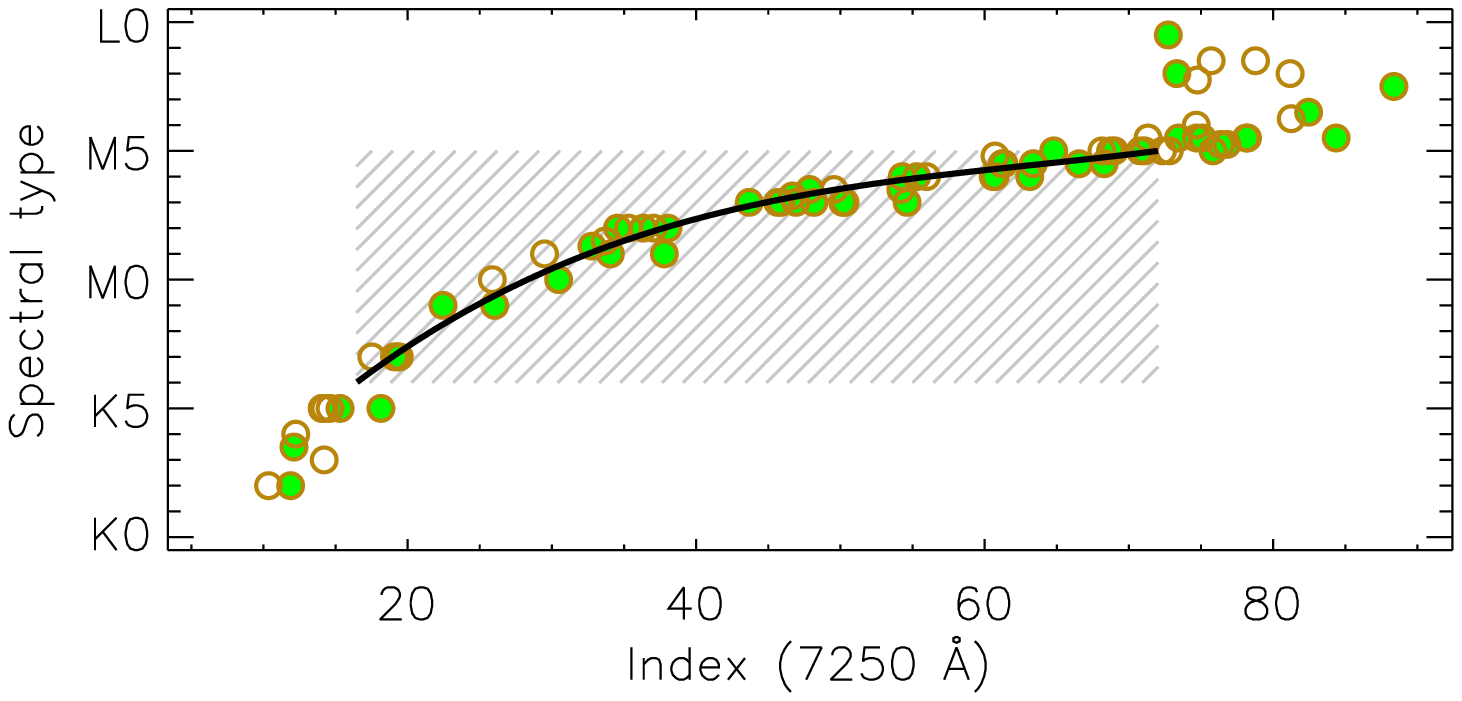}
\includegraphics[width=0.45\columnwidth]{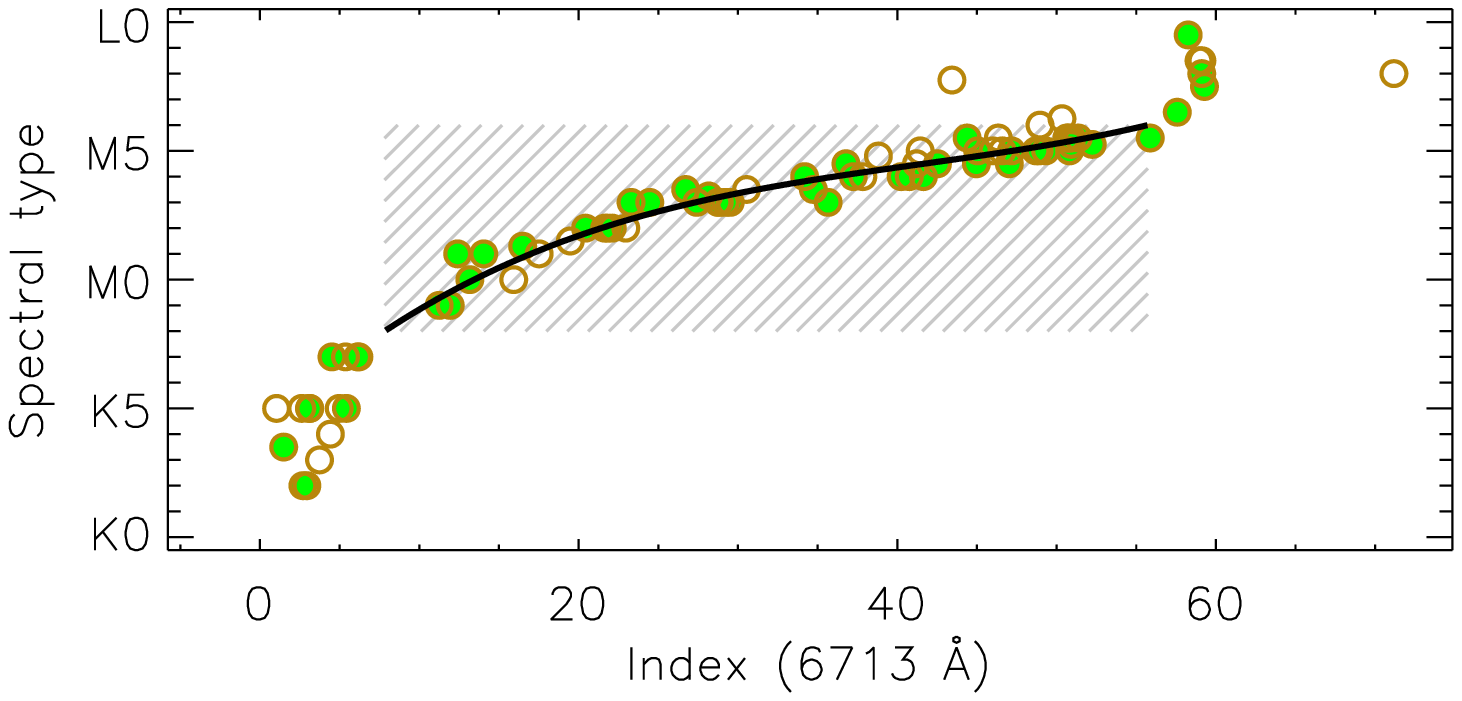}
\includegraphics[width=0.45\columnwidth]{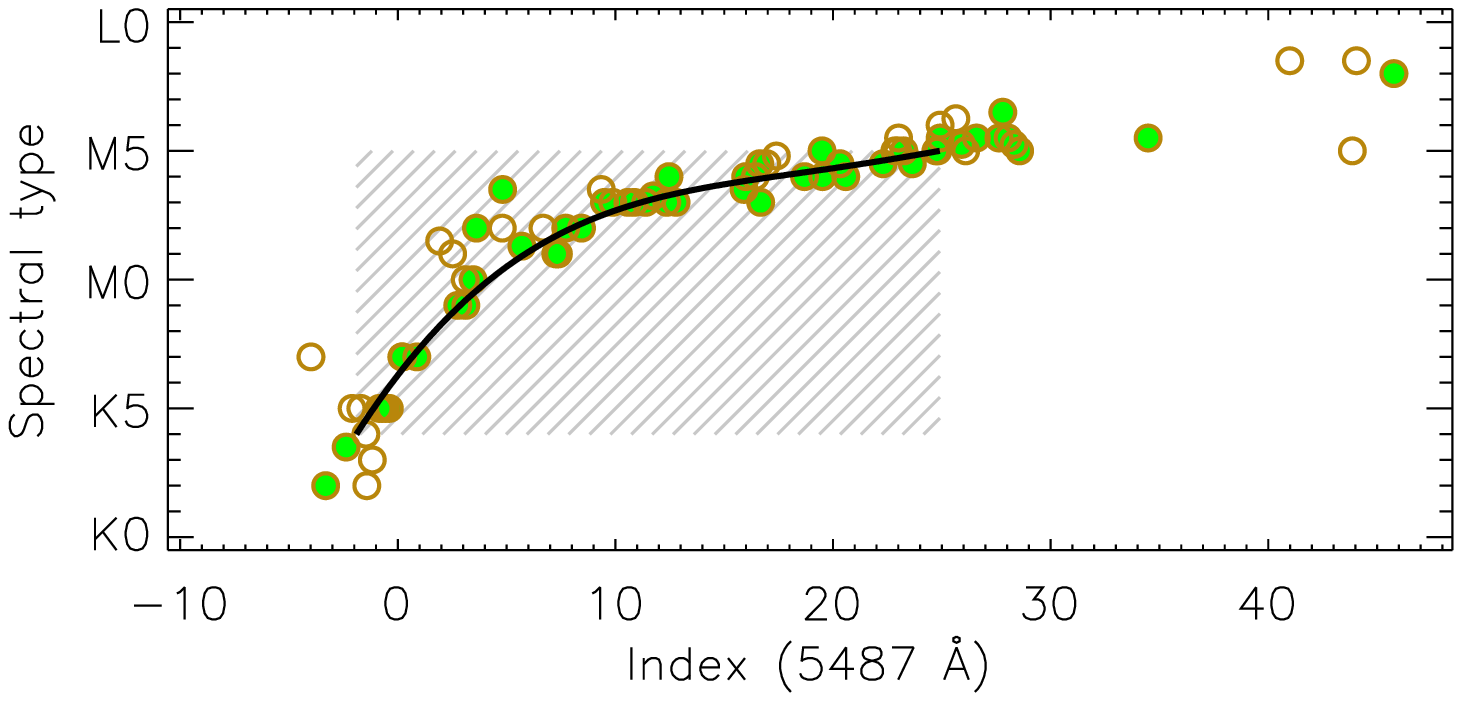}
\caption{Same as in Fig.~\ref{Fig:Index1}, but for different spectral features. \label{Fig:Index2}}
\end{figure*}

Where $\Delta\lambda_{\rm FB}$ is the width of the FB,  $F_{\rm FB}$ is the average flux over the width of the FB. We calculated the indices for 22 FBs. In Table~4, we list these features including $\lambda_{\rm FB}$, $\Delta\lambda_{\rm FB}$, $\lambda_{\rm BCB}$, $\lambda_{\rm RCB}$, and the widths of the BCBs and RCBs. In Fig.~\ref{Fig:Index1} and ~\ref{Fig:Index2}, we show the relations between the indices of individual spectral features and the spectral types. We fit these relations using polynomial functions, $Spt=C_{0}+C_{1}\times Index (FB)+C_{2}\times (Index (FB))^2+C_{3}\times (Index (FB))^3+C_{4}\times (Index (FB))^4+C_{5}\times (Index (FB))^5$, where $Spt$ is spectral-type number, and the numbers 0--19 are corresponding to K0 to M9. The results are listed in Table~4. To fully cover the range of spectral types, we have included some sources with low accretion rates. For these sources, their spectral features at wavelengths longer than 5500\,\AA\ are used for the fitting, to avoid the possible accretion-related veiling in the blue part of the optical spectra.

\renewcommand{\tabcolsep}{0.02cm}
\begin{landscape}
\begin{longtable}{cccccccccccccccccccccc}
\caption{The parameters for WTTSs \label{Tab:source}}\\
\hline \hline
    &    & RA          &DEC             &        &           &Adopted  &      &EW$_{\rm H\alpha}$ &      &                          &A$_{\rm V}$  &           \\
ID & Name    & (J2000)     &(J2000)         & Spt    &Ref        &Spt      &Ref   &(\AA)            &$L_{\rm acc}/L_{\star}$ &Accreting &(mag)   &Ref \\  
\hline
\endfirsthead
\hline
\hline
  &    & RA          &DEC             &        &           &Adopted  &      &EW$_{\rm H\alpha}$ &      &                          &A$_{\rm V}$  &           \\
ID & Name    & (J2000)     &(J2000)         & Spt    &Ref        &Spt      &Ref   &(\AA)            &$L_{\rm acc}/L_{\star}$ &Accreting &(mag)   &Ref \\  
\hline
\endhead
\hline\endfoot
1 & RECX\,1 &08 36 56.24  &$-$78 56 45.7   &K4, K5, K6, K7 &1, 2, 3, 8 &K5       &3     &$-$1.2     &\nodata               &N          &0   & \\
2  & RECX\,5 &08 42 27.088 &$-$78 57 47.93  &M3.8, M4   &2, 3, 8        &M4       &3     &$-$12.9    &\nodata               &N         &0   & \\
3&RECX\,6&08 42 38.770 &$-$78 54 42.75  &M2, M3  &2, 3, 8, 9         &M3       &3     &$-$4.8     &\nodata               &N            &0   & \\
4&RECX\,7 &8 43 07.239  &$-$79 04 52.49  &K3, K4, K5, K6.9 &2, 3, 8, 10   &K5  &3     &$-$1.0     &\nodata               &N            &0   & \\
5&RECX\,9 &08 44 16.41  &$-$78 59 08.04  &M4, M4.5   &2, 3, 8         &M4.5      &3     &$-$12.4    &\nodata               &N            &0   & \\
6&RECX\,10&08 44 31.90  &$-$78 46 31.2   &K7, K9, M0.3, M1  &2, 3, 8, 9  &K9   &3     &$-$1.7     &\nodata               &N            &0   & \\
7&RECX\,11&08 47 01.63  &$-$78 59 34.37  &K5, K5.5, K6.5   &2, 3, 8     &K5    &3     &$-$8.9     &0.007                 &Y            &0   &28 \\
8&RECX\,12  &08 47 56.766  &$-$78 54 53.19  &M2, M3.2, M3.25   &2, 3, 8, 9   &M3.25   &3  &$-$8.1 &\nodata               &N            &0   & \\
9&RECX\,17&08 38 51.50  &$-$79 16 13.7   &M5, M5.25, M5.75   &2, 3, 8     &M5.25    &3      &$-$12.1  &\nodata           &N            &0    & \\
10&RECX\,18&08 36 10.73  &$-$79 08 18.4   &M5.5       &2, 3, 8    &M5.5   &3                &$-$12.5    &\nodata          &N           &0    & \\
11&TWA\,2A &11 09 13.80  &$-$30 01 39.9   &M1.5, M2   &3, 4, 5  &M2       &3     &$-$2.5           &\nodata               &N           &0    & \\
12&TWA\,7  &10 42 30.064 &$-$33 40 16.62  &M2, M3     &3, 6     &M3       &3     &$-$5.2           &\nodata               &N           &0    & \\
13&TWA\,8A &11 32 41.25  &$-$26 51 55.9   &M2, M3     &3, 6, 7  &M3       &3     &$-$8.8           &\nodata               &N           &0    & \\
14&TWA\,8B &11 32 41.16  &$-$26 52 09.0   &M5, M5.5   &3, 7     &M5.5     &3     &$-$17.2          &\nodata               &N           &0    & \\
15&TWA\,9A &11 48 24.223 &$-$37 28 49.15  &K5, K7     &3, 7     &K7       &3     &$-$1.7           &\nodata               &N           &0    & \\
16&TWA\,9B &11 48 23.73  &$-$37 28 48.5   &M1, M3.5  &3, 4       &M3.5      &3   &$-$5.4          &\nodata                &N           &0    & \\
17&TWA\,13A &11 21 17.24  &$-$34 46 45.5   &M1           &3, 6       &M1      &3 &$-$8.4          &\nodata                &N           &0    & \\ 
18&TWA\,13B &11 21 17.24  &$-$34 46 45.5   &M1           &3, 6       &M1      &3 &$-$2.0          &\nodata                &N           &0    & \\ 
19&TWA\,22 &10 17 26.89  &$-$53 54 26.5   &M5         &3          &M5      &3  &$-$13.0           &\nodata                &N           &0    & \\
20&TWA\,25 &12 15 30.72  &$-$39 48 42.6   &K9, M0     &3, 4     &K9       &3     &$-$3.0          &\nodata                &N           &0.4  & \\
21&TWA\,26&11 39 51.140 &$-$31 59 21.50   &M8      &3, 6  &M8       &3      &$-$12.3              &\nodata                &N           &0    & \\
22&TWA\,27&12 07 33.467 &$-$39 32 54.00   &M8      &3, 6  &M8       &3      &$-$194.4             &0.003                  &Y           &0    &43 \\
23&TWA\,28&11 02 09.833 &$-$34 30 35.53   &M8.5    &3, 13     &M8.5 &3      &$-$101.4             &0.0093                 &Y           &0    &43 \\
24&TWA\,29 &12 45 14.16  & $-$44 29 07.7  &M9.5        &3, 4, 12    &M9.5  &3      &$-$10.6       &\nodata                &N           &0                & \\
25&TWA\,30A&11 32 18.314 &$-$30 19 51.85     &M5          &3, 18     &M5       &3      &$-$6.6    &\nodata                &N           &0                & \\
26&V4046\,sgr&18 14 10.466&$-$32 47 34.50 &K4, K5         &3, 17     &K4         &3      &$-$44.0 &0.096               &Y              &0          &42 \\
27&RECX\,4 &08 42 23.77  &$-$79 04 03.0   &M0, M1.3, M1.75  &2, 3, 8     &M1.3     &2     &$-$4.2 &\nodata             &N                &0       & \\
28&TWA\,3A &11 10 27.81  &$-$37 31 53.2   &M3, M4       &5, 6       &M4       &5     &$-$9.8      &\nodata        &N                   &0          & \\
29&TWA\,3B &11 10 27.88   &$-$37 31 52.0  &M3.5, M4     &5, 6       &M4       &5     &$-$6.2      &\nodata        &N                   &0         & \\
30&TWA\,6  &10 18 28.700 &$-$31 50 02.85  &K7, M0        &4, 11  &K7         &11      &$-$4.5     &\nodata        &N                   &0         & \\
31&TWA\,14 &11 13 26.221 &$-$45 23 42.74  &M0, M0.5     &4, 11      &M0      &11      &$-$6.1     &\nodata        &N                   &0         & \\
32&TWA\,15A &12 34 20.65 &$-$48 15 13.50   &M2           &  6       &M3      &this work &$-$12.8          &\nodata                &N        &0                 & \\ 
33&TWA\,15B &12 34 20.47  &$-$48 15 19.50   &M2           & 6       &M3      &this work &$-$9.5          &\nodata                &N         &0                & \\ 
34&Sz\,74  &15 48 05.228 &$-$35 15 52.83  &M1.5, M3.5   &14, 15   &M3.5      &14      &$-$22.8    &0.030          &Y                  &1.5             &14 \\
35&Sz\,84  &15 58 02.53  &$-$37 36 02.7   &M5, M5.5     &14, 15    &M4.5   &this work         &$-$113.2    &0.021          &Y               &0.8              &25 \\
36&Sz\,91  &16 07 11.592 &$-$39 03 47.54  &M0.5, M1, M1.5    &14, 15, 16    &M1    &14      &$-$139.6 &0.051       &Y                 &1.2           &14 \\
37&Sz\,94  &16 07 49.596 &$-$39 04 28.79  &M4        &4, 15      &M3.5         &this work      &$-$6.4      &\nodata         &N             &0.0            & \\
38&Sz\,97  &16 08 21.803 &$-$39 04 21.48  &M3, M4  &14, 15&M4       &14             &$-$38.6     &0.007            &Y                 &0           &14 \\
39&Sz\,100 &16 08 25.764 &$-$39 06 01.19  &M5, M5.5&14, 15&M5.5     &14             &$-$45.1      &0.006           &Y                 &0           &14 \\
40&Sz\,104 &16 08 30.815 &$-$39 05 48.87  &M5,     &14, 15&M5       &14             &$-$34.1      &0.006           &Y                 &0           &14 \\
41&Sz\,107 &16 08 41.799 &$-$39 01 37.02  &M5.5, M5.75 &4, 19     &M5.5    &4       &$-$12.8      &\nodata         &N                 &0           & \\
42&Sz\,111 &16 08 54.687 &$-$39 37 43.11  &M1, M1.5    &14, 15      &M0    &this work      &$-$79.7      &0.019            &Y               &0.7         &14 \\
43&Sz\,112 &16 08 55.530 &$-$39 02 33.95  &M4, M5, M6      &14, 15, 19      &M5   &14 &$-$17.4    &\nodata         &N                 &0           & \\
44&Sz\,114 &16 09 01.850 &$-$39 05 12.42  &M4, M4.8, M5.5      &14, 15, 20  &M4.8   &14 &$-$94.4  &0.010           &Y                 &0           &14 \\
45&Sz\,115 &16 09 06.214 &$-$39 08 51.86  &M4, M4.5  &14, 15   &M4.5  &14            &$-$10.9     &\nodata         &N                 &0           & \\
46&Sz\,121 &16 10 12.199 &$-$39 21 18.11  &M3, M4     &4, 15, 24     &M4     &24      &$-$7.8     &\nodata         &N                 &0           & \\
47&Sz\,122 &16 10 16.424 &$-$39 08 05.07  &M2         &4, 15, 24    &M2         &4      &$-$6.9   &\nodata         &N                 &0           & \\
48&SO\,587 &05 38 34.04  &$-$02 36 37.3   &M3.5, M4.5  &21, 22     &M4.5   &22         &$-$16.3   &\nodata         &N                 &0           & \\
49&SO\,641 &05 38 38.57  &$-$02 41 55.8   &M5        &4, 22      &M5   &22      &$-$8.7           &\nodata         &N                 &0           & \\
50&SO\,797 &05 38 54.91  &$-$02 28 58.19  &M4, M4.5   &4, 22        &M4.5       &22   &$-$7.7     &\nodata         &N                 &0           & \\
51&SO\,879 &05 39 05.42  &$-$02 32 30.34  &K5, K7     &4, 21, 24    &K7         &24   &$-$2.2     &\nodata        &N                  &0           & \\
52&SO\,925 &05 39 11.41  &$-$02 33 32.8   &M5.5    &4, 22  &M5.5    &22      &$-$9.6              &\nodata        &N                  &0           & \\
53&SO\,999 &05 39 20.25  &$-$02:38 25.8   &M5.5    &4, 22  &M5.5     &22      &$-$11.0            &\nodata         &N                 &0           & \\
54&Par-Lup3-1 &16 08 16.03 &$-$39 03 04.29&M6.5, M7.5&23, 24    &M6.5     &24      &$-$17.6       &\nodata         &N                 &0           & \\
55&Par-Lup3-2 &16 08 35.78 &$-$39 03 47.91 &M5, M6 &23, 24    &M5      &24      &$-$5.5           &\nodata         &N                 &0           & \\
56&Par-Lup3-3 &16 08 49.40	&$-$39 05 39.2  &M4, M4.5 &14, 23     &M3.5      &this work      &$-$28.2       &\nodata         &N         &3.5         & \\
57&SST-Lup3-1 &16 11 59.798 &$-$38 23 38.34 &M5        &14      &M5         &14      &$-$44.4     &0.004           &Y                 &0           &14 \\
58&Lup706     &16 08 37.30  &$-$39 23 10.8   &M7.75   &14        &M7.75   &14         &$-$222.5   &0.005           &Y                 &0           &14 \\
59&Lup604s    &16 08 00.20  &$-$39 02 59.7  &M5.25    &14        &M5.25   &14         &$-$16.7   &\nodata          &N                 &0           & \\
60&Lup818s    &16 09 56.29  &$-$38 59 51.7  &M6       &14        &M6      &14         &$-$52.3    &0.003           &Y                 &0           &14 \\
61&CrA75      &19 02 22.1   &$-$36 55 40.9  &K2       &25        &K2      &25         &$-$1.3    &\nodata          &N                 &0.3         & \\
62&ISO-217    &11 09 52.15  &$-$76 39 12.8  &M6.25    &26        &M6.25      &26      &$-$125.2  &0.007                 &Y            &2.6         &41 \\
63&AKC2006-19 &15 44 57.90 &$-$34 23 39.5 &M5  &14    &M5      &14                    &$-$29.8   &0.005            &Y                 &0           &14 \\
64&2MASSJ16085953-3856275 &16 08 59.53  &$-$38 56 27.6 &M8.5 &14      &M8.5 &14      &$-$111.5   &0.003            &Y                 &0           &14 \\
65&SSTc2d160901.4-392512 &16 09 01.40  &$-$39 25 11.9 &M4    &14, 19     &M3   &this work   &$-$42.7    &0.007            &Y                 &0.8           &14 \\
66&Cha\,H$\alpha$\,1   &11 07 16.68 &$-$77 35 53.2  &M7.5, M7.75   &26, 27       &M7.75      &26    &$-$125.9    &\nodata   &N         &0          &40 \\
67&Ass-Cha-T-2-51 &11 12 24.415 &$-$76 37 06.41 &K3.5, K4   &10, 26     &K3.5         &26      &$-$2.9  &\nodata   &N                  &0          & \\
68&LkCa\,15    &04 39 17.796  &+22 21 03.48       &K2, K5    &25, 29, 30           &K5      &29  &$-$28.3          &0.066  &Y          &0.6          &25\\
69&CS\,Cha    &11 02 24.912   &$-$77 33 35.72     &K2, K4, K5, K6     &6, 25, 26, 31  &K5    &31  &$-$29.5         &0.069  &Y          &0.3          &25\\
70&CHXR\,22E  &11 07 13.300   &$-$77 43 49.88     &M3.5, M4           &25, 26         &M4    &25  &$-$6.4          &\nodata  &N        &3.4           &    \\
71&Sz\,18     &11 07 19.154   &$-$76 03 04.85     &M2, M2.5           &25, 26         &M2    &25  &$-$25.0         &0.048         &Y   &0.6                &25    \\
72&Sz\,27     &11 08 39.051   &$-$77 16 04.24     &K7, K8             &25, 26, 29     &K7    &25  &$-$58.7         &0.076         &Y   &2.8                &25    \\
73&RX\,J1615$-$3255 &16 15 20.231 &$-$32 55 05.10 &K5, K7             &25, 32, 33     &K5    &32  &$-$35.6         &0.056         &Y   &0               &25 \\
74&Oph\,22    &16 22 45.40   &$-$24 31 23         &M3                 &25, 34         &M3    &25  &$-$5.0          &\nodata       &N   &1.4                &\\  
75&Oph\,24    &16 25 06.91   &$-$23 50 50.3       &M0, M3             &25, 34         &M2    &this work  &$-$5.1          &\nodata       &N   &0                &\\  
76&ISO-Oph196 &16 28 16.51   &$-$24 36 57.9       &M4.5, M5.5         &25, 35         &M5    &this work  &$-$106.8        &0.062         &Y   &2.0                 &25\\  
77&Ser\,29    &18 29 11.50   &00 20 38.6          &M0, M2             &25, 36             &M2    &25  &$-$13.9                &$<$0.004       &Y &2.9                 &25\\    
78&Ser\,34    &18 29 44.11   &00 33 56.0          &M0, M1             &25, 36            &M1.5    &this work  &$-$13.6                &0.008       &Y        &2.3            &25\\   
79&RX\,J1842.9$-$3532 &18 42 57.95  &$-$35 32 42.7  &K2               &25, 37            &K2    &37  &$-$32.9                &0.056              &Y     &1.7         &25\\
80&RX\,J1852.3$-$3700 &18 52 17.29  &$-$37 00 11.9  &K2, K3           &25, 37            &K3    &37  &$-$44.8                &0.051              &Y     &1.6          &25\\
81&LkH$\alpha$\,330 &03 45 48. &29 32 24 11.9    &G3, G4              &25, 38           &G4    &25   &$-$15.8       &0.024                   &Y   &2.9  &25\\         
82&SR\,21         &16 27 10.28 & $-$24 19 12.7   &G4                  &25               &G4    &25   &1.3      &0.017                        &Y   &5.2  &25\\
83&T21        &11 06 15.4  &$-$77 21 56.9        &G5                  &25, 26, 29          &G5    &25   &$-$0.5                       &\nodata    &N &3.9  & \\
84&IC348-127  &03 45 07.9  &32 04 01.8           &G4                  &25, 39              &G4    &25   &$-$2.8                       &\nodata     &N &6.2 &\\
85&Cha\,H$\alpha$\,9 &11 07 18.608  &$-$77 32 51.66&M5.5                &30                  &M5.5  &30   &$-$20.1                      &\nodata   &N &5.0  &40\\
\hline
\end{longtable}
\let\thefootnote\relax{$^{a}$ The reference for the available spectral types of individual sources, $^{b}$ The reference for the adopted spectral type, $^{c}$ The reference for the accretion luminosity.\\ References: 1. \citet{1997A&A...328..187C}, 2. \citet{2004MNRAS.355..363L}, 3. \citet{2013ApJS..208....9P}, 4. \citet{2013A&A...551A.107M}, 5. \citet{2006AJ....132..866R}, 6. \citet{2006A&A...460..695T}, 7. \citet{1999ApJ...512L..63W}, 8. \citet{2004ApJ...609..917L}, 9. \citet{1999ApJ...516L..77M}, 10. \citet{1997A&A...328..187C}, 11. \citet{2003AJ....125..825T}, 12. \citet{2007ApJ...669L..97L}, 13. \citet{2008A&A...489..825T}, 14. \citet{2014A&A...561A...2A}, 15. \citet{1994AJ....108.1071H}, 16. \citet{2012ApJ...749...79R}, 17. \citet{2008A&A...480..735F}, 18. \citet{2010ApJ...714...45L}, 19. \citet{2012ApJ...749...79R}, 20. \citet{2009A&A...500.1045C}, 21. \citet{2008A&A...488..167S}, 22. \citet{2012A&A...548A..56R}, 23. \citet{2003A&A...406.1001C}, 24. \citet{2013A&A...558A.141S}, 25. \citet{2014arXiv1406.1428M}, 26.\citet{2004ApJ...602..816L}, 27. \citet{2000A&A...359..269C}, 28. \citet{2013ApJ...767..112I}, 29. \citet{2012ApJ...745..119N}, 30. \citet{2001ApJ...556..265W}, 31. \citet{1988cels.book.....H}, 32. \citet{2010ApJ...724..835W}, 33. \citet{1997A&AS..123..329K}, 34. \citet{2010ApJ...712..925C}, 35. \citet{2005AJ....130.1733W}, 36. \citet{2009ApJ...691..672O}, 37. \citet{2007AJ....133.2524W}, 38. \citet{1979ApJS...41..743C}, 39. \citet{2007ApJ...667..308C}, 40. \citet{2005ApJ...626..498M}, 41. \citet{2005ApJ...625..906M}, 42. \citet{2011MNRAS.417.1747D}, 43. \citet{2008ApJ...681..594H}}

\clearpage
\end{landscape}
\normalsize


\renewcommand{\tabcolsep}{0.02cm}
\begin{table}
\caption{Optical feature for spectral typing \label{Tab:FB}}
\tiny
\begin{tabular}{cccccccccccccccccccccccc}
\hline\hline
&     &     Center       &Width       &$C_{\rm left}$  &Width    &$C_{\rm right}$ &Width  &Spt &Index    &           \\
Index &Feature &(\AA) &(\AA)          &(\AA)          &(\AA)    &(\AA)         &(\AA) &Range&Range   &$C_{\rm 0}$ &$C_{\rm 1}$ & $C_{\rm 2}$ &$C_{\rm 3}$ &$C_{\rm 4}$  &$C_{\rm 5}$\\
\hline
1     &VO      &7810     &110         &7570           &25       &8140          &20    &M0--M8  &11.51 to 66.23  & 5.44607&  0.507010 &  $-$1.06971$\times10^{-2}$ & 8.91401$\times10^{-5}$   &    0    &  0 \\
2     &VO     &7450     &40          &7540           &20       &7540          &20    &M3--M9.5&2.97 to 13.62 & 7.40348  &  2.63063  &  $-$0.284250  & 1.12802$\times10^{-2}$     &    0    &  0 \\ 
3     &VO      &7920     &110         &7562           &20       &8130          &20    &M2--M9.5&13.31 to 71.70& 8.56015  &  0.310141 &  $-$3.77074$\times10^{-3}$& $-$1.51972$\times10^{-5}$& 4.95808$\times10^{-7}$ &0\\
4     &VO      &8500     &110         &8408           &10       &8840          &10    &M3--M9  &12.60 to 61.72& 9.46791 &   0.377978 &  $-$8.78991$\times10^{-3}$& 8.24764$\times10^{-5}$   &    0    &  0 \\
5     &VO      &8675     &150         &8408           &12       &8840          &10    &M4--M9.5&13.54 to 58.77& 10.0762 &   0.410874 &  $-$1.03032$\times10^{-2}$ & 1.00317$\times10^{-4}$ &    0    &  0 \\ 
6     &VO      &8880     &35          &8850           &10       &9045          &10    &M3--M9  &2.87 to 11.90& 8.62001 &   2.16112  &  $-$0.256581  & 1.24604$\times10^{-2}$     &    0    &  0 \\
7     &TiO     &7108     &30          &7040           &15       &7326          &15    &K7--M6  &2.37 to 18.62& 3.61125 &   1.62902  &  $-$8.76069$\times10^{-2}$ & 1.92530$\times10^{-3}$    &    0    &  0 \\
8     &TiO     &6255     &45          &6112           &15       &6525          &15    &K4--M5  &4.52 to 25.74& $-$9.61363& 4.82202  &  $-$0.522211  & 3.05772$\times10^{-2}$     & $-$8.68462$\times10^{-4}$ & 9.40336$\times10^{-6}$\\
9     &TiO     &4812     &18          &4742           &10       &4938          &15    &K7--M6  &0.01 to 10.68& 6.98046 &   2.31755  &  $-$0.250489  &  1.05380$\times10^{-2}$    &    0    &  0 \\
10    &TiO     &5225     &75          &4940           &15       &5415          &15    &K4--M5  &8.65 to 20.53& $-$5.56706& 0.852030 &     4.54304$\times10^{-2}$ & $-$1.85754$\times10^{-3}$ &    0    & 0  \\
11    &CaH     &6964     &22          &6533           &10       &7032          &10    &K6--M4  &0.78 to 4.76& 3.19107 &   3.86115  &  $-$0.334198  &    0          &    0    &  0 \\
12    &CaH     &6840     &30          &6533           &10       &7032          &10    &K6--M5  &3.04 to 15.68& 1.23510 &   1.83375  &  $-$9.31380$\times10^{-2}$ &  2.05201$\times10^{-3}$   &    0    &  0 \\
13    &TiO     &4975     &40          &4940           &15       &5385          &15    &K5--M6  &2.53 to 20.70& $-$5.00123&  5.58890 &  $-$0.783747  &  5.93566$\times10^{-2}$    & $-$2.27083$\times10^{-3}$ &3.46253$\times10^{-5}$\\
14    &TiO     &7070     &30          &7045           &10       &7385          &15    &K4--M5  &1.90 to 13.59& $-$8.50191&   9.52190&  $-$1.87384   &  0.188538     &$-$9.13227$\times10^{-3}$  & 1.69261$\times10^{-4}$\\
15    &TiO     &5950     &60          &5810           &20       &6080          &30    &K6--M8.5&1.88 to 29.18& 3.54680 &   1.43835  &  $-$7.45602$\times10^{-2}$ &  1.44120$\times10^{-3}$   &    0    &  0 \\
16    &TiO     &6800     &50          &6530           &15       &7020          &30    &K4--M5  &2.80 to 26.72& $-$2.75493 & 3.21562 &  $-$0.336413  &  1.91640$\times10^{-2}$ &$-$5.29676$\times10^{-4}$ & 5.61103$\times10^{-6}$\\
17    &TiO     &7150     &50          &7045           &15       &7400          &20    &K7--M5  &7.14 to 33.44&2.22419  &  0.788315  &  $-$1.79554$\times10^{-2}$ &  1.73645$\times10^{-4}$  &    0    &  0 \\
18    &TiO     &7220     &50          &7045           &15       &7315          &15    &K4--M5  &3.53 to 23.97&$-$5.94651 & 3.86310  &  $-$0.351318  &  1.70392$\times10^{-2}$   & $-$4.04817$\times10^{-4}$ & 3.68664$\times10^{-6}$ \\ 
19    &TiO     &6330     &100         &6080           &15       &6458          &30    &K4--M2  &3.02 to 14.76&$-$6.93018 &  5.45853 &  $-$0.741166  &  4.67135$\times10^{-2}$    &$-$1.06148$\times10^{-3}$ &   0 \\
20    &TiO     &7250     &150         &7045           &15       &7531          &50    &K6--M5  &16.44 to 72.03&$-$3.22166 &  0.716094&  $-$1.03322$\times10^{-2}$ & 5.41808$\times10^{-5}$   &   0  &    0\\
21    &TiO     &6713     &100         &6530           &20       &6940          &30    &K8--M6  &7.82 to 55.73&4.18776    & 0.573689 &  $-$1.17938$\times10^{-2}$ & 9.51566$\times10^{-5}$   &   0  &    0\\
22    &TiO     &5487     &76          &5385           &10       &5385          &10    &K4--M5  &$-$1.90 to 24.91&6.28645    &  1.09733 &  $-$5.51680$\times10^{-2}$ & 7.34970$\times10^{-4}$   & 2.72954$\times10^{-5}$& $-$6.52506$\times10^{-7}$ \\ 
\hline\hline
\end{tabular}
\end{table}
\normalsize

\clearpage
\begin{center}
\begin{longtable}{lcccccccccccccccccccccccccccccccccccc}
\caption{Stellar properties for the YSOs}\label{Tab:spt}\\
\hline\hline \multicolumn{1}{c}{} &\multicolumn{1}{c}{RA} &\multicolumn{1}{c}{DEC} &\multicolumn{1}{c}{}&\multicolumn{1}{c}{Lum} &\multicolumn{1}{c}{Av} &\multicolumn{1}{c}{Mass} &\multicolumn{1}{c}{Age}&\multicolumn{1}{c}{$EW_{\rm Li \lambda6708\AA}$} &\multicolumn{1}{c}{$EW_{\rm H\alpha}$}  &\multicolumn{1}{c}{} &\multicolumn{1}{c}{} &\multicolumn{1}{c}{} &\multicolumn{1}{c}{} &\multicolumn{1}{c}{}&\multicolumn{1}{c}{}\\
\multicolumn{1}{c}{} &\multicolumn{1}{c}{(J2000)} &\multicolumn{1}{c}{(J2000)} &\multicolumn{1}{c}{Spt}&\multicolumn{1}{c}{($L_{\odot}$)} &\multicolumn{1}{c}{(mag)} &\multicolumn{1}{c}{($M_{\odot}$)} &\multicolumn{1}{c}{Myr}&\multicolumn{1}{c}{(\AA)} &\multicolumn{1}{c}{(\AA)} &\multicolumn{1}{c}{$\alpha_{3.6-8.0}$} &\multicolumn{1}{c}{$\alpha_{3.6-24}$}  &\multicolumn{1}{c}{TTS\tablenotemark{a}} &\multicolumn{1}{c}{Disk\tablenotemark{b}}  &\multicolumn{1}{c}{Criteria\tablenotemark{c}} \\ \hline
\endfirsthead
\multicolumn{12}{c}{{\bfseries \tablename\ \thetable{} -- continued}}\\
\hline\hline \multicolumn{1}{c}{} &\multicolumn{1}{c}{RA} &\multicolumn{1}{c}{DEC} &\multicolumn{1}{c}{}&\multicolumn{1}{c}{Lum} &\multicolumn{1}{c}{Av} &\multicolumn{1}{c}{Mass} &\multicolumn{1}{c}{Age}&\multicolumn{1}{c}{$EW_{\rm Li \lambda6708\AA}$} &\multicolumn{1}{c}{$EW_{\rm H\alpha}$}  &\multicolumn{1}{c}{} &\multicolumn{1}{c}{} &\multicolumn{1}{c}{} &\multicolumn{1}{c}{} &\multicolumn{1}{c}{}&\multicolumn{1}{c}{}\\
\multicolumn{1}{c}{} &\multicolumn{1}{c}{(J2000)} &\multicolumn{1}{c}{(J2000)} &\multicolumn{1}{c}{Spt}&\multicolumn{1}{c}{($L_{\odot}$)} &\multicolumn{1}{c}{(mag)} &\multicolumn{1}{c}{($M_{\odot}$)} &\multicolumn{1}{c}{Myr}&\multicolumn{1}{c}{(\AA)} &\multicolumn{1}{c}{(\AA)}  &\multicolumn{1}{c}{$\alpha_{3.6-8.0}$} &\multicolumn{1}{c}{$\alpha_{3.6-24}$}  &\multicolumn{1}{c}{TTS\tablenotemark{a}} &\multicolumn{1}{c}{Disk\tablenotemark{b}}  &\multicolumn{1}{c}{Criteria\tablenotemark{c}} \\ \hline
\endhead
\hline\endfoot
\\[-1.8ex] \hline \hline
\endlastfoot
1  & 05 31 50.98  &$-$07 00 05.7  &K8.7  &  0.610&  0.2&0.58&  1.5& 0.71 &$    -4.6$ &\nodata&\nodata&W&N&Li\\[0.3pt]
2  & 05 32 11.59  &$-$06 05 22.2  &M4.8  &  0.076&  0.0&0.12&  1.4& ... &$    -8.1$ &\nodata&\nodata&W&N&Na\\[0.3pt]
3  & 05 32 12.63  &$-$06 12 34.6  &M5.3  &  0.059&  0.0&0.096&  1.5& ... &$   -11.3$ &\nodata&\nodata&W&N&Na\\[0.3pt]
4  & 05 32 14.78  &$-$06 04 11.2  &M4.0  &  0.096&  0.0&0.18&  2.0& ... &$    -7.0$ &\nodata&\nodata&W&N&Na\\[0.3pt]
5  & 05 32 19.82  &$-$05 48 27.0  &M4.1  &  0.320&  0.0&0.18& $<0.5$& 0.33 &$    -6.2$ &\nodata&\nodata&W&N&Li, Na\\[0.3pt]
6  & 05 32 29.64  &$-$06 18 34.8  &M6.4  &  0.053&  0.0&0.065& $<0.5$& ... &$   -61.3$ &\nodata&\nodata&C&?&Na\\[0.3pt]
7  & 05 32 31.67  &$-$06 15 18.5  &M4.7  &  0.063&  0.0&0.13&  2.1& 0.32 &$    -6.4$ &\nodata&\nodata&W&N&Li, Na\\[0.3pt]
8  & 05 32 35.04  &$-$06 05 37.3  &M3.6  &  0.696&  0.0&0.21& $<0.5$& ... &$    -0.1$ &\nodata&\nodata&W&N&Na\\
\end{longtable}
\end{center}
\normalsize

\end{document}